\documentclass[final,5p,times,twocolumn]{elsarticle}
\usepackage{natbib}
\usepackage{array}
\usepackage{booktabs}
\usepackage{tabu}
\usepackage{dcolumn}
\usepackage{amsmath}
\usepackage{amsfonts}
\usepackage{amssymb}
\usepackage{graphicx}
\usepackage{subfigure}
\usepackage{epsfig}
\usepackage{dcolumn}% Align table columns on decimal point
\usepackage{bm}% bold math
\def\be{\begin{equation}}
\def\ee{\end{equation}}
\def\bea{\begin{eqnarray}}
\def\eea{\end{eqnarray}}

\journal{Physics of The Dark Universe}

\begin{document}

\begin{frontmatter}



\title{Irreversible thermodynamical description of warm inflationary cosmological models}


\author[AO,BBU,SYSU]{Tiberiu Harko}\ead{tiberiu.harko@aira.astro.ro}
\author[NWU,IPM]{Haidar Sheikhahmadi\corref{cor1}}\ead{h.sh.ahmadi@gmail.com;h.sheikhahmadi@ipm.ir}

\cortext[cor1]{Corresponding author}
\address[AO] {Astronomical Observatory, 19 Ciresilor Street, 400487 Cluj-Napoca, Romania,}
\address[BBU]{Department of Physics, Babes-Bolyai University, Kogalniceanu Street,
400084 Cluj-Napoca, Romania,}
\address[SYSU]{School of Physics, Sun Yat-Sen University, 510275 Guangzhou, People's
Republic of China}
\address[NWU]{Center for Space Research, North-West University, Mafikeng, South Africa}
\address[IPM]{School of Astronomy, Institute for Research in
Fundamental Sciences (IPM),  P. O. Box 19395-5531, Tehran, Iran}

\begin{abstract}
We investigate the interaction between scalar fields and radiation in the framework of warm inflationary models by using the irreversible thermodynamics of open systems with matter creation/annihilation.  We consider the scalar fields and radiation as an interacting  two component cosmological fluid in a homogeneous, spatially flat and isotropic Friedmann-Robertson-Walker (FRW) Universe. The thermodynamics of open systems as applied together with the gravitational field equations to the two component cosmological fluid leads to a generalization of the elementary scalar
field-radiation interaction model, which is the theoretical basis of warm inflationary models, with the decay (creation) pressures explicitly considered as parts of the cosmological fluid energy-momentum tensor. Specific models describing coherently oscillating scalar waves, scalar fields with a constant potential, and scalar fields with a Higgs type potential are considered in detail. For each case exact and numerical solutions of the gravitational field equations with scalar field-radiation interaction are obtained, and they show the transition from an accelerating inflationary phase to a decelerating one. The theoretical predictions of the warm inflationary scenario with irreversible matter creation are also compared in detail with the Planck 2018 observational data, and constraints on the free parameters of the model are obtained.
\end{abstract}

\begin{keyword}
Warm inflation \sep Particle creation \sep Irreversible thermodynamics
\PACS 02.30.Hq; 02.30.Mv; 02.30.Vv; 02.60.Cb

\end{keyword}

\end{frontmatter}

\tableofcontents


\section{Introduction}

One of the keystones of present day cosmology is represented by the inflationary paradigm, introduced in \cite{guth1981inflationary}, and representing one of the most influential theoretical models ever proposed in cosmology. Inflation requires the presence in the early Universe of a scalar field $\phi$, with a self-interaction potential $V(\phi)$, and having an energy density $\rho _{\phi}$, and a pressure $p_{\phi}$, respectively \cite{B6}. In the initial formulation of the inflationary model, the scalar field potential reaches a local minimum at $\phi=0$,  through supercooling from a phase transition. After that the Universe experiences an  exponential, de Sitter type expansion. But in this theoretical model, called "old inflation", there is no graceful exit to the rapidly accelerating, inflationary era. Hence several inflationary models, including the "new" and the chaotic inflationary models  have been proposed \cite{linde1982new, albrecht1982cosmology, linde1983chaotic}, with the explicit goal of solving the graceful exit problem. However, each of these models face their own theoretical problems. For recent reviews of different aspects of cosmology and the inflationary theory see \cite{revinfl5,revinfl6}.

Recently, high precision measurements of the Cosmic Microwave Background (CMB) radiation has given the possibility of  testing the crucial predictions of inflation on the primordial fluctuations, such as Gaussianity and scale independence \cite{Pl1,Pl2,Pl3,Pl4,Pl5,Haidar2}. From the CMB fluctuations one can determine the cosmological parameters, and then they can be used to constrain the inflationary models. Two important inflationary parameters, the slow-roll parameters $\epsilon$ and $\eta$ can be obtained immediately from the potentials of the scalar fields driving inflation. Since these parameters can be also determined observationally, it follows that the corresponding inflationary models can be fully tested \cite{liddle1992cobe, dodelson1997cosmic, kinney1998constraining, kinney2000new, kinney2008latest}.

A central result of inflationary scenarios has been the prediction of the statistical isotropy of the Universe. Nevertheless,  recent observations of the large scale structure of the Universe have suggested the prospect that the principles of homogeneity and isotropy may not be valid at all scales. Thus the presence of an inherent large scale anisotropy of cosmological origin in the Universe cannot be excluded {\it a priori} \cite{An}.  The CMB data may also indicate the existence of some tension between the predictions of the theoretical models and the observations. For example, the combined WMAP and Planck polarization data point towards  a numerical value of the index of the power spectrum as given  at the pivot scale $k_0 = 0.05$ Mpc$^{-1}$  by $n_s = 0.9603\pm 0.0073$ \cite{Pl2,Pl3}. This result, if correct, clearly excludes at more than $5\sigma$ the exact scale-invariance ($n_s = 1$), implied by some inflationary models \cite{Haw,Muk}.  The combined constraints on $r$ and $n_s$ impose strong limits on the inflationary theories. Thus, for example,  power-law potentials of the form  $\phi ^4$ cannot generate a satisfactory number of e-folds (of the order of  50-60) for inflation models in the restricted space  $r-n_s$ at a $2\sigma $ level \cite{Pl3}. In fact, presently the theories of inflation have not yet reached a general consensus and acceptance. One important reason for this is that not only the global inflationary description, but also the specific aspects of the theoretical models are based on scientific ideas beyond the Standard Model of particle physics.

The exponential growth of the size of the Universe during inflation leads to a homogeneous, isotropic but deserted Universe. Hence it is believed that radiation, elementary particles and different forms of matter have been created at the end of inflation,  in the period of reheating. Matter was created due to the transfer of energy from the inflationary scalar field to elementary particles. Reheating was initially suggested in the framework of the new inflationary scenario \cite{linde1982new}, and subsequently  developed in \cite{Alb, Kof, kofman1997towards}. The basic idea of reheating is as follows. After the rapid, de Sitter type expansion of the Universe, the inflationary scalar field reaches its minimum value. Then it begins to oscillate around the minimum of the potential, and it decays into matter, in the form of Standard Model elementary particles that interact with each other, finally attaining a thermal equilibrium state of temperature $T$.

 Quantum field theoretical approaches have led to a more complex view of reheating, involving the presence of initial, powerful particle creation through parametric resonance and inflaton, in a phase called preheating,  with the produced particles in a highly nonequilibrium distribution, which afterwards  relaxes to an equilibrium state. These explosive particle production processes are a result of the  spinodal instabilities in the broken symmetry phase, or of the parametric amplification of quantum fluctuations in the nonbroken symmetry case (as is the case for chaotic inflation) \cite{Pre1,Pre2,Pre3,Pre4}.

%The maximum temperature of the radiation dominated Universe is generally assumed to be the reheating temperature $T_{\text{reh}}$, reached after the end of the %reheating period. Nonetheless it  was claimed that the reheating temperature may not necessarily be the maximum temperature of the Universe %\cite{scherrer1985decaying}. In this approach reheating is just one of the several stages of particle production, and other matter components  with larger %freeze-out temperatures could have been created after inflation via other physical process \cite{giudice2001largest}.

In the study of reheating one of the widely investigated approaches is the phenomenological description introduced in \cite{Alb}. The basic idea  is the introduction of a specific loss term in the Klein-Gordon equation of the scalar field, which is also included as a source term for the energy density of the newly produced elementary particles. If suitably chosen, this loss/gain term is responsible for the reheating process that follows after the adiabatic supercooling during the inflationary era. Hence, in this  model,  the reheating dynamics is considered within a simple two component fluid model, the first
component being the scalar field, while the second component is represented by normal matter (radiation).  The transition process between the scalar field and the radiation component is described  phenomenologically  by a friction term,  characterizing the decay
of the inflaton field, and representing the source term for the newly created matter fluid. Different aspects of post inflationary reheating processes have been investigated for various gravitational theories in \cite{Re11,Re10,Re9,Re8,Re7,Re6,Re5,Re4,Re3,Re2,Re1, Reh1}. For reviews on the post inflationary reheating era of the evolution of the Universe see \cite{allahverdi2010reheating} and \cite{RehRev}, respectively.

The standard reheating scenario faces its own theoretical problems. One important issue is that the perturbative decay width can describe the decay of a scalar field close to the minimum of its potential only. This is obviously not the case during slow-roll inflation. Moreover, finite temperature effects can additionally enhance the rate at which the scalar field dissipates its energy to create new particles \cite{allahverdi2010reheating, RehRev}.

On the other hand the scalar field driving inflation  could have been coupled nonminimally to other components present in the early Universe, and therefore it could have dissipated its energy during the accelerated expansion, thus warming up the Universe. This version of inflation is called warm inflation, and it was initially proposed in \cite{w1,w2}. Hence, in the warm inflationary scenario,  dissipative effects and particle creation processes can generate a thermal bath during the accelerated expansion. In of the first warm inflation models \cite{w3} it was suggested that the physical parameters in an inflationary model may be randomly distributed. This lead to the so-called distributed-mass-model \cite{w4,w5,w6,w7}, which has been developed in the framework of string theory. Warm inflation has become a very active field of study, and it may represent a real alternative to the cold inflation/reheating paradigms. The physical properties and cosmological evolution in the warm inflationary models have been investigated in detail in \cite{w8}-\cite{w48}.
%\cite{w8,w9,w9a,w9b, w9c, w10,w11, w11a, w12, w12a, w13,w14,w15,w15a, w15b, w16,w17,w18,w19,w20,w21, w22, %w23,w24,w25,w26,w27,w28,w29,w30,w31,w32,w33,w34,w35,w36,w37,w38,w39,w40,w41,w42,w43,w44,w45,w46,w47, w48}.

  An interesting extension of the warm inflationary model, the warm vector inflation scenario was proposed in \cite{w11},  by using the intermediate inflation model. The constraints that Planck 2015 temperature, polarization and lensing data impose on the parameters of warm inflation were revisited in \cite{w39}.   Two-field warm inflation models with a double quadratic potential and a linear temperature dependent dissipative coefficient were studied in \cite{w46}.   The scalar spectral index $n_s$ and the tensor-to-scalar ratio $r$ were computed for several representative potentials. Warm inflationary scenarios in which the accelerated expansion of the early Universe is driven by chameleon-like scalar fields were investigated in \cite{w47}, and the model was constrained by using Planck 2018 data.

If the importance for the early cosmological evolution of the dissipative and nonequilibrium aspects of particle production during reheating has been already pointed out a long time ago \cite{Zim}, the open and  irreversible characteristics of these processes did not receive the attention they deserve. Thermodynamical systems in which matter creation/annihilation takes place belong to the particular class of open thermodynamical systems. In such systems the usual adiabatic conservation laws must be modified to explicitly include irreversible particle production processes, which can be modeled by means of a creation pressure \cite{prigogine1988thermodynamics}. The thermodynamics of open systems was applied for the first time to cosmology in \cite{prigogine1988thermodynamics}.  The phenomenological classical description  of \cite{prigogine1988thermodynamics}, was subsequently investigated and generalized in \cite{Calv, Calv1}, where a covariant formulation was developed. This approach allows for specific entropy variations, as expected for nonequilibrium and irreversible processes. The thermodynamics of open systems and irreversible processes has many implications for cosmology, which have been extensively studied in \cite{op0}-\cite{op36}. %\cite{op0,op0a,op1,op2,op3,op4,op5,op6,op7,op8,op9,op10,op11,op12,op13,op14,op15,op16,op17,op18,op19,op20,op21,op22,op23,op24,op25,op26,op27,op28,op29,op29a, op30,op31,op32,op33,op34,op35, %op36}.

It is the purpose of the present paper to investigate, by  using the thermodynamics of open
systems, as introduced in \cite{prigogine1988thermodynamics}  and developed in \cite{Calv}, the properties of a cosmological fluid mixture, consisting of two basic components: a scalar field, and matter, in the form of radiation, respectively. We will further assume that in this system  particle decay and production occur via the energy transfer from the scalar field to radiation. We assume that from  a cosmological point of view this physical system describes the warm inflation model.  The thermodynamical approach of open systems and irreversible  processes as applied to warm inflationary cosmological models leads to a self-consistent theoretical description of the particle  production processes, which in turn determine the whole dynamics and future evolution of the cosmic expansion, and structure formation. This novel approach to the study of the physical and dynamical properties of the early Universe may open some new perspectives in the understanding of the basic features of warm inflation, as analyzed, from another perspective, for example in \cite{w38,w39,w40,w41,w42}. From a theoretical point of view the present approach gives a systematic and consistent thermodynamic approach to the foundations of the warm inflationary models, and points out their irreversible, open, and nonequilibrium nature. In the present approach, as opposed to the standard scenarios, a new term, the creation pressure, describes the effects of the particle creation on the cosmological dynamics. The creation pressure is determined by both the scalar field and matter (radiation) content of the early Universe, and its presence can strongly enhance the decay of the scalar field, and the radiation creation processes.

 By applying the basic principles of the thermodynamics of open systems together with the cosmological Einstein equations for a flat, homogeneous and isotropic Universe we derive first a set of ordinary differential equations describing  radiation creation due to the decay of the scalar field.  In order to simplify the analysis of the basic equations describing warm inflation with irreversible particle creation we rewrite them in a dimensionless form, by introducing a set of appropriate dimensionless variables.

 As a cosmological application of the general formalism we investigate radiation creation from a scalar field by adopting different mathematical forms of the self-interaction potential $U$ of the field. More exactly, we will consider the case of the coherent scalar field, and of the Higgs type potentials, together with the zero-potential case. For each potential the evolution of the warm-inflationary Universe with radiation creation is investigated in detail by solving numerically the set of the warm inflationary cosmological  evolution equations.  The results display the process of inflationary scalar field decaying to radiation, and the effects of the expansion of the Universe on the decay. We concentrate on the evolution of the radiation component, which increase from zero to a maximum value, as well as on the scalar field decay. A simple model, which can be fully  solved analytically, is also presented.

 As a general result we find that the scalar field potential also plays an important role in the warm inflationary process, and in the decay of the scalar field. Some of the cosmological parameters of each model are also presented, and analyzed in detail.  The models are constrained by the observational parameters called the inflationary parameters, including the scalar spectral index $n_s$, the tensor to scalar ratio $r$, the number of e-folds $N$, and the reheating temperature $T_{\text{reh}}$. These parameters can be calculated  directly from the potentials used in the models, and they can be used together with the observational data to constrain the free parameters in the potentials.  The study of the warm inflationary models, and the development of new formalisms and constraints to the theories are undoubtedly an essential part of the completion to the theory. Analysis of the current observational data, including the estimation of the inflationary parameters, is able to test the warm inflationary models.

The present paper is organized as follows. The basic results in the thermodynamics of irreversible processes and open systems are briefly reviewed in Section~\ref{sect2}. The full set of equations describing warm inflationary models with irreversible radiation creation due to the decay of a scalar field are obtained in Section~\ref{sect3}. A detailed comparison of the standard warm inflationary model and of the irreversible warm inflation is also presented.  Several warm inflationary models, corresponding to different choices of the scalar field self-interaction potential are considered in Section~\ref{sect4}, by numerically solving the set of cosmological evolution equations. The time dynamics of the cosmological scalar field, and of the matter components are obtained, and analyzed in detail. An exact solution of the system of the irreversible warm inflation equations, corresponding to a simple form of the scalar field potential (coherent wave) is also obtained. We compare the theoretical predictions of our irreversible warm inflationary model, for the coherent wave case, as well as for two classes of Higgs type scalar field potentials, respectively, with the Planck 2018 observational data in Section~\ref{sect5a}. This comparison allows us to restrict the allowed range of model parameters, and to test the realistic nature of the model.  We conclude and discuss our results in Section~\ref{sect5}.

\section{Thermodynamics of irreversible cosmological matter creation}\label{sect2}

In the present Section we will briefly review the basic formalism of the thermodynamic of irreversible processes, taking place in open systems.  By open systems we understand thermodynamic systems that can transfer both energy and particles (matter) to their surroundings, via some dissipative processes. In the warm inflationary model the scalar field decays into radiation, and thus it transfers energy (and matter) to the cosmic environment. A closed thermodynamical system can exchange only energy (generally in the form of heat) but no matter with the environment. Moreover, a closed thermodynamic system has walls that are rigid and immovable, which cannot conduct heat, and reflect perfectly radiation. Therefore the walls of a closed system are impermeable to all forms of matter and all non-gravitational forces \cite{statphys}.  We also assume that the cosmological evolution is irreversible, that is, the scalar field can generate radiation (photons), but photons cannot decay into a scalar field (or bosonic particles). We will later fully use these results in our analysis of the warm inflationary cosmological models.

We begin our presentation with the discussion of some fundamental thermodynamic principles, and of the first two laws of thermodynamics, and the we proceed to the covariant formulation of the laws of the thermodynamic of open systems. As a last step we apply our results  to the case of homogeneous and isotropic cosmological models. In the present paper we use the natural system of units with $c=\hbar =k_B=1$, where $k_B$ is the Boltzmann constant. We also introduce the Planck mass defined as $M_P=\left(8\pi G\right)^{-1/2}$ For the metric signature we adopt the convention $\left(+,-,-,-\right)$.

\subsection{The laws of thermodynamics for open systems}\label{sect2a}

As  a starting point we consider a thermodynamic system in which we select a volume element $V$, containing $N$ particles.  Our approach to the description of the thermodynamic processes in the early Universe is based on the following principles (see, for example, \cite{statphys}),

a) In thermodynamical systems with matter creation/annihilation all the thermodynamic quantities are also functions of the (varying) particle number densities $n_i=N_i/V$.

b) In such systems the second law of thermodynamic is given by
\begin{equation}\label{seclaw}
TdS=dU+pdV-\sum_{i=1}^k{\mu _idN_i},
\end{equation}
where $\mu _i$, $i=1,...,k$ are the chemical potentials corresponding to the species of particles with particle numbers $N_i$, $i=1,2,...,k$.

c) The thermodynamic state $\Phi$ of a system with varying particle number is completely determined by the set of the (standard) thermodynamic variables $\left(U,p,T,S,N\right)$, so that  $\Phi\left(U,p,T,S,N\right)=0$. For example, for the energy $U$ of the system we can write $U=Nf\left(S/N,V/N,T/N\right)$, under the assumption that $U$ is a function of the entropy, of the temperature, and of the particle numbers.

d) The thermodynamic quantities are additive, and therefore they are homogeneous function of the first order with respect to the additive variables.

Generally, the second law of thermodynamics Eq.~(\ref{seclaw})  cannot be reformulated or interpreted as a standard energy conservation law.  For example, in the case of constant entropy, we obtain $\frac{d}{dt}\left(\rho V\right)+p\frac{dV}{dt}=\sum_{i=1}^k{\mu _i\frac{dN_i}{dt}}$, which shows that the energy is not conserved in the sense of the standard thermodynamics in the presence of constant particle numbers. However, in the case of systems with particle creation/annihilation, in  \cite{prigogine1988thermodynamics} an effective description of particle creation processes has been proposed, which is based on the reformulation of the second law of thermodynamics (\ref{seclaw}) in an effective form, as given by $d(\rho V)+\left(p+p_c\right)dV=0$, a result which can be achieved by the introduction of a new pressure term,  called the creation pressure $p_c$. Therefore, in systems with varying particle numbers and by taking into account the particle number dependence of the thermodynamical quantities typically particle creation/annihilation processes can be described effectively with the use of the creation pressure $p_c$.  In the presence of particle creation one has to also redefine the energy-momentum tensor $T_{\mu \nu}$ of the matter,  so that it includes the supplementary creation pressure term $p_c$, in addition to the true thermodynamical pressure $p$ term.

In the following we will present the main results of the thermodynamics of open systems in the presence of irreversible particle creation.

\subsubsection{The first law of thermodynamics}

 For a closed thermodynamic system,  $N$ is a constant, and we can express the conservation of the internal energy $E$ via the first law of thermodynamics as \cite{prigogine1988thermodynamics}
\be\label{encons}
dE=dQ-pdV,
\ee
where we have denoted by $dQ$ the heat received by the system in the time interval $dt$, by $p$ the thermodynamic pressure, while $V$ represents any comoving volume. We also introduce the energy density $\rho$ of the system, given by $\rho = E/V$, the particle number density $n$, defined as $n = N/V$, and the heat per unit particle $d\tilde{q}$, where $d\tilde{q} = dQ/N$. Then Eq.~(\ref{encons}) can be written as
\be\label{encons1}
d\left(\frac{\rho }{n}\right)=d\tilde{q}-pd\left(\frac{1}{n}\right).
\ee

Eq.~(\ref{encons1}) is also valid for open thermodynamical systems with $N$ a function of time, $N=N(t)$.

\subsubsection{The second law of thermodynamics}

To formulate the second law of thermodynamics for irreversible processes in open systems characterized by an entropy $S$ we introduce first the differential entropy flow $d_eS$, and the differential entropy creation $d_iS\geq 0$. The for the total entropy variation of the system we obtain the expression \cite{prigogine1988thermodynamics}
\begin{equation}
dS=d_eS+d_iS\geq 0.
\end{equation}
For adiabatic and closed systems we always have $d_iS\equiv 0$.  In the following we will investigate only open thermodynamic systems for which we assume that  matter is created in a thermal equilibrium state, so that $d_eS=0$. In such systems the entropy increases only due to the presence of particle production processes.

For the total differential of the entropy we find
\bea
Td\left(sV\right)&=&d\left(\rho V\right)-\frac{h}{n} d\left(nV\right)+Td\left(sV\right)=\nonumber\\
&&d\left(\rho V\right)-\left(h-Ts\right) dV
=d\left(\rho V\right)-\tilde{\mu} d\left(nV\right),\nonumber\\
\eea
where the entropy density $s$ is given by $s=S/V\geq 0$, and  we have defined the chemical potential as
\be
\tilde{\mu}=\frac{h-Ts}{n}\geq 0.
\ee

Hence for the entropy production due to irreversible processes in an open system we obtain
\bea\label{entropypro}
T d_iS&=&TdS=\left( \frac{h}{n}\right)  d\left(nV\right)-\tilde{\mu}d\left(nV\right)
=\nonumber\\
&&T\left( \frac{s}{n}\right) d\left(nV\right)\geq 0.
\eea

The matter creation processes in open systems can also be formulated in a covariant form in the framework of general relativity. To apply the thermodynamics of open systems to investigate warm inflationary cosmological models, in the following we formulate the thermodynamics of open systems  in a general relativistic covariant form by following the approach pioneered in \cite {Calv}. Then, we  particularize this general approach to the case of a homogeneous and isotropic Universe, for which we also derive the entropy evolution.

\subsection{Covariant general relativistic formulation of the thermodynamics of open systems}

The basic macroscopic variables that describe the thermodynamic phases of a general relativistic perfect fluid are the energy-momentum tensor $T_{\mu \nu}$, the particle flux vector $N^{\mu }$, and the entropy flux vector $s^{\mu }$, respectively.  In the case of open thermodynamical systems one must also take into account the variation of the particle numbers due to irreversible matter creation/decay. Hence for an open system the energy-momentum tensor must be  written as
 \be\label{tmunu}
 T^{\mu \nu}=\left(\rho +p+p_c\right)u^{\mu }u^{\nu }-\left(p+p_c\right)g^{\mu \nu},
 \ee
where $u^{\mu }$ is the four-velocity of the fluid, normalized according to $u_{\mu}u^{\mu}=1$, and  the creation  pressure $p_c$ takes into account particle creation and other dissipative thermodynamic effects. The energy-momentum tensor $T_{\mu \nu}$ is required to satisfy the covariant conservation law
\be\label{econs}
\nabla _{\nu }T^{\mu \nu}=0,
\ee
where $\nabla _{\nu }$ denotes the covariant derivative with respect to the metric $g_{\mu \nu}$, which defines the line element $d\tilde{s}^2=g_{\mu \nu}dx^{\mu}dx^{\nu}$.

The particle flux vector is defined as
\be\label{baparflux}
N^{\mu} =nu^{\mu },
\ee
and it satisfies the balance equation
\be\label{n}
\nabla _{\mu}N^{\mu }=\Psi \left(x^{\mu}\right) ,
\ee
where the function $\Psi  \left(x^{\mu}\right)$ is the particle creation rate. If $\Psi >0$, we have a particle source, while for $\Psi <0$ we have a particle sink. In standard cosmological models usually  $\Psi $ is assumed to vanish. We also define the entropy flux $s^{\mu }$, given by \cite{Calv},
\be\label{entropyflux}
s^{\mu }=n\sigma u^{\mu },
\ee
 where by $\sigma =s/n$ we have denoted the specific entropy per particle. The second law of thermodynamics requires that $\nabla _{\mu }s^{\mu }\geq 0$. In the presence of matter creation  the Gibbs equation for an open thermodynamic system with temperature $T$  is given by \cite{Calv}
\be\label{gibbseq}
nTd\sigma =d\rho -\frac{\rho +p}{n}dn.
\ee

To derive the energy balance equation in open systems in the presence of particle creation we multiply both sides of Eq.~(\ref{econs}) by the four-velocity vector $u^{\mu}$, thus obtaining
\bea
u_{\mu}\nabla_{\nu}T^{\mu\nu}&=&u_{\mu}\nabla_{\nu}\left(\rho+p+p_c\right)u^{\mu}u^{\nu}
+u_{\mu}\left(\rho+p+p_c\right)\times \nonumber\\
&&\nabla_{\nu}(u^{\mu}u^{\nu})
-u_{\mu}\nabla^{\mu}\left(p+p_c\right)\nonumber\\
&=&u^{\nu}\nabla_{\nu}(\rho+p+p_c)
+\left(\rho+p+p_c\right)\times\nonumber\\
&&(u_{\mu}u^{\nu}\nabla_{\nu}u^{\mu}+u_{\mu}u^{\mu}\nabla_{\nu}u^{\nu})
-\left(\dot{p}+\dot{p_c}\right)\nonumber\\
&=&\dot{\rho}+\left(\rho+p+p_c\right)\nabla_{\nu}u^{\nu}=0,
\eea
where we have introduced the notations $\dot{\rho}=u^{\mu}\nabla_{\mu}\rho=d\rho/d\tilde{s}$, $u_{\mu}u^{\mu}=1$, and $u_{\mu}u^{\nu}\nabla_{\nu}u^{\mu}=0$, respectively.  Therefore we have obtained the energy conservation equation for open systems in the form
\begin{equation}\label{dotrho}
\dot{\rho}+(\rho+p+p_c)\nabla_{\mu}u^{\mu}=0.
\end{equation}

In order to find the entropy variation in open systems we substitute the relation (\ref{dotrho}) into the Gibbs equation (\ref{gibbseq}), and we use Eqs.~(\ref{entropyflux}), (\ref{baparflux}) and (\ref{n}) to obtain first
\bea
0&=&\dot{\rho}-\frac{\rho+p}{n}u^{\mu}\nabla_{\mu}n+
T\nabla_{\mu}s^{\mu}-T\sigma\nabla_{\mu}N^{\mu} \nonumber\\
nTu^{\mu}\nabla_{\mu}\sigma&=&-p_c\nabla_{\mu}u^{\mu}-\left( \frac{\rho+p}{n}-T\sigma\right) \nabla_{\mu}N^{\mu}.
\eea
Thus we obtain the entropy balance equation as \cite{Calv}
\begin{equation}\label{baentro}
\nabla_{\mu}s^{\mu}=-\frac{p_c\Theta}{T}-\frac{\tilde{\mu}\Psi}{T},
\end{equation}
where $\tilde{\mu}=(\rho+p)/n-T\sigma$ is the chemical potential, and $\Theta=\nabla_{\mu}u^{\mu}$ is the expansion of the fluid.

In the following we will adopt the physical scenario according to which in a given geometry particles are created so that they are in thermal equilibrium with the previously
existing ones. In this case the entropy production is a result of matter creation processes only. As for the creation pressure  $p_c $ associated to particle creation in the following we will assume the following phenomenological ansatz \cite{prigogine1988thermodynamics, Calv}
\be
p_c =-\tilde{\alpha} \left(x^{\mu}\right)\frac{\Psi}{\Theta},
\ee
where $\tilde{\alpha }$ is a function satisfying the condition $\tilde{\alpha }\left(x^{\mu}\right)>0$, $\forall x^{\mu}$. This choice gives the entropy balance equations as
\bea
\nabla _{\mu }s^{\mu }&=&\frac{\Psi}{T}\left[\tilde{\alpha} \left(x^{\mu}\right) -\tilde{\mu} \right]=\Psi \sigma +\left[\tilde{\alpha} \left(x^{\mu}\right) -\frac{\rho +p}{n}\right]\frac{\Psi}{T}=\nonumber\\
&&\Psi \sigma +n\dot{\sigma },
\eea
where we have defined $\dot{\sigma}=u^{\mu }\nabla _{\mu }\sigma =d\sigma /d\tilde{s}$.  Hence with the use of Eq.~(\ref{baentro}) we obtain for the specific entropy production $\dot{\sigma }$ the relation \cite{Calv}
\be\label{eq2}
\dot{\sigma }=\frac{\Psi}{nT}\left[\tilde{\alpha }\left(x^{\mu}\right) -\frac{\rho +p}{n}\right].
\ee

Now we restrict our general thermodynamic formalism by imposing the condition that the specific entropy of the newly created particle is a constant, $\sigma ={\rm constant}$.  Then Eq.~(\ref{eq2}) gives for $\tilde{\alpha }\left(x^{\mu}\right)$ the expression $\tilde{\alpha } \left(x^{\mu}\right)=\left(\rho +p\right)/n$.  Thus we obtain for the creation pressure induced by the irreversible particle creation in open systems  the expression \cite{Calv}
\be\label{pc}
p_c=-\frac{\rho +p}{n\Theta }\Psi .
\ee

Similarly, after taking into account the condition of the constancy of the specific entropy, the Gibbs equation takes the form
\be\label{ad}
\dot{\rho }=\left(\rho +p\right)\frac{\dot{n}}{n}.
\ee

\subsection{Irreversibly thermodynamics of matter creation in homogeneous and isotropic cosmological models}

We will apply now the previously developed general thermodynamic formalism to the specific cosmological case of a homogeneous and isotropic space-time. To simplify the description of the cosmological model we adopt a comoving frame in which the components of the four-velocity are given by $u^{\mu }=\left(1,0,0,0\right)$. Moreover, in order to satisfy the cosmological principle, we suppose that all the thermodynamical and the geometric quantities are functions of the time coordinate $t$ only. Then it follows that the derivative of any function $f(t)$ with respect to the affine parameter $s$ are identical with the time derivative, $\dot{f}(t)=u^{\mu }\nabla _{\mu }f(t)=df(t)/dt$. On the other hand in the comoving reference frame the expansion of the cosmological fluid is given by $\nabla _{\mu }u^{\mu }=\dot{V}/V$. In the comoving frame
Eq.~(\ref{ad}) can be written as
\be\label{rhodot}
\dot{\rho }=\left(\frac{h}{n}\right)\dot{n},
\ee
where  we have denoted by $h = \rho + p$ the enthalpy (per unit volume) of the fluid, or, equivalently,
\be
p=\dot{\rho }-\rho \frac{\dot{n}}{n}.
\ee

In general relativity the geometry of the space-time is determined by the matter content via the Einstein gravitational field equations
\be
R_{\mu \nu }-\frac{1}{2}g_{\mu \nu }R=T_{\mu \nu },
\ee
where for the macroscopic energy-momentum tensor $T_{\mu \nu }$, we will adopt  in
the cosmological case the perfect fluid form as given by Eq.~(\ref{tmunu}). Phenomenologically, the energy - momentum tensor is described by the energy density $\rho $ and the total pressure $\bar{p}$ of the fluids, and in the comoving frame its components are given by
\be
T_0^0=\rho, T_1^1=T_2^2=T_3^3=-\bar{p}.
\ee
A direct consequence of the Einstein field equations is the conservation of the energy-momentum tensor $\nabla _{\nu }T_{\mu }^{\nu }=0$,  which is equivalent to the relation
\be\label{cons2}
d(\rho V ) = -\bar{p}dV.
\ee

In the presence of matter creation processes the correct analysis
must be done by using the irreversible thermodynamics of open systems. Hence in this case one must include in the pressure an
additional creation/annihilation pressure term $p_c$, which leads to a reformulation of Eq.~(\ref{ad}) in a form similar to Eq.~(\ref{cons2}), or specifically \cite{prigogine1988thermodynamics}
\be\label{cons3}
d(\rho V ) = -\left(p+p_c\right)dV.
\ee
From Eq.~(\ref{pc}) it follows that the cosmological creation pressure $p_c$ can be obtained  in the comoving frame as
\be\label{pc1}
p_c = -\left(\frac{h}{n}\right)\frac{d(nV)}{dV}=-\left(\frac{h}{n}\right)\frac{V}{\dot{V}}\left(\dot{n}+\frac{\dot{V}}{V}n\right).
\ee

Hence matter creation  induces a (negative) supplementary pressure $p_c$, which must be additively included in the total cosmological pressure $\bar{p}$ appearing in the energy-momentum tensor, and consequently in the Einstein field equations,
\be
\bar{p} = p + p_c.
\ee
Note that decaying of matter leads to a positive decay pressure. The variation of the entropy $dS$ in an open thermodynamic system in the presence of irreversible processes can be decomposed into two components: an entropy flow $d_0S$, and an entropy creation $d_iS$, respectively, so that
\be\label{ent1}
dS = d_0S + d_iS,
\ee
where in order to satisfy the second law of thermodynamics we must have $d_iS \geq 0$. To calculate $dS$ we consider the total differential of the entropy, given by
\be\label{ent2}
Td(sV ) = d(\rho V ) + pdV -\tilde{\mu} d(nV ),
\ee
where we have defined  $ s= S/V\geq 0$ and $\tilde{\mu }n = h - Ts$, $\tilde{\mu} \geq 0$ is the chemical potential. In a homogeneous system $d_0S = 0$, and hence only matter creation provides a contribution to the entropy production. From Eqs.~(\ref{ent1}) and (\ref{ent2}) we find \cite{prigogine1988thermodynamics}
\be
T\frac{dS}{dt}= T\frac{d_iS}{dt}= T\frac{s}{n}\frac{d(nV)}{dt}.
\ee

To close the problem of the thermodynamic description of matter creation processes we need one more equation that relates the particle number $n$ and the comoving volume  $V$. This relation should describe the time variation of $n$ as a result of irreversible matter creation (or decay) processes. Thus a relation can be obtained from Eq.~(\ref{n}), which, for a homogeneous and isotropic cosmological model, can be written as
\be\label{Psi}
\frac{1}{V}\frac{d(nV )}{dt}=\Psi (t),
\ee
where $\Psi(t)$ is the irreversible matter creation (or decay) rate ($\Psi (t) > 0$ describes particle creation, while $\Psi (t) < 0$ describes  particle decay) \cite{prigogine1988thermodynamics, Calv}. Hence the creation pressure (\ref{pc}) is essentially determined by the matter creation (or decay) rate. Hence Eqs.~(\ref{pc}) and (\ref{Psi}) are coupled to each other, and both of them enter into the energy conservation law (\ref{cons3}), which is a consequence of the Einstein field equations themselves.

We can also express the entropy production as a function of the irreversible matter creation rate according to
\be
S(t)=S\left(t_0\right)e^{\int_{t_0}^{t}{\frac{\Psi \left(t'\right)}{n}dt'}},
\ee
where $S\left(t_0\right)$ is an arbitrary integration constant.

\section{Warm inflation in a Universe with irreversible scalar field-radiation interaction }\label{sect3}

In the present Section we will introduce the description of the warm inflationary models by using the physical and mathematical formalism of the thermodynamics of open systems, as described in the previous Section. We begin our analysis by a brief presentation of the standard formulation of the warm inflationary models. Then, we will proceed to the reformulation and extension of the warm inflation theory by taking into account the thermodynamic aspects generated by the irreversible particle production in the cosmological fluid.

In the following we assume that the geometry of the spacetime is described at the cosmological level by the flat Friedmann-Robertson-Walker line element, given by
\be
d\tilde{s}^2=dt^2-a^2(t)\left(dx^2+dy^2+dz^2\right),
\ee
where $a$ is the dimensionless scale factor. We also introduce the Hubble function $H$, defined as $H=\dot{a}/a$.

We will assume that the  very early Universe can be modeled as an open thermodynamical system, consisting of a scalar field and of radiation, forming a two-component perfect fluid, with the corresponding particle number densities denoted by $n_{\phi }$, and  $n_{rad}$, respectively. $n_{\phi }$ represents the ``particles'' of the scalar field, while $n_{rad}$ is the radiation particle number density. The corresponding energy densities of the two components are denoted by $\rho _{\phi }$ and $\rho _{rad}$, respectively. The energy-momentum tensor of the two-component cosmological perfect fluid is obtained as
\be
T_{\mu }^{\nu }=T_{\mu }^{(\phi )\nu }+T_{\mu }^{(rad)\nu }=\left(\bar{\rho}+\bar{p} \right)u_{\mu }u^{\nu }-\bar{p}\;\delta _{\mu }^{\nu },
\ee
where $u^{\mu }=dx^{\mu }/d\tilde{s}$ is the four-velocity, and
\be
\bar{\rho} =\rho _{\phi }+\rho _{rad}, \qquad \bar{p}= p_{\phi }+p_{rad}.
\ee

In the comoving frame the energy density $\rho _{\phi }$ and pressure $p _{\phi }$ of the scalar field are given by
\be\label{enscf}
\rho _{\phi }=\frac{\dot{\phi }^2}{2}+U(\phi ),
\ee
 and
 \be\label{pscf}
 p_{\phi }=\frac{\dot{\phi }^2}{2}-U(\phi ),
 \ee
 respectively, where $U(\phi )$ is the self-interaction potential of the field.

 \subsection{The standard warm inflationary scenario}

 The warm inflationary model \cite{w1,w2} is an interesting theoretical alternative to the cold inflation and reheating theories. Similarly to standard inflationary theories, in warm inflation the Universe also experiences an accelerated very early expansion stage,  which is triggered by the presence of  scalar field, representing the dominant cosmological component. But, as opposed to the cold inflation scenario, besides a scalar field, a matter component of the cosmological fluid, usually assumed to the radiation,  is also present, being generated by the scalar field. During the cosmological evolution these two components interact dynamically. The cosmological evolution is described by the Friedmann equations,
 \be\label{w01}
 3H^2=\frac{1}{M_P^2}\left(\rho _{\phi }+\rho _{rad}\right),
 \ee
 \be\label{w02}
 2\dot{H}=-\frac{1}{M_P^2}\left(\dot{\phi}^2+\frac{4}{3}\rho_{rad}\right),
 \ee
where by $M_P$ we have denoted the Planck mass. Due to the decay of the scalar field, which is essentially a dissipative process, energy is transferred from the field to radiation, and this process is described by the following energy balance equations,
\be\label{w02a}
\dot{\rho}_{\phi}+3H\left(\rho _{\phi}+p_{\phi}\right)=-\Gamma \dot{\phi}^2,
\ee
\be\label{w03}
\dot{\rho}_{rad}+3H\left(\rho _{rad}+p_{rad}\right)=\Gamma \dot{\phi}^2,
\ee
where $\Gamma$ is the dissipation coefficient. By using the explicit expressions of the energy density and pressure of the scalar field, Eq.~(\ref{KG0}) can be reformulated as the generalized Klein-Gordon equation for the scalar field,
\be\label{KG0}
\ddot{\phi}+3H\left(1+Q\right)\dot{\phi}+U'(\phi)=0,
\ee
where $Q=\Gamma /3H$. By assuming that the cosmological expansion is quasi-de Sitter, that the scalar field energy density is much bigger than the energy density of the radiation, $\rho _{\phi}>>\rho _{rad}$, and that the potential term of the scalar field energy density dominates the kinetic one, so that $\rho _{\phi}\approx U(\phi)$,  Eqs.~(\ref{w01}), (\ref{w03}), and (\ref{KG0}) can be approximated as
\be
3H^2\approx \frac{1}{M_P^2}U(\phi), \dot{\phi}\approx -\frac{U'(\phi)}{3H(1+Q)},
\ee
\be\label{w04}
\rho _{rad}=\frac{\pi^2 g_*}{30}T^4\approx \frac{\Gamma}{4H}\dot{\phi}^2,
\ee
where $g_*$ is the number of degrees of freedom of the photon fluid. To obtain Eq.~(\ref{w04}) we also used the approximations $\dot{\rho}_{rad}<<H\rho_{rad}$, and $\dot{\rho}_{rad}<<\Gamma \dot{\phi}^2$, respectively.

As an indicator of the inflationary behavior we introduce the deceleration parameter $q$, defined as
\be
q=\frac{d}{dt}\frac{1}{H}-1.
\ee
Negative values of $q$ indicate accelerating expansion, while positive values of the deceleration parameter correspond to decelerating cosmological dynamics. With the use of Eqs.~(\ref{w01}) and (\ref{w02}) we immediately obtain for the deceleration parameter of the standard warm inflation theory the expression
\be
q=\frac{1}{2}\left[1+\frac{3\left(p_{\phi}+p_{rad}\right)}{\rho _{\phi}+\rho_{rad}}\right].
\ee

 Similar information as the one contained in the deceleration parameter $q$ can be obtained from the quantity  $\epsilon_{ H} \equiv d \ln H/d\mathcal{N} $, where $\mathcal{N}$ is the number of e-folds.  $\epsilon_{ H}$ is related to the parameter $\epsilon$, which provides a useful description of the slow-roll approximation in inflationary scenarios.

 \subsection{Warm inflation in the presence of irreversible particle creation}

 In the presence of particle creation neither the particle numbers nor the energy-momentum of the components of the cosmological fluids are independently conserved. In such a mixture, in which the particle numbers are not conserved, energy and momentum exchange between the two components can take place.  The cosmological fluid mixture of scalar field and radiation is described, besides its total energy density $\bar{\rho }= \rho _{\phi } + \rho _{rad}$, and total thermodynamic pressure $\bar{p} = p_{\phi }
 + p_{rad}$, also by a total particle number $\bar{n} = n_{\phi } + n_{rad}$.

We suppose that the particle number densities $n_{\phi }$ and $n_{rad}$ of each component
of the scalar field - radiation mixture fluid obey the following balance laws,
\bea \label{rate1}
\dot{n}_{\phi}  + 3Hn_{\phi } &=& -\Gamma _1\frac{\rho _{\phi}}{m_{\phi}},
%\ee
%\be
   \\
\label{rate2}
\dot{n}_{rad}  + 3Hn_{rad } &=& \Gamma _2\frac{\rho _{\phi }}{m_{\phi}},
\eea
respectively, where $m_{\phi}$ is the mass of the scalar field particle, and $\Gamma _1\neq 0$ and $\Gamma _2\neq 0$ (the dissipation coefficients) are arbitrary functions of the thermodynamic parameters, to be determined from physical considerations. The functional form we have adopted for the particle source terms is based on the thermodynamic principle introduced in Section~\ref{sect2a}, according to which the thermodynamic states as well as the thermodynamic quantities can be functions only of the full set of the basic thermodynamic variables $\left(\rho, T, S,N\right)$. In a thermodynamical system the particle numbers are generally functions of the particle energies, temperatures, and chemical potentials, as one can be seen easily from the investigation of the standard particle distribution functions (Fermi-Dirac, Bose-Einstein etc.) \cite{statphys}. Therefore in a consistent thermodynamic approach the particle creation/annihilation source terms should depend on the same quantities. In our present approach we have considered the simplest possible choice,  but of course other more general forms of the source terms than the ones considered are also possible. The coefficients $\Gamma _1$ and $\Gamma _2$ can be generally taken as functions of the temperature, matter and scalar field energy densities, and the particle numbers, so that $\Gamma _i=\Gamma _i\left(T, \rho_{\phi}, \rho_{rad}, n_{\phi},n_{rad},... \right)$.

Eqs.~(\ref{rate1}) and (\ref{rate2}) describe the decay of the scalar field particles, and the creation of photons, with both the scalar field decay rate $\Psi _{\phi}(t)$ and the photon creation
rate $\Psi _{rad}(t)$ proportional to the energy density of the scalar field, $ \Psi _{\phi}(t)\propto \rho _{\phi }$, and $ \Psi _{rad}(t)\propto \rho _{\phi }$. Hence, the time evolution  of the decay of the scalar field particles and the creation rates of the photons is controlled in the present approach by the energy density of the scalar field. From Eqs.~(\ref{rate1}) and (\ref{rate2}) it follows that the total particle number $\bar{n}=n_{\phi}+n_{rad}$ is described by the balance equation
\be
\dot{\bar{n}} + 3H\bar{n} = \left(\Gamma _2-\Gamma _1\right)\frac{\rho _{\phi }}{m_{\phi}}.
\ee
Hence in the case of an interacting mixture of scalar field and radiation and the total particle number conservation, implying $\dot{\bar{n}} + 3H\bar{n}=0$  exists only in some particular cases, when $\Gamma _1=\Gamma_2$. If this condition is not satisfied some other channels for particle production from the scalar field may exist.   For the sake of generality in the following we shall suppose that $\Gamma _1\neq \Gamma _2$.

In the theoretical formalism of the thermodynamics of irreversible processes in system with particle creation
and decay a corresponding creation or decay thermodynamic pressure does appear as a natural consequence of the variations of the particle numbers. By taking into account that in the Friedmann-Robertson-Walker geometry the comoving volume is given by $V=a^3$, and that $\dot{V}/V=3\dot{a}/a=3H$, Eq.~(\ref{pc1}) gives the general expression of the creation pressure as
\be
p_c=-\frac{h}{n}\frac{1}{3H}\left(\dot{n}+3nH\right).
\ee
The enthalpies $h$ of the scalar field and of the radiation fluid are given by $h^{(\phi )}=\rho _{\phi}+p_{\phi}$, and $h^{(rad)}=\rho _{rad}+p_{rad}$, respectively. Then, with the use of the particle number balance equations (\ref{rate1}) and (\ref{rate2}), respectively, it follows that
in the scalar field - radiation fluid mixture the creation pressures are given by
\be
p_c^{(\phi )}=\frac{\Gamma _1\left(\rho _{\phi }+p_{\phi }\right)\rho _{\phi }}{3m_{\phi}n_{\phi }H},
\ee
and
\be
p_c^{(rad)}=-\frac{\Gamma _2\left(\rho _{rad}+p_{rad}\right)\rho _{\phi }}{3m_{\phi}n_{rad}H},
\ee
respectively, with the total creation pressure expressed as
\bea
\hspace{-0.7cm}p_c^{(total)}&=&p_c^{(\phi )}+p_c^{(rad)}=
   \nonumber\\
\hspace{-0.7cm}&&\frac{\rho _{\phi }}{3m_{\phi}H}\left[\frac{\Gamma _1\left(\rho _{\phi }+p_{\phi }\right)}{n_{\phi }}-\frac{\Gamma _2\left(\rho _{rad}+p_{rad}\right)}{n_{rad}}\right].
\eea

Using the results obtained above it follows that the complete Friedmann equations,  describing the cosmological evolution of a flat Friedmann-Robertson-Walker spacetime
filled with a mixture of interacting scalar field and radiation are given by
\bea\label{Freq}
3H^2&=&\frac{1}{M_P^2}\left(\rho _{\phi }+\rho _{rad}\right),
%\ee
%\bea
   \\
2\dot{H}+3H^2&=&-\frac{1}{M_P^2}\Bigg\{p_{\phi}+p_{rad}+\frac{\rho _{\phi }}{3m_{\phi}H}\times
    \nonumber\\
&&\hspace{-1.25cm}\left[\frac{\Gamma _1\left(\rho _{\phi }+p_{\phi }\right)}{n_{\phi }}-\frac{\Gamma _2\left(\rho _{rad}+p_{rad}\right)}{n_{rad}}\right]\Bigg\},\label{freq1}
\eea
{where $\rho _{rad}=\rho _{rad}\left(n_{rad}\right)$ and $p_{rad}=p_{rad}\left(n_{rad}\right)$}. The dynamical evolution of the scalar field and radiation particles $n_{\phi}$ and $n_{rad}$ is given by Eqs.~(\ref{rate1}) and (\ref{rate2}), respectively, while the energy density and pressure of the cosmological scalar field are given by Eqs.~(\ref{enscf}) and (\ref{pscf}), respectively. As for the newly created radiation fluid we assume that its thermodynamic properties are described by the relations
\be
p_{rad}=\frac{\rho _{rad}}{3}, \rho _{rad}=\frac{8\pi ^5}{15}T^4, n_{rad}=16\pi \zeta (3)T^3,
\ee
where $\zeta (n)$ is the Riemann zeta function.

As applied to each component of the mixture of the scalar field and radiation  fluid,  Eq.~(\ref{rhodot}), representing the second law of thermodynamics in the presence of irreversible processes, gives the relationships
\be\label{scalf}
\dot{\rho }_{\phi }+3H\left(\rho _{\phi }+p_{\phi }\right)+\frac{\Gamma _1\left(\rho _{\phi }+p_{\phi }\right)\rho _{\phi }}{m_{\phi}n_{\phi }}=0,
\ee
and
\be\label{dm}
\dot{\rho }_{rad }+3H\left(\rho _{rad }+p_{rad }\right)=\frac{\Gamma _2\left(\rho _{rad }+p_{rad }\right)\rho _{\phi }}{m_{\phi}n_{rad }}\,,
\ee
respectively. Eq.~(\ref{dm}) can be equivalently reformulated as giving the evolution of the temperature of the radiation fluid as
\be\label{temp51}
\dot{T}+HT=\frac{\Gamma _2}{48\pi \zeta (3)m_{\phi}}\frac{\rho _{\phi}}{T^2}.
\ee

Eq.~(\ref{scalf}), describing the dynamical evolution of the scalar field during the generation of photons, can be rewritten as
\be\label{scaleq}
\ddot{\phi }+3H\dot{\phi }+\Gamma \left(T,\phi, \dot{\phi },U\right)\dot{\phi }+U'(\phi )=0,
\ee
where we have denoted by $\Gamma \left(T,\phi, \dot{\phi }, U\right)=\Gamma _1\rho _{\phi }/m_{\phi}n_{\phi }$ the dissipation function,  and the prime denotes the derivative with respect to the scalar field. Therefore in the framework of the thermodynamical description of open systems, in the presence of irreversible
processes in the scalar field evolution equation Eq.~(\ref{scaleq}) a friction term naturally appears in a general form, and it is a direct outcome of the second law of thermodynamics as it is formulated for fluid mixtures with interacting components.

Adding Eqs.~(\ref{scalf}) and (\ref{dm}), the evolution of the total energy density $\bar{\rho}  =
\rho _{\phi } + \rho _{rad}$ of the cosmological fluid consisting of a mixture of scalar field and radiation is governed by the balance equation
\bea\label{62}
&&\dot{\bar{\rho}}+3H\left(\rho_{\phi}+\rho _{rad} +p_{\phi }+p_{rad}\right)=\nonumber\\
&&\frac{\rho _{\phi }}{m_{\phi }}\left[\frac{\Gamma _2\left(\rho _{rad}+p_{rad}\right)}{n_{rad}}-\frac{\Gamma _1\left(\rho _{\phi }+p_{\phi }\right)}{n_{\phi }}\right].
\eea

For the entropy of the newly created photons we obtain the expression
\be\label{42}
S_{rad}(t)=S_{rad}\left(t_0\right)\exp\left({\int_{t_0}^{t}{\Gamma _2\frac{\rho _{\phi }\left(t'\right)}{n_{rad}\left(t'\right)}dt'}}\right),
\ee
where $S_{rad}\left(t_0\right)$ is an arbitrary integration constant. In the present approach
irreversible photon creation from the scalar field is an adiabatic process. The entropy produced during radiation generation from the scalar field is entirely due to the increase in the number of photons in the cosmological fluid, and we do not take into account any increase in the entropy per particle due to the presence of other dissipative processes, like, for example, photon fluid viscosity.

With the use of the Friedmann equations (\ref{Freq}) the deceleration parameter can be obtained as
\bea\label{decpar}
q&=&\frac{1}{2}\Bigg\{ 1+\frac{3\rho _{\phi }}{\rho _{\phi }+\rho _{rad}}\Bigg[
\frac{p_{\phi }+p_{rad}}{\rho _{\phi }}+\frac{1}{\sqrt{3}\sqrt{\rho _{\phi
}+\rho _{rad}}}\times \nonumber\\
&&\left( \frac{\Gamma _{1}\dot{\phi}^{2}}{n_{\phi }}-\frac{2\pi
^{4}\Gamma _{2}}{45\zeta \left( 3\right) }T\right) \Bigg] \Bigg\} .
\eea

\subsection{Standard warm inflation versus irreversible warm inflation}

The inclusion of the mathematical and physical formalism of the thermodynamics of irreversible processes, as developed in \cite{prigogine1988thermodynamics}, \cite{Calv},  and \cite{Calv1}, respectively, leads to significant modifications in the equations describing the dynamical evolution of the warm inflationary cosmological theories, and consequently, in their physical and cosmological implications. From a physical point of view the basic physical variables in our formalism become the particle numbers associated to the scalar field and radiation, $\left(n_{\phi},n_{rad}\right)$. For the particle numbers we impose the evolution equations (\ref{rate1}) and (\ref{rate2}), respectively, which assume that the decay and creation of the particle is proportional to the energy density of the scalar field. In standard warm inflation similar balance equations are imposed at the levels of energy densities, as in Eqs.~(\ref{w02a}) and (\ref{w03}). But in the present approach the source terms in the balance equations are different from those of standard warm inflation. While in Eqs.~(\ref{w02a}) and (\ref{w03}) the creation and decay is determined by the kinetic part of the scalar field energy density only, in the present approach the decay and creation rates are assumed to depend on the full energy density of the scalar field. As we have already pointed out, if the creation and decay rates of the particle numbers are equal, which implies $\Gamma _1=\Gamma _2$, the total number of particles is conserved, satisfying the usual conservation law $\dot{\bar{n}}+3H\bar{n}=0$. In the cosmological applications of the irreversible warm inflation theory we will generally assume that the condition of the conservation of the total particle numbers holds.

Once the particle number balance equations are known, the dynamical evolution of the energy densities are given, in the framework of the thermodynamics of irreversible processes, by Eq.~(\ref{rhodot}). The time evolutions of the energy densities are determined by the enthalpies of the cosmological fluid components, and by the variations of the particle numbers. Hence from the basic equation of the thermodynamics of the irreversible processes we obtain the energy densities balance equations, Eqs.~(\ref{scalf}) and (\ref{dm}), which fix the scalar field and radiation fluid decay/creation rates, without any supplementary physical assumptions. They follow directly from the particle number balance equations, and they have a very different form from the ones used in standard warm inflation models. Due to their explicit dependence on the particle numbers, the decay rate of the scalar field is different from the creation rate of the radiation. For the scalar field we have the decay rate of the energy density as given by
\be
R_{\phi}=-\frac{\Gamma _1\left(\rho _{\phi}+p_{\phi}\right)\rho _{\phi}}{m_{\phi}n_{\phi}}=-\frac{\Gamma _1\dot{\phi}^2\rho _{\phi}}{m_{\phi}n_{\phi}},
\ee
while the creation rate of the radiation energy is given by
\be
R_{rad}=\frac{\Gamma _2\left(\rho _{rad}+p_{rad}\right)\rho _{\phi}}{m_{\phi}n_{rad}}=\frac{4\Gamma _2}{3}\frac{\rho _{rad}\rho_{\phi}}{m_{\phi}n_{rad}}=\frac{2\pi ^4\Gamma _2}{45\zeta (3)m_{\phi}}T\rho _{\phi}.
\ee

In their general form these expressions are obviously different from the simple creation/decay rates of standard warm inflation, given by $\pm \Gamma \dot{\phi}^2$. In the present approach the source terms are constructed from the variables admitted by the general laws of thermodynamic.
However, our present choice of the source terms is arbitrary, and it is mainly motivated by the requirement of obtaining relatively simple and solvable theoretical models. On the other hand, the standard form $\pm \Gamma \dot{\phi}^2$ of the source terms in the standard warm inflationary scenario is actually
simpler than the present choice suggested by the thermodynamics of the irreversible processes, it is also physically well motivated, and can give a very good theoretical description of the observational data. But the present approach may represent a first step in building more realistic warm inflationary models.

However, the standard warm inflationary scenario can be recovered under the assumption that the enthalpies per particles of the scalar field and of the radiation fluid are constants, and the kinetic energy of the scalar field is much larger than the potential one,
\bea\label{69}
\hspace{-0.5cm}W^{(\phi)}&=&\frac{h^{(\phi)}}{m_{\phi}n_{\phi}}=\frac{\left(\rho _{\phi}+p_{\phi}\right)}{m_{\phi}n_{\phi}}=C_1, \nonumber\\ \hspace{-0.5cm}W^{(rad)}&=&\frac{h^{(rad)}}{m_{\phi}n_{rad}}=\frac{\left(\rho _{rad}+p_{rad}\right)}{m_{\phi}n_{rad}}=C_2, \frac{\dot{\phi}^2}{2}>>U(\phi),
\eea
where $C_1$ and $C_2$ are constants.
Then, after a rescaling of the factors $\Gamma$, we immediately obtain for the energy decay and creation rates the expressions $R_{\phi}=-\Gamma \dot{\phi}^2$ and $R_{rad}=\Gamma \dot{\phi}^2$. Under these conditions Eqs.~(\ref{scalf}) and (\ref{dm}) reduce to the form of the energy balance equations in the standard warm inflationary scenario, Eqs.~(\ref{w02a}) and (\ref{w03}), respectively.  The condition of the constancy of the enthalpy can be interpreted thermodynamically as follows. By reintroducing the total particle numbers $N_{\phi}$ and $N_{rad}$, and the total energies $E_{\phi}=\rho _{\phi}V$, $E_{rad}=\rho_{rad}V$, the first two of the conditions (\ref{69}) can be reformulated as $E_{\phi}+p_{\phi}V=C_1m_{\phi}N_{\phi}$, and $E_{rad}+p_{rad}V=C_2m_{\phi}N_{rad}$, respectively. Then, under the assumption that the pressures are roughly constant during the cosmological evolution, from the second law of thermodynamics we obtain \cite{statphys}
\bea
\hspace{-0.4cm}TdS_{\phi}&=&dQ_{\phi}=d\left(E_{\phi}+p_{\phi}V\right)=C_1m_{\phi}dN_{\phi}, \nonumber\\
\hspace{-0.4cm}TdS_{rad}&=&dQ_{rad}=d\left(E_{rad}+p_{rad}V\right)=C_2m_{\phi}dN_{rad},
\eea
giving $TdS_{tot}=T\left(dS_{\phi}+dS_{rad}\right)=dQ_{tot}=CdN_{tot}$, where $C=\left(C_1+C_2\right)m_{\phi}$, $Q_{tot}=Q_{\phi}+Q_{rad}$, and $N_{tot}=N_{\phi}+N_{rad}$. From a thermodynamic point of view this means that in the standard warm inflationary scenario, characterized by a constant enthalpy per particle, the total heat generated in the system is linearly proportional to the total particle number $N_{tot}$. On the other hand in the irreversible warm inflationary model the entropy increase is given by a more general expression as given by Eq.~(\ref{42}). As for the creation pressures of the standard warm inflationary model they are given by $p_c^{(\phi)}=-\Gamma \dot{\phi}^2/3H$, and $p_c^{(rad)}=\Gamma \dot{\phi}^2/3H$, respectively.

It is interesting to note that in the general case the creation rate of the radiation fluid is proportional not only to the scalar field energy density, but also with the temperature $T$ of the radiation fluid, indicating that at high temperatures even a relatively low scalar field energy density can still lead to a significant creation of photons.

If the potential term of the scalar field dominates the kinetic one, $\dot{\phi}^2/2<<U(\phi)$, the decay and the creation rates of the scalar field and radiation become
\be
R_{\phi}\approx -\frac{\Gamma \dot{\phi}^2U(\phi)}{m_{\phi}n_{\phi}},
\ee
and
\be
R_{rad}\approx \frac{2\pi ^4\Gamma TU(\phi)}{45\zeta (3)m_{\phi}},
\ee
respectively, where we have assumed $\Gamma _1=\Gamma _2=\Gamma$. If the condition $\Gamma U(\phi)/m_{\phi}n_{\phi}\approx {\rm constant}$ holds, we recover the decay rate of the energy density of the scalar field in the standard warm inflationary scenario, $R_{\phi}\propto -\dot{\phi}^2$. This condition can be realized, for example, by potentials for which $\rho _{\phi}\approx U(\phi)\approx m_{\phi}n_{\phi}$ (coherent scalar fields) \cite{Bar}. Under the same condition the creation rate of the radiation becomes $R_{rad}\propto Tn_{\phi}$. At the beginning of the inflation the radiation temperature is negligible small, and hence the rate of radiation creation is negligibly small. Similarly, the initial kinetic energy density of the scalar field can be considered very small. Hence we can assume that in the scalar field potential dominated phase the decay rate of the scalar field, as well as the radiation creation rates are negligibly small, and no significant amount of radiation is created. On the other hand a very large mass of the scalar field $m_{\phi}$ can also reduce the creation rates during the slow-roll phase of the inflationary era.

In the opposite limit of the dominance of the kinetic energy term of the scalar field the decay and creation rates become
\be
R_{\phi}\approx -\frac{\Gamma }{2m_{\phi}n_{\phi}}\dot{\phi}^4,
\ee
and
\be
R_{rad}\approx \frac{2\pi ^4\Gamma T}{45\zeta (3)m_{\phi}}\dot{\phi}^2,
\ee
respectively. If $n_{\phi}\propto \dot{\phi}^2$, and $T/m_{\phi}\approx {\rm constant}$, we recover the evolution laws of the standard warm inflationary scenario. In our above qualitative estimations we have assumed that $\Gamma $ is a constant. A possible dependence of $\Gamma $ on the temperature or on the scalar field energy density could lead to a significant change in the functional form of the decay and creation rates of $\rho _{\phi}$ and $\rho _{rad}$, respectively.

The total energy in the systems with matter decay/creation is generally conserved only in the effective sense, which involves the introduction of the creation pressure, so that $d(\rho V)+\left(p+p_c\right)dV=0$ \cite{prigogine1988thermodynamics}. Consequently, in the present model the energy conservation law does not hold in the standard sense, as one can see from Eq.~(\ref{62}).
However, conservative (in the usual interpretation) cosmological scenarios can be constructed if the parameters of the model satisfies the relation
\be\label{67}
\frac{2\Gamma _2\pi ^4}{45\zeta (3)}T=\Gamma _1\frac{\dot{\phi}^2}{n_{\phi}}.
\ee

If $\Gamma _1=\Gamma_2$, the total energy of the cosmological fluid is conserved as
\be
\dot{\bar{\rho}}+3H\left(\rho_{\phi}+\rho _{rad} +p_{\phi }+p_{rad}\right)=0.
\ee
With the use of condition (\ref{67}) the energy and the pressure of the scalar field can be expressed as
\be
\rho_{\phi}=\frac{\pi^4}{45\zeta (3)}n_{\phi}T+U(\phi),p_{\phi}=\frac{\pi^4}{45\zeta (3)}n_{\phi}T-U(\phi).
\ee

Hence we can formulate the following "conservative" version of the irreversible warm inflationary model, with standard energy conservation,
\be\label{70}
\dot{n}_{\phi}+3H\left[1+\frac{\Gamma _1\pi ^4}{135\zeta (3)m_{\phi}}\frac{T}{H}\right]=-\frac{\Gamma _1}{m_{\phi}}U(\phi),
\ee
\be\label{71}
\dot{T}+HT=\frac{\Gamma _1n_{\phi}}{48\pi \zeta (3)m_{\phi}}\left[\frac{\pi ^4}{45\zeta (3)}T+\frac{U(\phi)}{n_{\phi}}\right],
\ee
\be\label{72a}
\dot{H}=-\frac{1}{M_P^2}\frac{\pi ^4}{45}T\left[\frac{n_{\phi}}{\zeta (3)}+16\pi T^3\right].
\ee

Eqs.~(\ref{70})-(\ref{72a}) represent a closed system of three differential equations for the unknowns $\left(n_{\phi},T,H\right)$. However, in the present paper we will focus our investigations on the full formulation of irreversible warm inflation, and therefore we will not consider its "conservative" version.

In the presence of irreversible matter decay/creation processes, the generalized Klein-Gordon equation (\ref{scaleq}) describing the evolution of the scalar field is given by
\be
\ddot{\phi}+3H\left(1+\frac{\Gamma _1\rho _{\phi}}{3m_{\phi}n_{\phi}H}\right)\dot{\phi}+U'(\phi)=0,
\ee
which gives for the definition of the parameter $Q$ the expression
\be
Q=\frac{\Gamma }{3H}\frac{\Gamma _1\rho _{\phi}}{3m_{\phi}n_{\phi}H}.
\ee

It is interesting to note that the expression of the creation pressure associated to the scalar field can be written as
\be\label{Qpc}
p_c^{(\phi)}=Q\left(\rho _{\phi}+p_{\phi}\right)=Q\dot{\phi}^2.
\ee

The expression of $Q$ thus obtained is very different from the one used in the standard warm inflationary scenario, $Q=\Gamma /3H$, especially under the assumption that $\Gamma $ is a constant. The evolution of the scalar field is also much more complex than in the standard warm inflation scenario, leading to the possibilities of the construction of more realistic cosmological evolution models. The evolution of the field is coupled with the evolution of the particle numbers of the scalar field (the two equations must be solved together), and explicitly they take the form
\be
\ddot{\phi}+3H\dot{\phi}+\frac{3}{2}\frac{\Gamma _1}{m_{\phi}n_{\phi}}\dot{\phi}^3+U'(\phi)\left[1+\frac{3\Gamma _1}{m_{\phi}n_{\phi}}\frac{U(\phi)}{U'(\phi)}\right]=0,
\ee
\be
\dot{n}_{\phi}+3Hn_{\phi}+\frac{\Gamma _1}{m_{\phi}}\left[\frac{\dot{\phi}^2}{2}+U(\phi)\right]=0.
\ee
This evolution of the scalar field is significantly different from the one given by the simple Klein-Gordon equation (\ref{KG0}) of standard warm inflation. Hence, the inclusion in the formalism of warm inflation of the thermodynamics of open systems considerably enlarges and diversifies the possibilities of cosmological evolution.

In the presence of irreversible particle decay/creation, described by the formalism of irreversible thermodynamics, important differences do appear in the cosmological evolution. From Eqs.~(\ref{Freq}) and (\ref{freq1}) we obtain for the time variation of the Hubble function the equation
\be
2\dot{H}=-\frac{1}{M_P^2}\Bigg \{\dot{\phi}^2+\frac{32\pi ^5}{45}T^4+\frac{\rho _{\phi}}{3m_{\phi}H}\Bigg[\frac{\Gamma _1\dot{\phi}^2}{n_{\phi}}-\frac{2\pi ^4\Gamma _2}{45\zeta (3)}T\Bigg] \Bigg\}.
\ee

The evolution equation for $H$ contains a number of new terms that does not appear in the standard warm inflation evolution equation of the Hubble function. In particular, there is a supplementary dependence on the energy density of the scalar field, and on the temperature. Moreover, there is an explicit dependence on the scalar field decay factor $\Gamma_1$, which indicates a clear dependence of the cosmological expansion on the decay mechanism of the scalar field. A similar pattern does appear in the behavior of the deceleration parameter, which, in the case of the irreversible thermodynamic formulation or warm inflation, takes the form given by Eq.~(\ref{decpar}). The nature of the accelerated expansion also depends on the decay rate of the scalar field, and on the temperature of the radiation fluid. As compared to the standard case the deceleration parameter has a new term proportional to the energy density of the scalar field. In the conservative case, in which the model parameters satisfy Eq.~(\ref{67}), the deceleration parameter reduces to the form given by the standard warm inflationary model.

Finally, under the approximation $\dot{T}/T<<H$, the temperature of the photon fluid is obtained as
\be
T\approx \left(\frac{\Gamma _2}{48\pi \zeta (3)m_{\phi}}\frac{\rho _{\phi}}{H}\right)^{1/3}.
\ee

The prediction of the temperature of the radiation fluid in irreversible warm inflation is different as compared to standard warm inflation model, with $T\propto \left(\dot{\phi}^2/H\right)^{1/4}$, and involves again a full dependence on the energy density of the scalar field. This kind of dependence is a general feature of the present model, in which all physical and geometrical quantities involve a dependence on $\rho _{\phi}$, and, therefore, on both the kinetic and potential terms of the scalar field. Another particular feature of the irreversible warm inflation is the dependence of all cosmological results on the mass $m_{\phi}$ of the initial scalar field, a dependence that does not appear in standard warm inflation. The explicit dependence on the microscopic parameters of the scalar field (mass $m_{\phi}$ and number density $n_{\phi}$) increases the number of degrees of freedom of the irreversible warm inflation model, thus leading to the possibility of constructing more realistic descriptions of the cosmological evolution of the very early Universe. In the thermodynamic limit of the constant enthalpy per particle, from Eq.~(\ref{dm}) we reobtain the law of evolution of the temperature of the standard warm inflationary model. If the potential energy of the scalar field is much bigger than the kinetic one, the temperature of the Universe varies according to $T\approx \left(\sqrt{3}\Gamma _2/48\pi \zeta(3)m_{\phi}\right)^{1/3}U^{1/6}(\phi)$, where we have assumed $3H^2\approx U(\phi)$. If $U(\phi)$ is approximately constant, and for a large enough mass of the scalar field particle, the temperature of the cosmic fluid is practically a constant, with the particle creation processes contributing very little to the overall warming of the Universe.

\section{Warm inflationary models with irreversible radiation generation}\label{sect4}

In the present Section we will investigate, in the framework of the irreversible thermodynamics of open systems with matter creation/annihilation, a number of specific warm inflationary cosmological models in which radiation is generated by the decay of the scalar field, due to the scalar field-photon interaction. As a first example we analyze the case for which the density of the radiation fluid is much smaller than the energy density of the scalar field. This case corresponds to an Universe dominated by the scalar field component of the cosmological fluid. Moreover, we assume that the scalar field is represented by a coherent wave of $\phi$-particles, in which the kinetic energy term dominates in the total energy of the scalar field. As a second case we consider warm inflation driven by a potential energy dominated decaying scalar field, and we study the evolution of the radiation fluid. The general dynamics of a warm inflationary cosmological model with decaying scalar field and radiation creation is also considered, with the cosmological evolution equations studied numerically.

The role of the dissipative processes in warm inflation has been already been pointed out in several studies, without the use of the thermodynamics of open systems. The observed baryon asymmetry, and a possibly asymmetry in the dark matter sector, can be produced through dissipative particle production during inflation \cite{w12}. In \cite{w15a} it was shown that inflation can naturally occur at a finite temperature $T>H$ if sustained by dissipative effects,  with the inflaton field corresponding to a pseudo Nambu-Goldstone boson of a broken gauge symmetry. The inflationary expansion driven by a standard scalar field whose decay ratio $\Gamma $ has a generic power law dependence with the scalar field $\phi$ and the temperature of the thermal bath $T$ was considered in \cite{w21}. By assuming an exponential power law dependence in the cosmic time for the scale factor $a(t)$, corresponding to the intermediate inflation model, the background and perturbative dynamics was solved in both weak dissipative and strong dissipative regimes. Non-equilibrium solutions of the Boltzmann equation in warm inflation were considered in \cite{w37}, and it was shown that,  even if thermal equilibrium cannot be maintained, the nearly constant Hubble rate and temperature lead to an adiabatic evolution of the number density of particles interacting with the thermal bath. A dissipative  mechanism necessary consisting of the interactions between the inflaton and a tower of chiral multiplets with a mass gap was considered in \cite{w41}.

\subsection{Warm inflationary models with $\rho _{rad} \ll \rho _{\phi }$}

We begin our investigation of warm inflationary cosmological models by considering the limiting case in which the energy density and the particle number of the newly generated radiation fluid is much smaller than the particle number density and the energy density of the inflationary scalar field  component, that is, the approximations $n_{rad} \ll n_{\phi }$, $\rho _{rad} \ll \rho _{\phi }$, and $p_{rad}\ll p_{\phi }$ are valid. In this case the inflationary early Universe is dominated by the scalar field energy density, and its pressure, and its dynamical expansion is not influenced
by the matter content. However, we assume that during this phase generation of a radiation fluid takes place, and the scalar field decays correspondingly. We shall also assume that both scalar field and radiation fluid quantities are finite at the initial moment $t = t_0$. The interaction between the scalar field and the radiation fluid is introduced through the balance equations of the two fluids, and the creation/decay of the particles is determined by the scalar field energy density.  Hence, in this approximation,  the basic equations describing the warm inflationary dynamics of a flat Friedmann-Robertson-Walker Universe filled with a decaying scalar field component and radiation creation are given by
\bea
3H^2&=&\frac{1}{M_P^2}\rho _{\phi }
   \\
%\ee
%\be
2\dot{H}+3H^2&=&-\frac{1}{M_P^2}\left[p_{\phi }+\frac{\Gamma _1\left(\rho _{\phi }+p_{\phi }\right)\rho _{\phi }}{3Hm_{\phi}n_{\phi}}\right],
\eea
and
\be \label{45}
\dot{\rho} _{\phi}=\left(\rho _{\phi }+p_{\phi }\right)\frac{\dot{n}_{\phi }}{n_{\phi }},
\ee
\bea
\dot{n}_{\phi }+3Hn_{\phi }&=&-\Gamma _1\frac{\rho _{\phi }}{m_{\phi}},
%\ee
%\be
   \\
\dot{n}_{rad }+3Hn_{rad }&=&\Gamma _2\frac{\rho _{\phi }}{m_{\phi}},
\eea
respectively.

\subsubsection{Coherent scalar field-radiation interaction}\label{Coherent scalar field-radiation}

We assume now that the inflationary field is a  homogeneous scalar field, oscillating with a frequency $m_{\phi }$. Such a field can be interpreted as a coherent wave of scalar ``particles'' having zero momenta, and with the corresponding particle number density given by \cite{op14}
\be\label{bar1}
n_{\phi }=\frac{\rho _{\phi}}{m_{\phi }},\qquad m_{\phi }={\rm constant}.
\ee
From a physical point of view we can consider that $n_{\phi}$ oscillators having the same frequency $m_{\phi }$, all oscillating coherently with the same phase, can be described as a single homogeneous scalar wave $\phi (t)$. By using the energy density of the scalar field as given by Eq.~(\ref{bar1}) in Eq.~(\ref{45}) we obtain the condition
\be
p_{\phi }=0,
\ee
or, equivalently,
\be
U\left(\dot{\phi }\right)=\frac{\dot{\phi }^2}{2}.
\ee
Hence, a homogeneous oscillating scalar field is described by a Barrow-Saich
type potential, for which the potential self-interaction energy of the scalar field is proportional
to its kinetic energy \cite{Bar}. Then the energy density of the scalar field
can be written as  $\rho _{\phi } = \dot{\phi }^2$.  This equation, obtained naturally in the framework of the thermodynamics of irreversible processes in open systems, is very similar to the relation $\rho _{\phi }=\left<\dot{\phi }^2\right> $, which can be found
by substituting $\dot{\phi }^2$ by its average value per cycle \cite{Kolb}.

In this case the equations describing the dynamics of the FRW type spacetime
filled by the decaying oscillating homogeneous scalar field in the presence of
dark matter creation become
\be\label{51}
3H^2 =\frac{1}{M_P^2}\dot{\phi }^2,
\ee
\be\label{52}
2\dot{H}+3H^2=-\Gamma _1 H,
\ee
\be\label{53}
\dot{n}_{rad}+3Hn_{rad}=3\frac{\Gamma _2}{m_{\phi}}H^2,
\ee
respectively.
 For simplicity, in the following we will assume, that $\Gamma _1$ and $\Gamma _2$ are constants. To simplify the mathematical formalism we introduce a set of dimensionless variables $\tau $, $\tilde{h}$ and $\theta _{rad}, \Phi$ by means of the
transformations
\bea\label{Parameters}
t&=&\frac{2}{\Gamma _1}\tau,\quad H=\frac{\Gamma _1}{3}\tilde{h}, \nonumber\\
 n_{rad}&=&\frac{2\Gamma _1\Gamma _2M_P^2}{3m_{\phi}}\theta _{rad},\quad \Phi= M_P\phi.
\eea
Then Eqs.~(\ref{51}), (\ref{52}) and (\ref{53}) take the simple form
\be\label{phicoh}
\tilde{h}^2=\frac{3}{4}\left(\frac{d\Phi}{d\tau}\right)^2,
\ee
\be\label{56}
\frac{d\tilde{h}}{d\tau }=-\tilde{h}\left(1+\tilde{h}\right),
\ee
\be\label{57}
\frac{d\theta _{rad}}{d\tau }+2\tilde{h}\theta _{rad}=\tilde{h}^2.
\ee
Eqs.~(\ref{56}) and (\ref{57}) must be integrated with the initial conditions $\tilde{h}\left(\tau _0\right)=\tilde{h}_0$ and $\theta _{rad}\left(\tau _0\right)=\theta _{rad}^{(0)}$, and they have the following general solutions,
\be\label{72}
\tilde{h}(\tau)=\frac{1}{\left(1+1/\tilde{h}_0\right)e^{\tau -\tau _0}-1},
\ee
and
\be\label{thetacoh}
\theta _{rad}\left( \tau \right) =\frac{\left( \tilde{h}_{0}^{2}+2\theta
_{rad}^{(0)}\right) e^{2\left( \tau -\tau _{0}\right) }-\tilde{h}_{0}^{2}}{2\left[
\left( 1+\tilde{h}_{0}\right) e^{\tau -\tau _{0}}-\tilde{h}_{0}\right] ^{2}},
\ee
respectively.

The evolution of the scale factor is given by
\be
a(\tau )=a_0\frac{\left(1+\tilde{h}_0\right)e^{\tau -\tau _0}-\tilde{h}_0}{e^{\tau }},
\ee
where $a_0$ is an arbitrary constant of integration, while the time variation of the energy density of the scalar field can be obtained as
\be
\rho _{\phi}(\tau)=\frac{4}{3}\tilde{h}^2=\frac{4}{3}\left[\frac{1}{\left(1+1/\tilde{h}_0\right)e^{\tau -\tau _0}-1}\right]^2.
\ee
For the deceleration parameter $q=d(1/H)/dt-1$ we find
\be\label{qcoh}
q=\frac{3}{2}\left(1+\frac{1}{\tilde{h}_0}\right)e^{\tau -\tau _0 }-1.
\ee

At the beginning of the oscillatory period corresponding to
the radiation fluid  production, corresponding to time intervals satisfying the condition $\left(t- t_0\right) \ll \Gamma _1^{-1}$, the approximate solution of the generalized Friedmann equations is given by
\bea
h(\tau)&\approx & \tilde{h}_0-\tilde{h}_0 \left(1+\tilde{h}_0\right) \left(\tau -\tau_0\right)+\frac{1}{2} \tilde{h}_0 \left(1+\tilde{h}_0\right)\times \nonumber\\
&& (
   1+2\tilde{h}_0) \left(\tau-\tau_0\right)^2+O\left(\tau-\tau _0)^3\right),
\eea
\be
a(\tau)\approx a_0\Bigg[1+\tilde{h}_0 \left(\tau -\tau _0\right)-\frac{1}{2} \tilde{h}_0 \left(\tau -\tau_0\right)^2+O\left(\left(\tau
   -\tau _0\right)^3\right)\Bigg],
\ee
\bea
\theta _{rad}(\tau)&\approx &\theta _{rad}^{(0)}+\tilde{h}_0 (\tilde{h}_0-2 \theta _{rad}^{(0)}) \left(\tau-\tau _0\right)+\nonumber\\
&&\tilde{h}_0
   \left[\theta _{rad}^{(0)}+\tilde{h}_0 (-2 \tilde{h}_0+3 \theta _{rad}^{(0)}-1)\right]
   \left(\tau-\tau _0\right)^2+\nonumber\\
 &&  O\left(\left(\tau-\tau _0\right)^3\right),
\eea
\bea
q(\tau)&\approx& \frac{1}{\tilde{h}_0}+\left(\frac{1}{\tilde{h}_0}+1\right) \left(\tau-\tau _0\right)+\frac{(\tilde{h}_0+1)
   \left(\tau-\tau _0\right)^2}{2 \tilde{h}_0}+\nonumber\\
 &&  O\left(\left(\tau-\tau _0\right)^3\right).
\eea
This phase corresponds to a power law (non-inflationary) expansion of the Universe, with decaying scalar field, and radiation creation.

During the initial oscillating period of the scalar field with Barrow-Saich type potential, there is a rapid increase of the radiation content of the Universe. The photon number reaches a maximum value at the moment
\be
\tau _{max}= \ln \left[\frac{\tilde{h}_0 (\tilde{h}_0+1) e^{\tau _0}}{\tilde{h}_0^2+2
   \theta _{rad}^{(0)}}\right],
\ee
with the maximum created dimensionless photon  number given by
\be
\theta _{rad}^{(max)}=\frac{\tilde{h}_0^2+2 \theta _{rad}^{(0)}}{4 \tilde{h}_0-4 \theta _{rad}^{(0)}+2}.
\ee

The maximum temperature $T_{max}$ reached by the radiation fluid (assumed to be in thermodynamical equilibrium with the scalar field) can be obtained as
\be
T_{max}=\left\{ \frac{M_P^2\Gamma _{1}\Gamma _{2}\left( \tilde{h}_{0}^{2}+2\theta
_{rad}^{(0)}\right) }{48\pi \zeta (3)m_{\phi }\left[ 1+2\left( \tilde{h}%
_{0}-\theta _{rad}^{(0)}\right) \right] }\right\} ^{1/3}.
\ee

The entropy produced during warm inflation through photon creation  can be easily obtained from Eq.~(\ref{42}) and it is given by
\begin{equation}
S_{rad}(\tau )=S_{rad}\left( \tau _{0}\right) \frac{\left( \tilde{h}%
_{0}^{2}+2\theta _{rad}^{(0)}\right) e^{2\left( \tau -\tau _{0}\right) }-%
\tilde{h}_{0}^{2}}{2\theta _{rad}^{(0)}\;e^{2\left( \tau -\tau _{0}\right) }}.
\end{equation}

In the first approximation we obtain
\be
\frac{S_{rad}(\tau)}{S_{rad}\left( \tau _{0}\right)}=1+\frac{\tilde{h}_0^2 (\tau -\tau _0)}{\theta _{rad}^{(0)}}-\frac{\tilde{h}_0^2 (\tau
   -\tau _0)^2}{\theta _{rad}^{(0)}}+O\left(\left(\tau -\tau _0\right)^3\right).
\ee
Therefore, in the small time limit, there is a linear time increase of the entropy of the radiation created by the decay of the wave scalar field.

It is interesting to compare the predictions of the irreversible warm inflationary model in the presence of the Barrow-Saich potential with the standard warm inflationary model. In the presence of the Barrow-Saich potential, by assuming that $\rho _{rad}<<\rho _{\phi}$,  the evolution equations of the standard warm inflationary model take the form
\be
3H^2 =\frac{1}{M_P^2}\dot{\phi }^2,
2\dot{H}+3H^2=0,
\ee
\be
\dot{\rho}_{rad}+4H\rho _{rad}=\Gamma \dot{\phi}^2.
\ee
For $H$, $\phi$ and the scale factor $a$ we immediately obtain
\be
H(t)=\frac{H_0}{1+3H_0\left(t-t_0\right)/2},
\ee
\be
\phi (t)=\frac{2M_P}{\sqrt{3}}\ln\left[1+3H_0\left(t-t_0\right)\right]+\phi _0,
\ee
\be
a(t)=a_0\left[1+3H_0\left(t-t_0\right)\right]^{2/3},
\ee
where we have used the initial condition $H\left(t_0\right)=H_0$, while $\phi _0$ and $a_0$ are arbitrary constants of integration. For the radiation energy density we find
\be
\rho_{rad}(t)=\frac{\sqrt{3}M_P\Gamma}{4}+\frac{C}{\left[1+3H_0\left(t-t_0\right)\right]^{8/3}},
\ee
where $C$ is an arbitrary integration constant. For the deceleration parameter of the model we obtain $q=1/2$. Obviously this warm inflationary model cannot describe an inflationary Universe, since it is decelerating for all times, with the energy density of the radiation decreasing in time. However, the inclusion of the thermodynamical description of irreversible matter creation/decay processes leads, at least in this case, to the possibility of obtaining a realistic cosmological model that could describe the very early stages of the evolution of the Universe.

\subsubsection{Coherent scalar field driven warm inflation with temperature dependent dissipative coefficient}

In our previous discussion of the warm inflationary scenario in which the evolution of the Universe is driven by the interaction between a coherent scalar field and radiation we have assumed that the dissipation coefficient $\Gamma $ is a constant. Even that such an approximation may be valid for some periods of the cosmological dynamics, generally the dissipation coefficient $\Gamma $ may be a function of the thermodynamic parameters describing the state of the system. In the following, for simplicity we assume that $\Gamma _1=\Gamma _2$, that is, the dissipation coefficient is a function of the temperature only. In order to build specific models, we further assume a power-law form for $\Gamma $, so that $\Gamma _1(T)=\Gamma _0T^{\alpha}$, where $\Gamma _0$ and $\alpha$ are constants. Then the basic cosmological evolution equations describing the dynamical evolution of a Universe filled with a coherent scalar field dissipating into radiation in the presence of the Barrow-Saich potential are given by
  \be\label{51a}
3H^2 =\frac{1}{M_P^2}\dot{\phi }^2,
\ee
\be\label{52a}
2\dot{H}+3H^2=-\Gamma _0T^{\alpha} H,
\ee
\be\label{53a}
\dot{n}_{rad}+3Hn_{rad}=3\frac{\Gamma _0T^{\alpha}}{m_{\phi}}H^2.
\ee

By taking into account the explicit dependence of the radiation fluid particle number on the temperature, Eq.~(\ref{53a}) gives the following differential equation for the temperature evolution,
\be\label{54a}
\dot{T}+HT=\frac{\Gamma _0}{16\pi \zeta (3)m_{\phi}}T^{\alpha -2}H^2.
\ee

Generally, the system of equations (\ref{52a}) and (\ref{54a}) can be solved only numerically. However, an approximate analytical solution can be obtained if we assume the condition $\dot{T}/T<<H$, which allows to obtain the temperature from Eq.~(\ref{54a}) in terms of the Hubble function as
\be
T\approx \left(\frac{16\pi \zeta (3)m_{\phi}}{\Gamma _0}\right)^{1/\left(\alpha -3\right)}H^{1/\left(3-\alpha\right)}.
\ee
Then Eq.~(\ref{52a}) takes the form
\be\label{55a}
2\dot{H}+3H^2=-\beta H^{\xi},
\ee
where we have denoted $\beta =\left(16\pi \zeta(3)m_{\phi}\Gamma _0^{-3/\alpha}\right)^{\alpha /(\alpha -3)}$, and $\xi =3/(3-\alpha)$, respectively. In the cosmological regime with $2\dot{H}>>3H^2$, Eq.~(\ref{55a}) has the general solution given by
\be
H(t)= \left[\frac{\alpha  \left(\beta  t-2 H_0\right)}{2\left(3-\alpha\right)
   }\right]^{(\alpha -3)/\alpha },
\ee
where $H_0$ is an arbitrary constant of integration. In order to have an expanding Universe, the Hubble function must be a monotonically decreasing function of time, a condition which imposes the constraint $\alpha <3$. For the scale factor of the Universe we obtain the expression
\be
a(t)=a_0\exp \left\{\frac{\alpha ^2 \left(\beta  t-2 H_0\right)^2 \left[\frac{\alpha  \left(\beta
   t-2 H_0\right)}{6-2 \alpha }\right]^{-3/\alpha }}{2 (3-\alpha ) (2 \alpha -3) \beta
   }\right\},
\ee
with $a_0$ a constant of integration. The scalar field varies as
\be
\phi(t)=\sqrt{3}M_P\left\{\frac{\alpha ^2 \left(\beta  t-2 H_0\right)^2 \left[\frac{\alpha  \left(\beta  t-2
   c_1\right)}{6-2 \alpha }\right]^{-3/\alpha }}{2 (3-\alpha) (2 \alpha -3) \beta }\right\}+\phi_0,
\ee
where $\phi_0$ is an arbitrary integration constant, while for the temperature evolution we find
\be
T(t)\approx  \left(\frac{16\pi \zeta (3)m_{\phi}}{\Gamma _0}\right)^{1/\left(\alpha -3\right)}\left[\frac{\alpha  \left(\beta  t-2 H_0\right)}{2(3-\alpha) }\right]^{-1/\alpha }.
\ee
For $\alpha <3$ and for time intervals so that $\beta t>2H_0$, the temperature is a monotonically decreasing function of time. Despite qualitative in nature, the above considerations indicate that the functional dependence of the dissipation parameter on the thermodynamic quantities characterizing the interacting scalar field-radiation fluid system (temperature, energy density or particle number) may play a fundamental role in the description of the dynamical evolution of the early Universe. The explicit form of $\Gamma $ must be obtained from a fundamental quantum theory that gives a firm theoretical basis to the physical processes that take place during the decay of the scalar field into radiation, or other elementary particles.

\subsection{Constant potential scalar field and radiation creation}

As a second example in the study of the warm inflationary models in the framework of the thermodynamics of the irreversible processes involving an interaction between the scalar field and radiation we consider the approximation in which the scalar field potential can be approximated, for a certain time interval, as a positive constant, so that $U(\phi )=\Lambda ={\rm constant}>0$. Hence, in this case the energy density and the pressure of the scalar field can be written as
\be
\rho _{\phi }=\frac{1}{2}\dot{\phi }^2+\Lambda, \qquad P _{\phi }=\frac{1}{2}\dot{\phi }^2-\Lambda .
\ee

\subsubsection{Radiation creation in the absence of background cosmological expansion}

 We introduce now a second approximation in the description of the warm inflationary models by assuming that the variation in the scalar field energy density, due to the cosmological expansion, can be neglected in the scalar field evolution equations. From a mathematical point of view this approximation implies  that in  Eqs.~(\ref{scalf}) and (\ref{dm}), giving the energy density and the particle number of the scalar field, the terms containing $3H$  can be neglected. Hence the main contribution to the time evolution of the scalar field is provided by its decay into photons, and not by the background cosmological expansion. However, we assume that the feedback of the newly created radiation gas on the cosmological evolution cannot be neglected.
Within the framework of these approximations, from Eq.~(\ref{rate1}), describing the scalar field particles decay,  we first find
\be
\rho _{\phi}=-\frac{m_{\phi}}{\Gamma _1}\dot{n}_{\phi}.
\ee
This relation gives the energy density of the scalar field as a function of the particle numbers. By substituting  $\rho _{\phi }$ from the above equation into Eq.~(\ref{scalf}) we obtain
\be
\ddot{n}_{\phi }+\Gamma _1\frac{\dot{\phi}^2}{n_{\phi }}\dot{n}_{\phi }=0.
\ee
By taking into account that the square of the time derivative of the scalar field $\dot{\phi}^2$ is given by
\be
\dot{\phi }^2=2\left(\rho _{\phi }-\Lambda\right)=2\left(-
\frac{m_{\phi}\dot{n}_{\phi }}{\Gamma_1}-\Lambda \right) \,,
\ee
 the equation describing the dynamics of the scalar field particles can be obtained as
\be\label{eqn}
n_{\phi}\ddot{n}_{\phi}-2n_{\phi }^2-2\frac{\Gamma _1}{m_{\phi}}\Lambda \dot {n}_{\phi}=0.
\ee
By denoting $\dot{n}_{\phi }=u$, $\ddot{n}_{\phi }=udu/dn_{\phi }$, Eq.~(\ref{eqn}) takes the form,
\be
n_{\phi}\frac{du}{dn_{\phi }}=2\left(u+\frac{\Gamma _1}{m_{\phi}}\Lambda \right),
\ee
immediately giving
\be
\dot{n}_{\phi}=N_1n_{\phi }^2-\frac{\Gamma _1}{m_{\phi}}\Lambda,
\ee
where $N_1$ is an arbitrary integration constant. By using the initial conditions $\dot{n}_{\phi }\left(t_0\right)=-\left(\Gamma _1/m_{\phi}\right)\rho _{\phi }\left(t_0\right)=-\left(\Gamma _1/m_{\phi}\right)\rho _{\phi 0}$ and $n_{\phi}\left(t_0\right)=n_{\phi 0}$ we obtain
\be
N_1=\frac{\left(\Gamma _1/m_{\phi}\right)\left(\Lambda -\rho _{\phi 0}\right)}{n_{\phi 0}^2}.
\ee

Therefore the general solution of Eq.~(\ref{eqn}), describing the variation of the scalar particle number, is given by
\bea
n_{\phi }(t)&=&\sqrt{\frac{\Gamma _{1}\Lambda }{m_{\phi }N_{1}}}\tanh \Bigg[
\sqrt{\frac{\Gamma _{1}\Lambda N_{1}}{m_{\phi }}}\left( t_{0}-t\right)+\nonumber\\
&&\tanh ^{-1}\left( \sqrt{\frac{m_{\phi }N_{1}}{\Gamma _{1}\Lambda }}n_{\phi
0}\right) \Bigg] .
\eea

The evolution equation for the temperature of the radiation follows from Eq.~(\ref{temp51}, and can be written as
\bea
\dot{T}+4HT&=&-\frac{\Gamma _{2}}{48\pi \zeta (3)\Gamma _{1}}\frac{\dot{n}%
_{\phi }}{T^{2}}=\nonumber\\
&&-\frac{\Gamma _{2}}{48\pi \zeta (3)\Gamma _{1}}\frac{1}{%
T^{2}}\left( N_{1}n_{\phi }^{2}-\frac{\Gamma _{1}}{m_{\phi }}\Lambda \right).
\eea

By neglecting the effects of the expansion of the Universe we obtain for $T$ the equation
\begin{equation}
\frac{d}{dt}T^{3}=\frac{\Gamma _{2}\Lambda }{16\pi \zeta (3)m_{\phi }}-\frac{%
\Gamma _{2}N_{1}}{16\pi \zeta (3)\Gamma _{1}}n_{\phi }^{2},
\end{equation}%
with the general solution given by
\bea
\hspace{-1.1cm}&&T^{3}=C_{0}^{3}-\frac{\Gamma _{2}\sqrt{\Lambda }}{48\pi \zeta (3)\sqrt{%
\Gamma _{1}m_{\phi }N_{1}}}\times \nonumber\\
\hspace{-1.1cm}&&\tanh \left[ \sqrt{\frac{\Gamma _{1}\Lambda N_{1}%
}{m_{\phi }}}\left( t_{0}-t\right) +\tanh ^{-1}\left( \sqrt{\frac{m_{\phi
}N_{1}}{\Gamma _{1}\Lambda }}n_{\phi 0}\right) \right] ,
\eea
where $C_{0}$ is an arbitrary constant of integration, which must be
determined from the initial condition $T\left( t_{0}\right) =T_{0}$. Thus we
obtain
\begin{equation}
C_{0}^{3}=T_{0}^{3}+\frac{\Gamma _{2}}{\Gamma _{1}}\frac{n_{\phi 0}}{48\pi
\zeta (3)}.
\end{equation}

In the limit of large times we obtain
\begin{equation}
\lim_{t\rightarrow \infty }T^{3}=T_{0}^{3}+\frac{\Gamma _{2}}{48\pi \zeta (3)%
}\left( \frac{n_{\phi 0}}{\Gamma _{1}}+\sqrt{\frac{\Lambda }{\Gamma
_{1}m_{\phi }N_{1}}}\right) .
\end{equation}
However, this result was obtained by neglecting the expansion of the Universe.

\subsubsection{Warm inflationary cosmological models with irreversible radiation fluid creation}

In the general case of warm inflationary models with constant scalar field potential, by taking into account the expansion of the Universe, the system of equations describing the irreversible photon fluid creation from the scalar field are given by
\be\label{f1}
3H^2=\frac{1}{M_P^2}\left(\rho _{\phi }+\rho _{rad}\right),
\ee
\be
\dot{n}_{\phi }+3Hn_{\phi }=-\frac{\Gamma _1}{m_{\phi}}\rho _{\phi },
\ee
\be
\dot{\rho} _{\phi }+6H\left(\rho _{\phi }-\Lambda \right)+\frac{2\Gamma _1\left(\rho _{\phi }-\Lambda \right)\rho _{\phi }}{m_{\phi}n_{\phi }}=0,
\ee
\be\label{f4}
\dot{\rho }_{rad}+   4H\rho _{rad}=\frac{4}{3}\Gamma _2\frac{\rho _{rad}}{m_{\phi}n_{rad }}\rho _{\phi },
\ee
\be\label{f5}
\dot{T}+HT=\frac{\Gamma _2}{48\pi \zeta (3)m_{\phi}}\frac{\rho _{\phi}}{T^2}.
\ee
For simplicity, in the following we will assume $\Gamma _1=\Gamma _2={\rm constant}$.

We introduce now a set of dimensionless variables $\left(r_{\phi}, N_{\phi },r_{rad}, \tau\right)$, defined as
\bea
\tau &=&\Gamma _1 t, \rho _{\phi}=\Lambda r_{\phi }, n_{\phi}=\Lambda \frac{N_{\phi }}{m_{\phi}}, \nonumber\\
     \rho _{rad}&=&\Lambda r_{rad}, T=\left(\frac{\Lambda}{48\pi \zeta (3)m_{\phi}}\right)^{1/3}\theta.
\eea
Then the system of field equations Eqs.~(\ref{f1})-(\ref{f5}) can be written in a dimensionless form as
\be
\frac{1}{a}\frac{da}{d\tau}=\lambda \sqrt{r_{\phi }+r_{rad}},
\ee
\be
\frac{dN_{\phi }}{d\tau }+3\lambda \sqrt{r_{\phi }+r_{rad}}N_{\phi }=-r_{\phi },
\ee
\be
\frac{dr_{\phi}}{d\tau }+6\lambda \sqrt{r_{\phi }+r_{rad}}\left(r_{\phi }-1\right)+\frac{2r_{\phi }\left(r_{\phi }-1\right)}{N_{\phi }}=0,
\ee
\be
\frac{dr_{rad}}{d\tau}+4\lambda \sqrt{r_{\phi }+r_{rad}}r_{rad}=\gamma \theta r_{\phi },
\ee
and
\be
\frac{d\theta}{d\tau}+\lambda \sqrt{r_{\phi }+r_{rad}}=\frac{r_{\phi}}{\theta ^2},
\ee
respectively, where we have denoted
\be
\lambda =\frac{1}{\Gamma _1}\sqrt{\frac{\Lambda}{3M_P^2}},
\ee
and
\be
\gamma =\left(\frac{32}{3}\right)^{1/3}\frac{\pi ^{11/3}}{135\zeta ^{4/3}(3)}\left(\frac{\Lambda }{m_{\phi}^4}\right)^{1/3},
\ee
respectively.

The density parameters $\Omega _{\phi}$ and $\Omega _{DM}$ of the scalar field and of the radiation are given by
\be
\Omega _{\phi }=\frac{r_{\phi }}{r_{\phi}+r_{rad}},
\ee
and
\be
\Omega _{rad}=\frac{r_{rad}}{r_{\phi }+r_{rad}},
\ee
respectively, and they satisfy the relation $\Omega _{\phi }+ \Omega _{rad}=1$.

The dynamics of the interacting scalar field-radiation system is determined by two control parameters, $\lambda $, and $\gamma $, depending on the mass $m_{\phi}$ of the scalar field particle.

For the initial conditions used for the numerical integration of the cosmological
 evolution equations we have chosen the numerical values  $a(0)=10^{-3}$, $N_{\phi }(0)=10^5$, $r_{\phi }(0)=10^{5.3}$, $r_{rad}(0)=10^{-6}$, and $\theta (0)=10^{-3/2}$, respectively.

\begin{figure}
 \centering
 \includegraphics[scale=0.70]{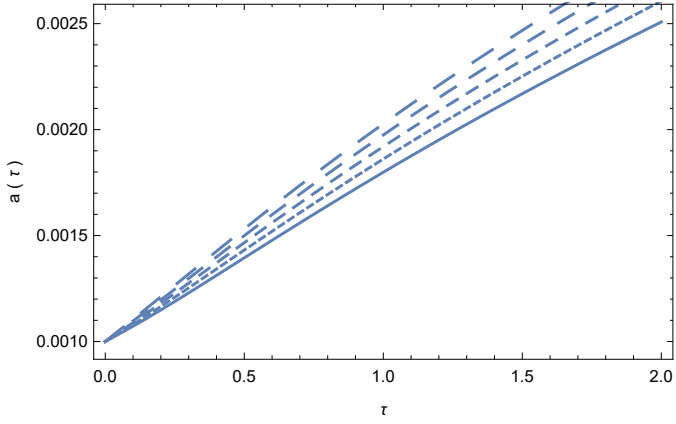}
 \caption{Time variation of the scale factor of the interacting scalar field - radiation fluid filled Universe for a constant potential scalar field, and for different values of the parameter
 $\lambda $: $\lambda =0.0016$ (solid curve), $\lambda =0.0018$ (dotted curve),
 $\lambda =0.0020$ (dashed curve), $\lambda =0.0022$ (long dashed curve), and $\lambda =0.0024$ (ultra long dashed curve), respectively.
 %The initial conditions used for the numerical integration of the cosmological
 %evolution equations are $a(0)=10^{-3}$, $N_{\phi }(0)=10^5$, $r_{\phi }(0)=10^{5.3}$, $r_{rad}(0)=10^{-6}$, and $\theta (0)=10^{-3/2}$, respectively.
 The numerical value of the parameter $\gamma $ was fixed to $\gamma =0.16$. }
 \label{a}
\end{figure}

\begin{figure}
 \centering
 \includegraphics[scale=0.70]{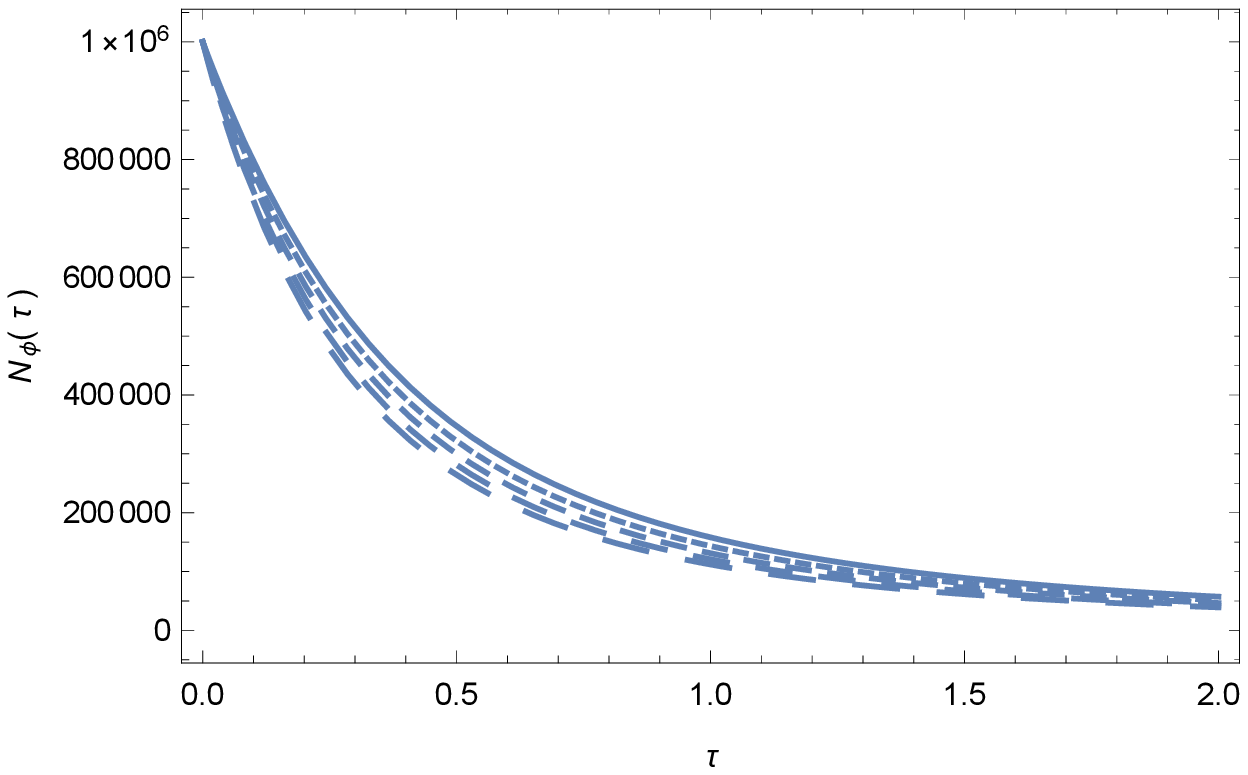}
 \caption{Time variation of the scalar field particle number $N_{\phi}$ in the warm inflationary scenario with
 an interacting scalar field - radiation fluid filled Universe for a constant potential scalar field, and  for different values of the parameter
 $\lambda $: $\lambda =0.0016$ (solid curve), $\lambda =0.0018$ (dotted curve),
 $\lambda =0.0020$ (dashed curve), $\lambda =0.0022$ (long dashed curve), and $\lambda =0.0024$ (ultra long dashed curve), respectively.
 %The initial conditions used for the numerical integration of the cosmological
 %evolution equations are $a(0)=10^{-3}$, $N_{\phi }(0)=10^5$, $r_{\phi }(0)=10^{5.3}$, $r_{rad}(0)=10^{-6}$, and $\theta (0)=10^{-3/2}$, respectively.
 The numerical value of the parameter $\gamma $ was fixed to $\gamma =0.16$.}
 \label{N}
\end{figure}

\begin{figure}
 \centering
 \includegraphics[scale=0.70]{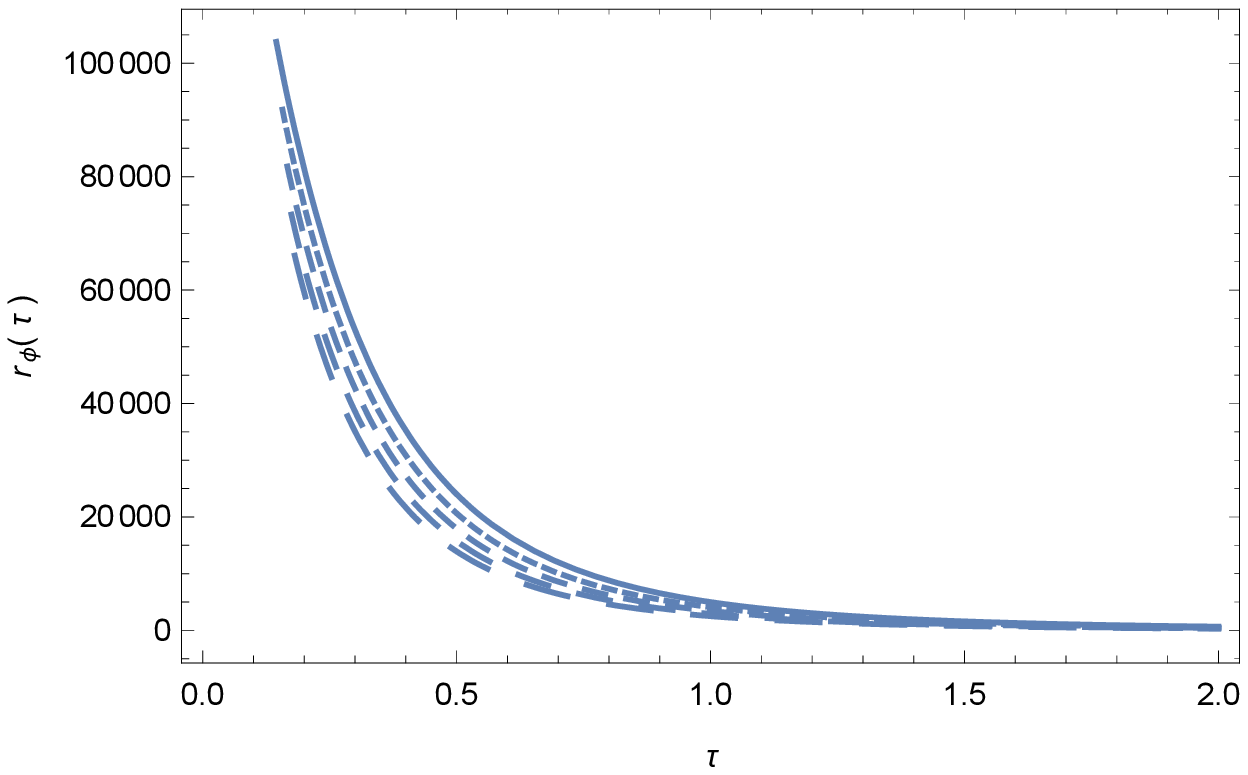}
 \caption{Time variation of the dimensionless scalar field energy $r_{\phi}$
 in the warm inflationary scenario with
 an interacting scalar field - radiation fluid filled Universe for a constant potential scalar field, and for different values of the parameter
 $\lambda $: $\lambda =0.0016$ (solid curve), $\lambda =0.0018$ (dotted curve),
 $\lambda =0.0020$ (dashed curve), $\lambda =0.0022$ (long dashed curve), and $\lambda =0.0024$ (ultra long dashed curve), respectively.
 %The initial conditions used for the numerical integration of the cosmological
 %evolution equations are $a(0)=10^{-3}$, $N_{\phi }(0)=10^5$, $r_{\phi }(0)=10^{5.3}$, $r_{rad}(0)=10^{-6}$, and $\theta (0)=10^{-3/2}$, respectively.
 The numerical value of the parameter $\gamma $ was fixed to $\gamma =0.16$.}
 \label{rf}
\end{figure}

\begin{figure}
 \centering
 \includegraphics[scale=0.70]{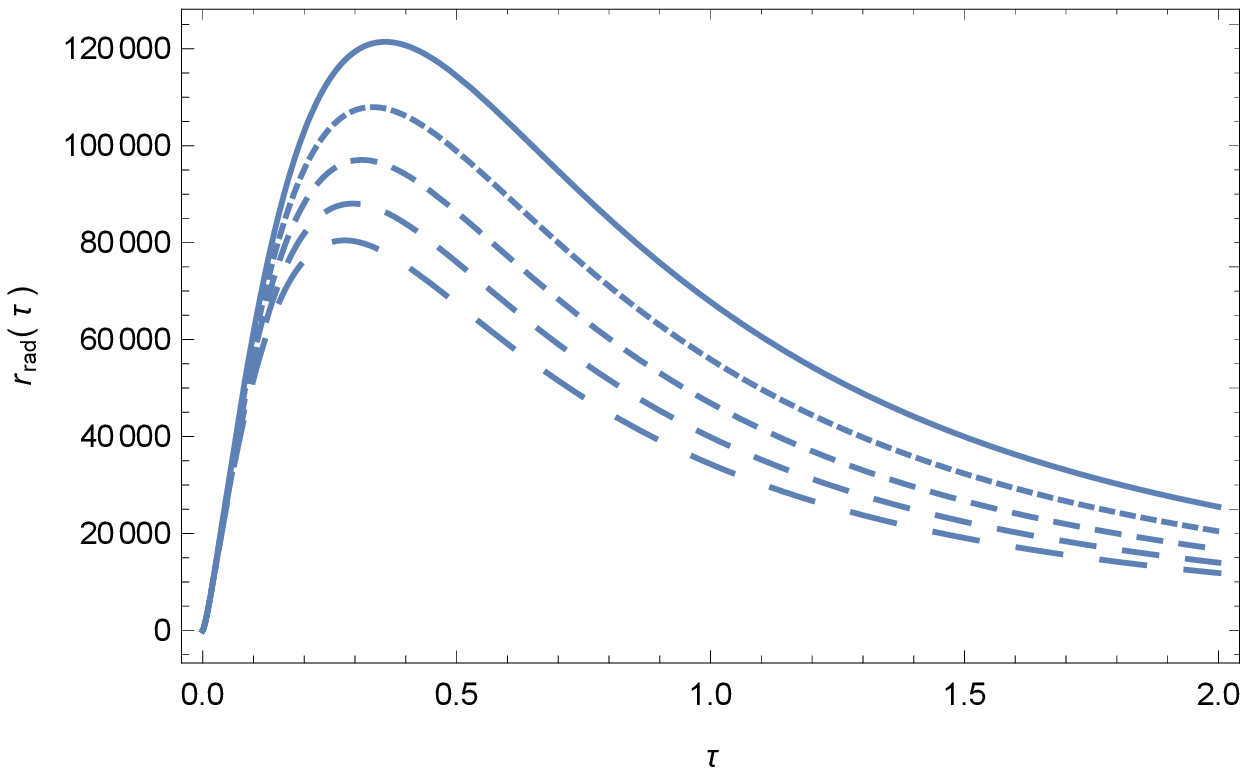}
 \caption{Time variation of the dimensionless radiation fluid  energy $r_{rad}$ in the warm inflationary scenario with
 an interacting scalar field - radiation fluid filled Universe for a constant potential scalar field, and for different values of the parameter
 $\lambda $: $\lambda =0.0016$ (solid curve), $\lambda =0.0018$ (dotted curve),
 $\lambda =0.0020$ (dashed curve), $\lambda =0.0022$ (long dashed curve), and $\lambda =0.0024$ (ultra long dashed curve), respectively.
 %The initial conditions used for the numerical integration of the cosmological
 %evolution equations are $a(0)=10^{-3}$, $N_{\phi }(0)=10^5$, $r_{\phi }(0)=10^{5.3}$, $r_{rad}(0)=10^{-6}$, and $\theta (0)=10^{-3/2}$, respectively.
 The numerical value of the parameter $\gamma $ was fixed to $\gamma =0.16$.}\label{rr}
\end{figure}

\begin{figure}
 \centering
 \includegraphics[scale=0.70]{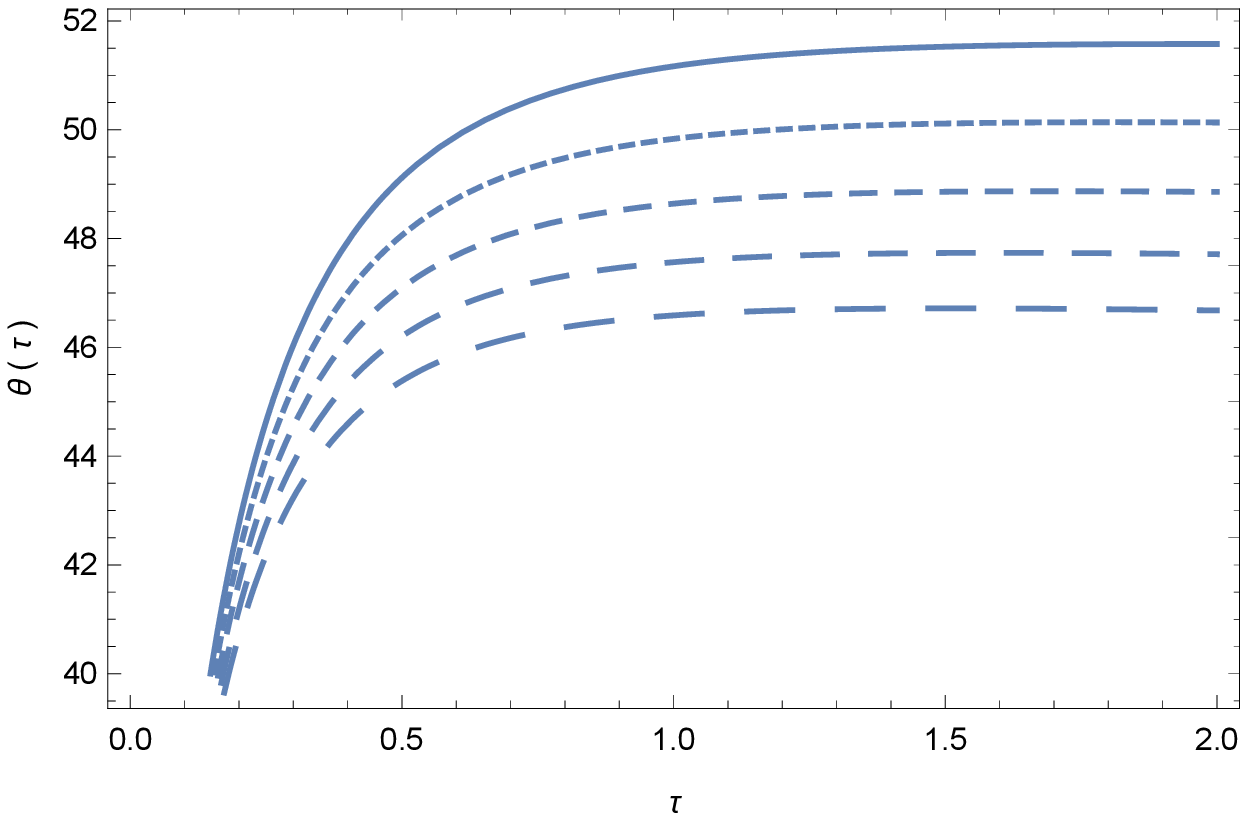}
 \caption{Time variation of the dimensionless temperature $\theta $ in the warm inflationary scenario with
 an interacting scalar field - radiation fluid filled Universe for a constant potential scalar field, and for different values of the parameter
 $\lambda $: $\lambda =0.0016$ (solid curve), $\lambda =0.0018$ (dotted curve),
 $\lambda =0.0020$ (dashed curve), $\lambda =0.0022$ (long dashed curve), and $\lambda =0.0024$ (ultra long dashed curve), respectively.
 %The initial conditions used for the numerical integration of the cosmological
 %evolution equations are $a(0)=10^{-3}$, $N_{\phi }(0)=10^5$, $r_{\phi }(0)=10^{5.3}$, $r_{rad}(0)=10^{-6}$, and $\theta (0)=10^{-3/2}$, respectively.
 The numerical value of the parameter $\gamma $ was fixed to $\gamma =0.16$.}\label{T}
\end{figure}

\begin{figure}
 \centering
 \includegraphics[scale=0.70]{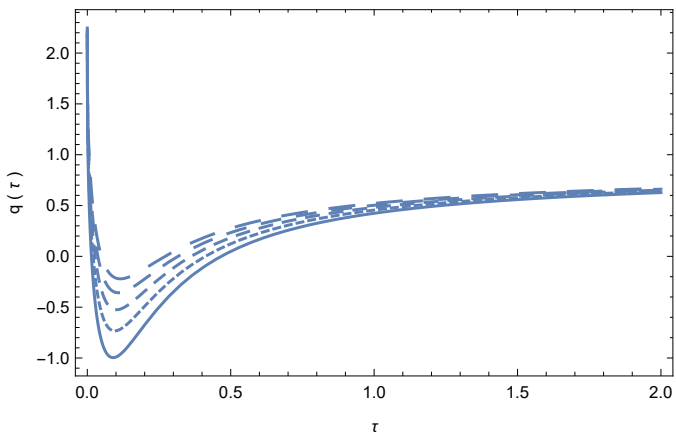}
 \caption{Time variation of the deceleration parameter $q $ in the warm inflationary scenario with
 an interacting scalar field - radiation fluid filled Universe for a constant potential scalar field, and for different values of the parameter
 $\lambda $: $\lambda =0.0016$ (solid curve), $\lambda =0.0018$ (dotted curve),
 $\lambda =0.0020$ (dashed curve), $\lambda =0.0022$ (long dashed curve), and $\lambda =0.0024$ (ultra long dashed curve), respectively.
 %The initial conditions used for the numerical integration of the cosmological
 %evolution equations are $a(0)=10^{-3}$, $N_{\phi }(0)=10^5$, $r_{\phi }(0)=10^{5.3}$, $r_{rad}(0)=10^{-6}$, and $\theta (0)=10^{-3/2}$, respectively.
 The numerical value of the parameter $\gamma $ was fixed to $\gamma =0.16$.}\label{q}
\end{figure}

 The time variations of the scale factor $a$, of the scalar field particle number $N_{\phi }$, of the scalar field energy $r_{\phi }$, of the radiation energy density $r_{rad}$, of the temperature $\theta$, and of the deceleration parameter $q$ are represented, for the above initial conditions, for different values of the parameter $\lambda $, and for $\gamma =0.16$,  in Figs.~\ref{a}-\ref{q}.

As one can see from Fig.~\ref{a}, the warm inflationary Universe filled with a mixture of interacting scalar field and radiation is an expansionary state, with the rate of the expansion, and the scale factor evolution, strongly dependent on the numerical values of the dimensionless parameter $\lambda $, which is a function of the ratio of the (constant) scalar field potential $\Lambda $, and of the scalar field decay rate $\Gamma _1$. Accelerated expansion can also be obtained in the framework of the present model, the creation pressure, corresponding to the irreversible decay of the scalar field, and the matter creation, can drive the Universe into a de Sitter type phase. As shown in Fig.~\ref{N}, the particle number of the scalar field decreases during the cosmological evolution, due to the creation of the photons. The decay rate strongly depends on the numerical value of the parameter $\lambda $. The dimensionless energy density of the scalar field $r_{\phi }$, shown in Fig.~\ref{rf}, tends in the large time limit to the value 1, corresponding to $\rho _{\phi }=\Lambda$, and to a de Sitter type expansion. This shows that the decay of the scalar field is determined and controlled by the kinetic energy term of the field $\dot{\phi }^2/2$, which is the source of the radiation  creation. When the energy and the pressure of the scalar field are dominated by the scalar field potential $\Lambda $, $\rho _{\phi}=-p_{\phi }=\Lambda $, then $\rho _{\phi }+p_{\phi }=0$, and from Eq.~(\ref{scalf}) it follows that $\rho _{\phi}={\rm constant}$, and the scalar field energy cannot be converted any more into other type of particles. The energy density of the radiation fluid, consisting of photons, and presented in Fig.~\ref{rr}, increases in time due to the decay of the scalar field particles. After a finite time interval $\tau _{max}$, the radiation energy density reaches a maximum value, and for time intervals $\tau >\tau_{max}$, it decreases monotonically, indicating a significant decrease in the number of the produced photons. The temperature of the radiation fluid, depicted in Fig.~\ref{T}, shows a similar evolutionary pattern, with the temperature of the Universe reaching its maximum value $\theta _{max}$ at $\tau _{max}$.

The evolution of the deceleration parameter, represented in Fig.~\ref{q}, indicates the existence of a complex dynamics of the interacting scalar field-radiation system. The evolution of the Universe begins at $\tau =0$ from a decelerating phase, with $q>0$ having values of around $q\approx 2$. This initial value is relatively independent on the adopted initial conditions, and the numerical values of the model parameters. Due to the irreversible radiation creation the expansion of the Universe accelerates, and very quickly the deceleration parameter becomes negative, reaching, for some parameter values, the de Sitter phase. The Universe remains in the accelerating state with $q<0$ a finite time interval, reaching the value $q=0$ in a parameter-dependent way, at a time interval $\tau _e$. The change of sign of $q$ indicates the transition from the accelerating to the decelerating phase (end of inflation), and, for enough large time intervals $q$ reaches the value $q\approx 0.5$, a numerical value approximately independent on the numerical values of the model parameters.

\subsection{Warm inflationary models with Higgs type potential and irreversible radiation creation}

As a second warm inflationary model radiation creation we will consider the case when the self-interaction potential $U(\phi)$ of the scalar field is  of the Higgs type,
\be
U(\phi) =\pm \frac{\mu^2}{2}\phi ^2+\frac{\xi }{4}\phi ^4,
\ee
where $\mu ^2$  and $\xi $ are constants. If, similarly to the standard approach in elementary particle physics, we assume that the constant $\mu ^2 < 0$ is related to the mass of the scalar particle by the relation $m_{\phi}^2= 2\xi v^2 = -2\mu ^2$, then $v^2 = -\mu ^2/\xi $ gives the
minimum value of the potential. In the case of strong interactions the Higgs self-coupling constant $\xi $ can be obtained from the determination of the mass of the Higgs boson from accelerator experiments, and it has the numerical value of the order of $\xi  \approx 1/8$ \cite{Higgs}. In the presence of the Higgs potential, the cosmological evolution equations of the warm inflationary scenario with irreversible radiation creation take the following form,
\be\label{fa}
3H^2=\frac{1}{M_P^2}\left(\frac{\dot{\phi}^2}{2}\pm \frac{\mu^2}{2}\phi ^2+\frac{\xi }{4}\phi ^4+\rho _{rad}\right),
\ee
\be\label{fb}
\dot{n}_{\phi}+3Hn_{\phi}=-\frac{\Gamma _1}{m_{\phi}}\left(\frac{\dot{\phi}^2}{2}\pm \frac{\mu^2}{2}\phi ^2+\frac{\xi }{4}\phi ^4\right),
\ee
\be\label{fc}
\ddot{\phi}+3H\dot{\phi}\pm \mu^2\phi+\xi \phi^3=-\frac{\Gamma _1}{m_{\phi}n_{\phi}}\dot{\phi}\left(\frac{\dot{\phi}^2}{2}\pm\frac{\mu^2}{2}\phi ^2+\frac{\xi }{4}\phi ^4\right),
\ee
\be\label{fd}
\dot{\rho}_{rad}+4H\rho _{rad}=\frac{2\pi^4}{45\zeta (3)}\frac{\Gamma _1}{m_{\phi}}T\left(\frac{\dot{\phi}^2}{2}\pm \frac{\mu^2}{2}\phi ^2+\frac{\xi }{4}\phi ^4\right),
\ee
\be\label{fe}
\dot{T}+HT=\frac{\Gamma _1}{48\pi \zeta (3)m_{\phi}}\frac{1}{T^2}\left(\frac{\dot{\phi}^2}{2}\pm \frac{\mu^2}{2}\phi ^2+\frac{\xi }{4}\phi ^4\right).
\ee

We reparameterize now the scalar field $\phi$ according to
\be
\phi\rightarrow M_P\phi,
\ee
and we introduce a set of dimensionless variable $\left(\tau, r_{rad}, N_{\phi},\theta\right)$ defined as
\bea
t&=&\frac{1}{\Gamma _1}\tau, \;\;\rho_{rad}=\Gamma _1^2 M_P^2\;r_{rad}, \nonumber\\
n_{\phi}&=&\frac{\Gamma _1^2M_P^2}{m_{\phi}}N_{\phi}, \;\;T =\left(\frac{\Gamma _1^2M_P^2}{48\pi \zeta (3)m_{\phi}}\right)^{1/3}\theta,
\eea

Moreover,  we denote
\be
\epsilon ^2=\frac{\mu ^2}{2\Gamma _1^2}, \sigma =\frac{\xi M_P^2}{4\Gamma _1^2}, \gamma =\frac{\pi ^{11/3}M_P^2\Gamma _1^{2/3}}{45\times 6^{1/3}\zeta ^{4/3}(3)m_{\phi}^{4/3}}.
\ee

Then the system of Eqs.~(\ref{fa})-(\ref{fe}) can be reformulated as a first order dynamical system given by
\be\label{f1a}
 \hspace{-0.5cm}\frac{d\phi}{d\tau}=u,
\ee
\be\label{f1b}
 \hspace{-0.5cm}\frac{da}{d\tau}=\frac{1}{\sqrt{3}}\sqrt{\frac{u^2}{2}\pm \epsilon ^2\phi ^2+\sigma \phi ^4+r_{rad}}\;a,
\ee
\bea\label{f1c}
 \hspace{-0.5cm}&&\frac{dN_{\phi}}{d\tau}+\sqrt{3}\sqrt{\frac{u^2}{2}\pm \epsilon ^2\phi ^2+\sigma \phi ^4+r_{rad}}\;N_{\phi}=\nonumber\\
 \hspace{-0.5cm}&&-\left(\frac{u^2}{2}-\epsilon ^2\phi ^2+\sigma \phi ^4\right),
\eea
\bea\label{f1d}
 \hspace{-0.5cm}&&\frac{du}{d\tau}+\sqrt{3}\sqrt{\frac{u^2}{2}\pm \epsilon ^2\phi ^2+\sigma \phi ^4+r_{rad}}\;u\pm 2\epsilon ^2\phi+4\sigma \phi^3=\nonumber\\
 \hspace{-0.5cm}&&-\frac{u}{N_{\phi}}\left(\frac{u^2}{2}\pm \epsilon ^2\phi ^2+\sigma \phi ^4\right),
\eea
\bea\label{f1e}
 \hspace{-0.5cm}&&\frac{dr_{rad}}{d\tau}+\frac{4}{\sqrt{3}}\sqrt{\frac{u^2}{2}\pm \epsilon ^2\phi ^2+\sigma \phi ^4+r_{rad}}\;r_{rad}=\nonumber\\
 \hspace{-0.5cm}&&\gamma \theta \left(\frac{u^2}{2}\pm \epsilon ^2\phi ^2+\sigma \phi ^4\right),
\eea
\bea\label{f1f}
 \hspace{-0.5cm}&&\frac{d\theta }{d\tau}+\frac{1}{\sqrt{3}}\sqrt{\frac{u^2}{2}\pm \epsilon ^2\phi ^2+\sigma \phi ^4+r_{rad}}\;\theta =\nonumber\\
 \hspace{-0.5cm}&&\frac{1}{\theta ^2}\left(\frac{u^2}{2}\pm \epsilon ^2\phi ^2+\sigma \phi ^4\right)
\eea

The system of equations (\ref{f1a})-(\ref{f1f}) must be integrated with the initial conditions $\phi (0)=\phi_0$, $u(0)=u_0$, $N_{\phi}(0)=N_{\phi 0}$, $r_{rad}(0)=r_{rad0}$, and $\theta (0)=\theta _0$. In the present study we adopt as the initial conditions used for the numerical integration of the cosmological
 evolution equations in the presence of a Higgs type potential the numerical values  $a(0)=10^{-3}$, $N_{\phi }(0)=10^5$, $\phi (0)=5$ for the case $\mu ^2<0$, $\phi (0)=1.4$ for the case $\mu ^2>0$, $u(0)=-1.25$, $r_{rad}(0)=10^{-6}$, and $\theta (0)=10^{-3/2}$, respectively.

In the warm inflationary cosmological model, in which the self-interaction potential of the scalar field is of Higgs type, the evolution of the Universe is determined by three parameters $\epsilon ^2$, $\sigma$, and $\gamma $, respectively, which are the dimensionless combinations of the parameters of the potential, the field decay rate, and the mass of the Higgs boson, respectively. In the following we will investigate the cosmological evolution for both signs of $\mu ^2$ in the Higgs potential.

\subsubsection{$U(\phi) =-\frac{\mu^2}{2}\phi ^2+\frac{\xi }{4}\phi ^4$}

\begin{figure}
 \centering
 \includegraphics[scale=0.70]{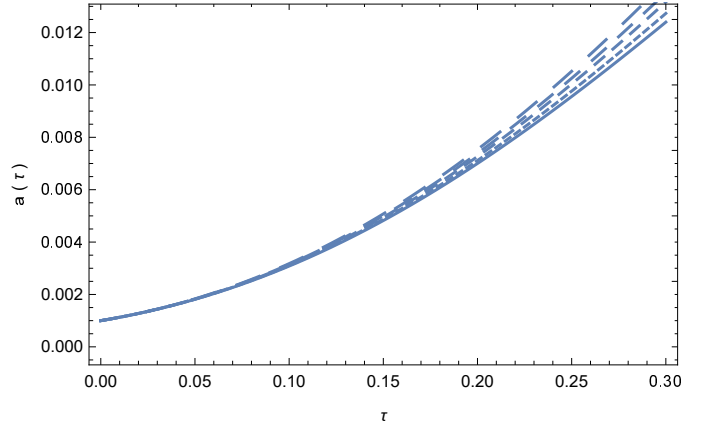}
 \caption{Time variation of the scale factor of the interacting scalar field - radiation fluid filled warm inflationary Universe, for the case of the Higgs type potential of the scalar field with $\mu ^2<0$, and for different values of the parameter
 $\gamma $: $\gamma =0.99$ (solid curve), $\gamma  =1.24$ (dotted curve),
 $\gamma =1.57$ (dashed curve), $\gamma =1.84$ (long dashed curve), and $\gamma =2.17$ (ultra long dashed curve), respectively.
 %The initial conditions used for the numerical integration of the cosmological
 %evolution equations are $a(0)=10^{-3}$, $N_{\phi }(0)=10^5$, $\phi (0)=5$, $u(0)=-1.25$, $r_{rad}(0)=10^{-6}$, and $\theta (0)=10^{-3/2}$, %respectively.
 The numerical value of the parameters $\epsilon  $ and $\sigma $ were fixed to $\epsilon =1.27$ and $\sigma =0.76$, respectively.}
 \label{a1}
\end{figure}

\begin{figure}
 \centering
 \includegraphics[scale=0.70]{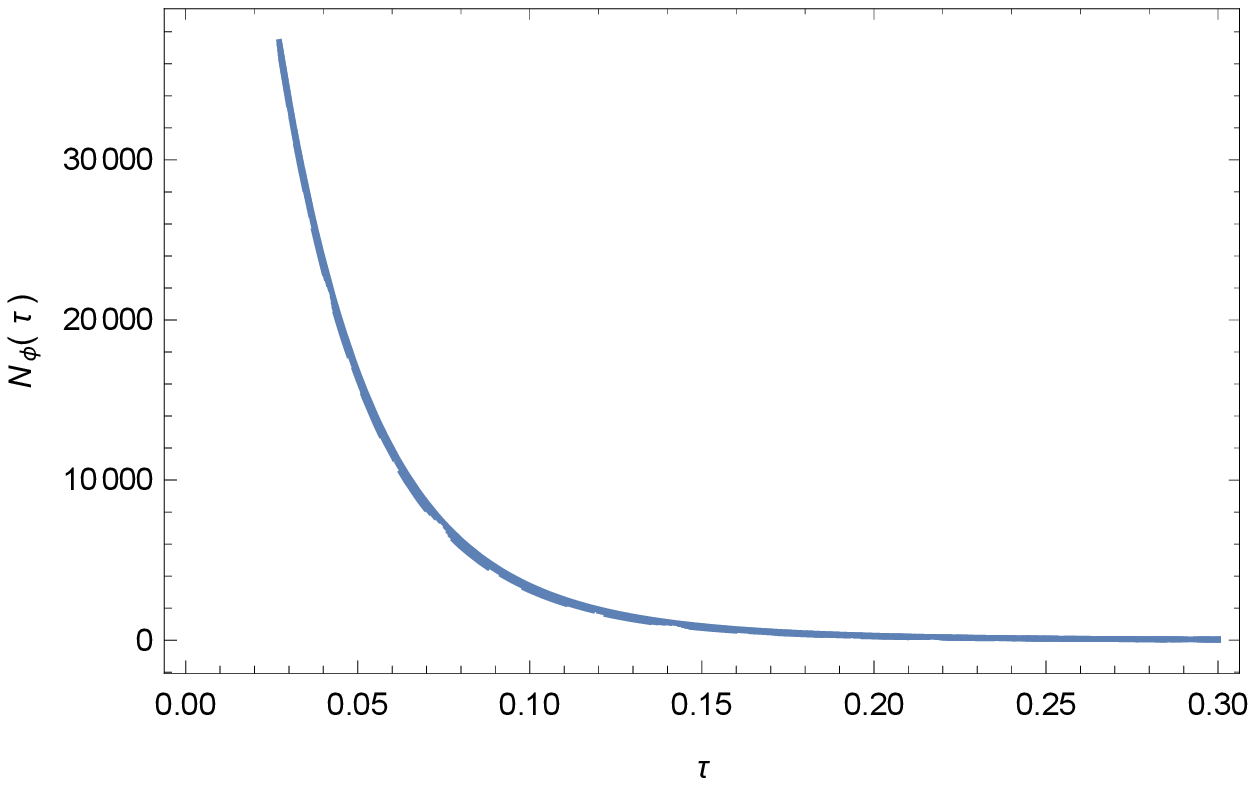}
 \caption{Time variation of the scalar field particle number $N_{\phi}$ in the warm inflationary scenario with
 an interacting scalar field - radiation fluid filled Universe, for the case of the Higgs type potential of the scalar field  with $\mu ^2<0$, and for different values of the parameter  $\gamma $: $\gamma =0.99$ (solid curve), $\gamma  =1.24$ (dotted curve),
 $\gamma =1.57$ (dashed curve), $\gamma =1.84$ (long dashed curve), and $\gamma =2.17$ (ultra long dashed curve), respectively.
 %The initial conditions used for the numerical integration of the cosmological
 %evolution equations are $a(0)=10^{-3}$, $N_{\phi }(0)=10^5$, $\phi (0)=5$, $u(0)=-1.25$, $r_{rad}(0)=10^{-6}$, and $\theta (0)=10^{-3/2}$, %respectively.
 The numerical value of the parameters $\epsilon  $ and $\sigma $ were fixed to $\epsilon =1.27$ and $\sigma =0.76$, respectively.}
 \label{N1}
\end{figure}

\begin{figure}
 \centering
 \includegraphics[scale=0.70]{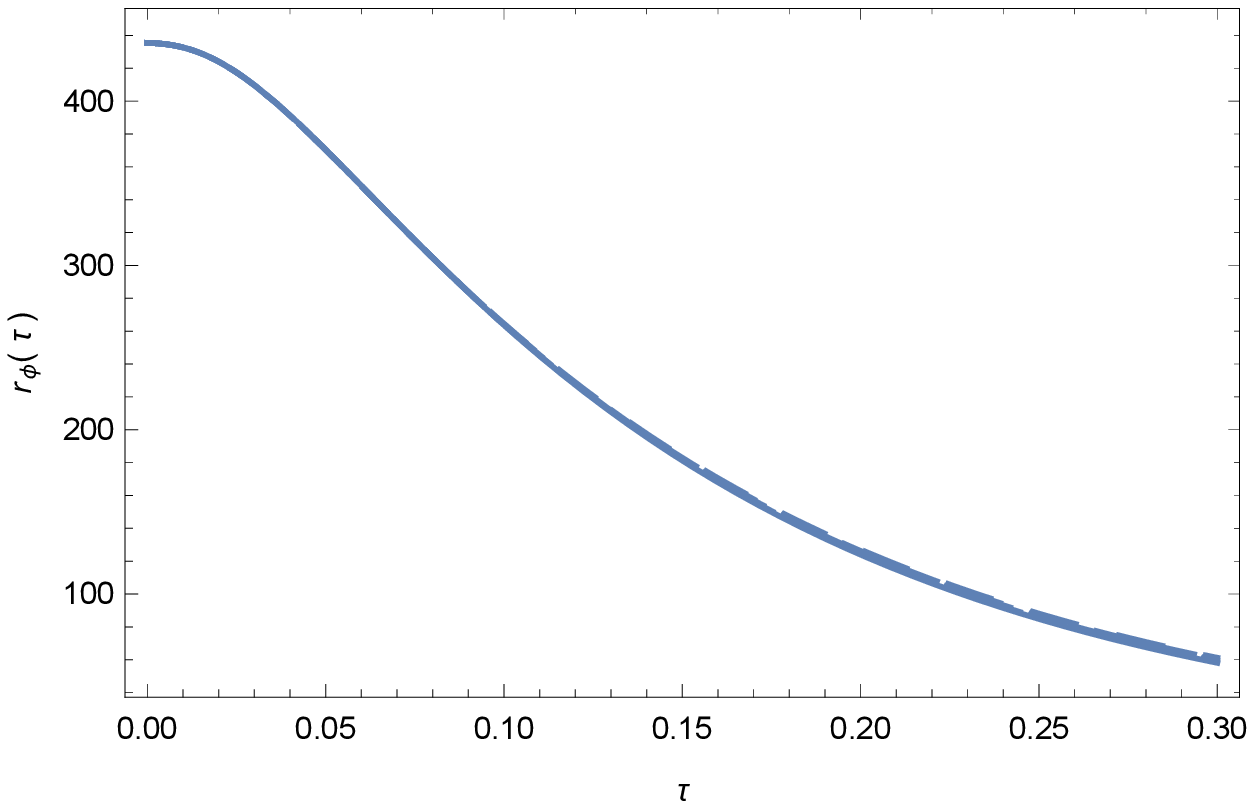}
 \caption{Time variation of the dimensionless scalar field energy $r_{\phi}$  in the warm inflationary scenario with  an interacting scalar field - radiation fluid filled Universe, for the case of the Higgs type potential of the scalar field  with $\mu ^2<0$, and for different values of the parameter  $\gamma $: $\gamma =0.99$ (solid curve), $\gamma  =1.24$ (dotted curve),  $\gamma =1.57$ (dashed curve), $\gamma =1.84$ (long dashed curve), and $\gamma =2.17$ (ultra long dashed curve), respectively.
 %The initial conditions used for the numerical integration of the cosmological  evolution equations are $a(0)=10^{-3}$, $N_{\phi }(0)=10^5$, $\phi %(0)=5$, $u(0)=-1.25$, $r_{rad}(0)=10^{-6}$, and $\theta (0)=10^{-3/2}$, respectively.
 The numerical value of the parameters $\epsilon  $ and $\sigma $ were fixed to $\epsilon =1.27$ and $\sigma =0.76$, respectively.}
 \label{rf1}
\end{figure}

\begin{figure}
 \centering
 \includegraphics[scale=0.70]{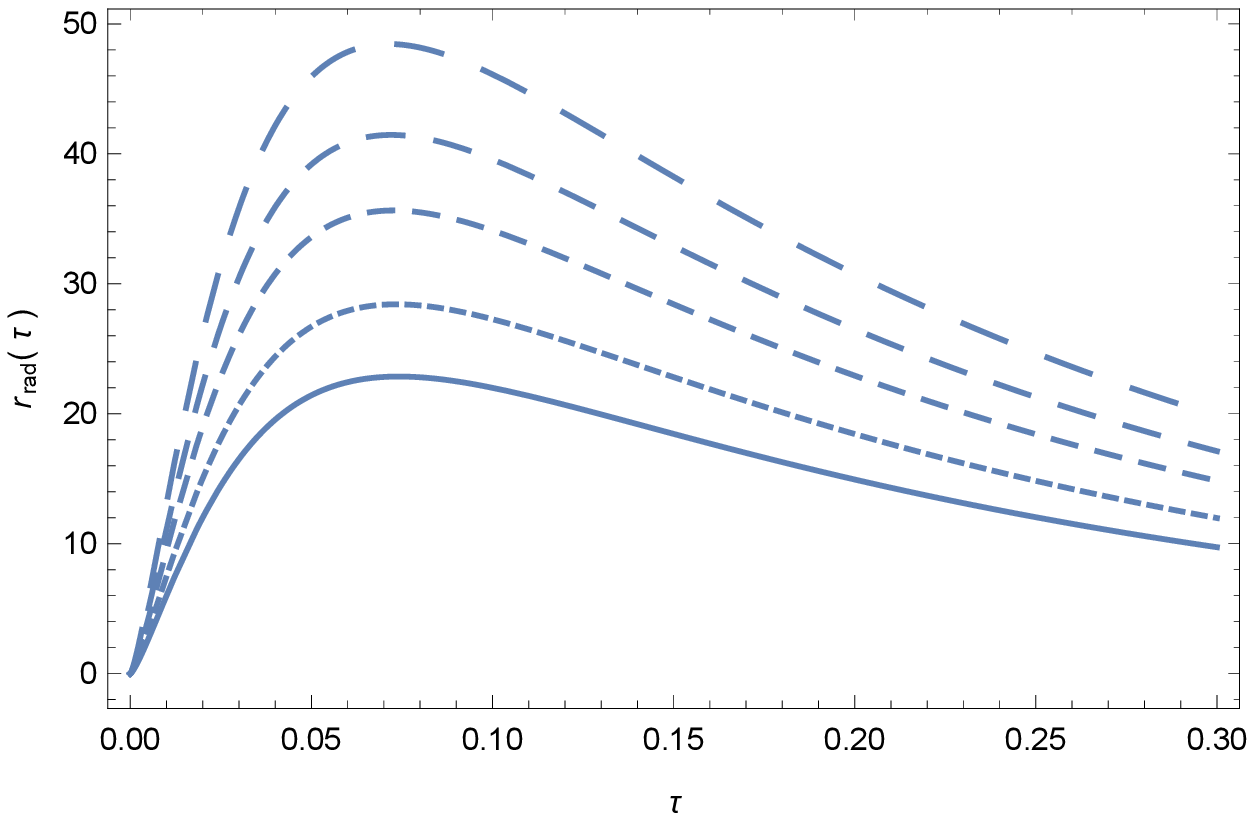}
 \caption{Time variation of the dimensionless radiation fluid  energy density $r_{rad}$ in the warm inflationary scenario with
 an interacting scalar field - radiation fluid filled Universe, for the case of the Higgs type potential of the scalar field  with $\mu ^2<0$, for different values of the parameter  $\gamma $: $\gamma =0.99$ (solid curve), $\gamma  =1.24$ (dotted curve),  $\gamma =1.57$ (dashed curve), $\gamma =1.84$ (long dashed curve), and $\gamma =2.17$ (ultra long dashed curve), respectively.
 %The initial conditions used for the numerical integration of the cosmological
 %evolution equations are $a(0)=10^{-3}$, $N_{\phi }(0)=10^5$, $\phi (0)=5$, $u(0)=-1.25$, $r_{rad}(0)=10^{-6}$, and $\theta (0)=10^{-3/2}$, % respectively.
 The numerical value of the parameters $\epsilon  $ and $\sigma $ were fixed to $\epsilon =1.27$ and $\sigma =0.76$, respectively.}\label{rad1}
\end{figure}

\begin{figure}
 \centering
 \includegraphics[scale=0.70]{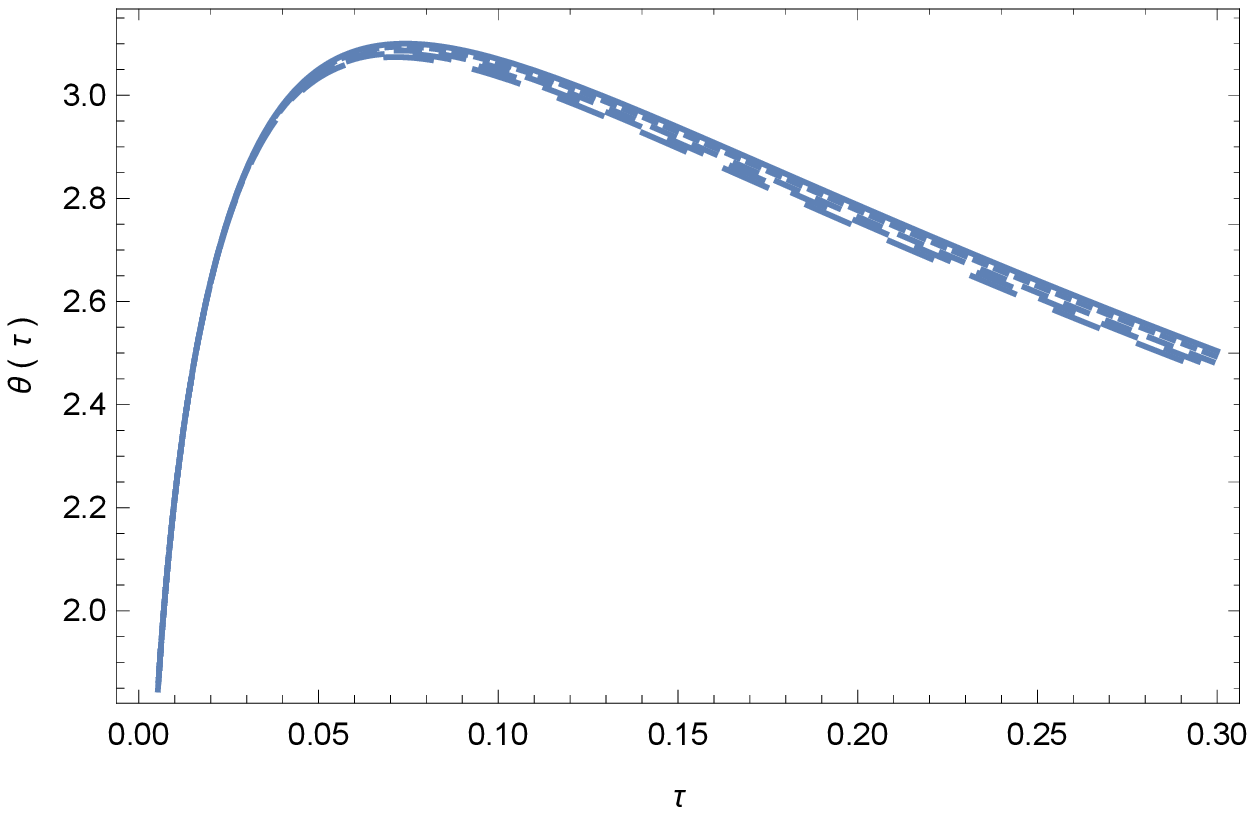}
 \caption{Time variation of the dimensionless temperature $\theta $  in the warm inflationary scenario with
 an interacting scalar field - radiation fluid filled Universe, for the case of the Higgs type potential of the scalar field  with $\mu ^2<0$, for different values of the parameter  $\gamma $: $\gamma =0.99$ (solid curve), $\gamma  =1.24$ (dotted curve),  $\gamma =1.57$ (dashed curve), $\gamma =1.84$ (long dashed curve), and $\gamma =2.17$ (ultra long dashed curve), respectively.
 %The initial conditions used for the numerical integration of the cosmological
 %evolution equations are $a(0)=10^{-3}$, $N_{\phi }(0)=10^5$, $\phi (0)=5$, $u(0)=-1.25$, $r_{rad}(0)=10^{-6}$, and $\theta (0)=10^{-3/2}$, %respectively.
 The numerical value of the parameters $\epsilon  $ and $\sigma $ were fixed to $\epsilon =1.27$ and $\sigma =0.76$, respectively.}\label{temp1}
\end{figure}

\begin{figure}
 \centering
 \includegraphics[scale=0.70]{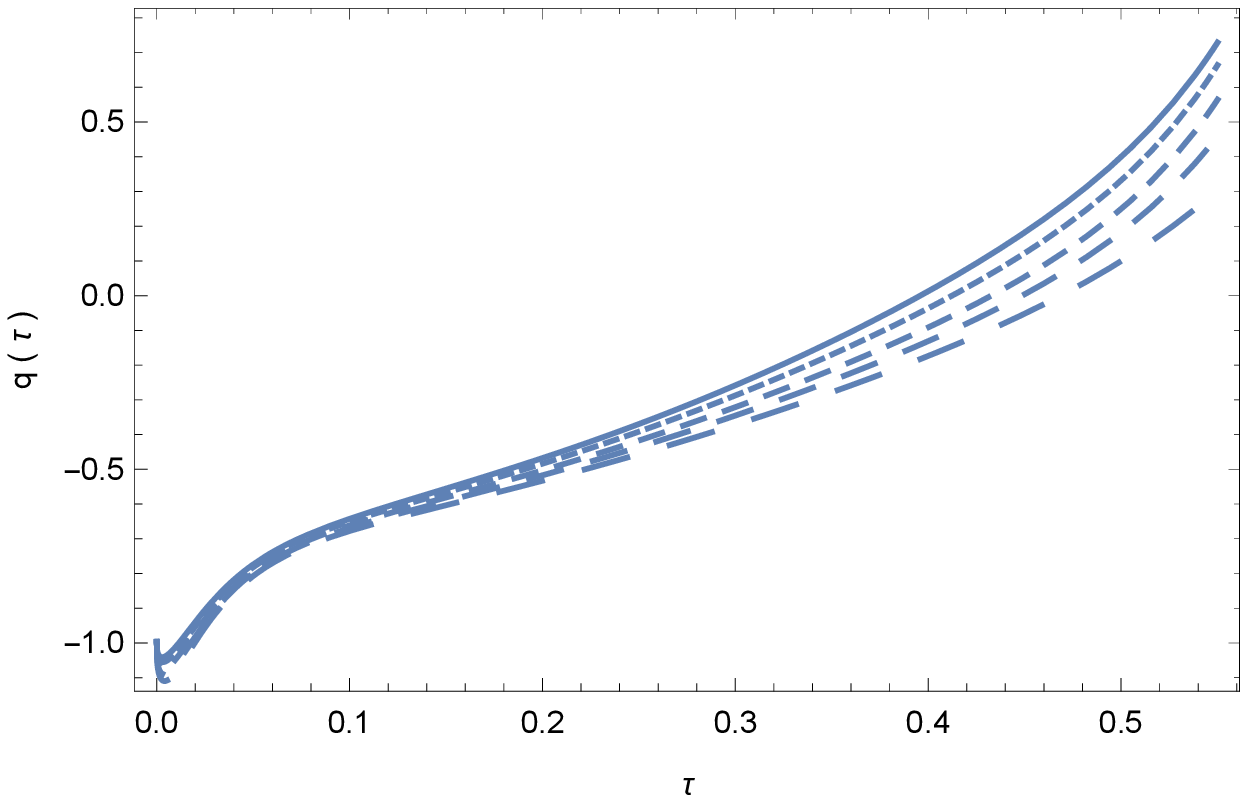}
 \caption{Time variation of the deceleration parameter $q $ in the warm inflationary scenario with
 an interacting scalar field - radiation fluid filled Universe, for the case of the Higgs type potential of the scalar field  with $\mu ^2<0$, for different values of the parameter  $\gamma $: $\gamma =0.99$ (solid curve), $\gamma  =1.24$ (dotted curve),  $\gamma =1.57$ (dashed curve), $\gamma =1.84$ (long dashed curve), and $\gamma =2.17$ (ultra long dashed curve), respectively.
 %The initial conditions used for the numerical integration of the cosmological
 %evolution equations are $a(0)=10^{-3}$, $N_{\phi }(0)=10^5$, $\phi (0)=5$, $u(0)=-1.25$, $r_{rad}(0)=10^{-6}$, and $\theta (0)=10^{-3/2}$, respectively.
 The numerical value of the parameters $\epsilon  $ and $\sigma $ were fixed to $\epsilon =1.27$ and $\sigma =0.76$, respectively.}\label{q1}
\end{figure}

First we consider the warm inflationary evolution for a negative $\mu ^2$, that is, we adopt for the Higgs potential the expression $U(\phi) =- \mu^2\phi ^2/2+\xi \phi ^4/4$, with $\mu ^2>0$. The time variations of the scale factor, scalar field particle number, scalar field energy density, radiation energy density, and of the temperature of the radiation fluid are represented for this form of the Higgs potential in Figs.~\ref{a1}-\ref{q1}.

As one can see from Fig.~\ref{a1}, the  scalar field-radiation interacting Universe in the presence of a Higgs potential is in an expansionary state, with the scale factor a monotonically increasing function of time. In the initial stages of expansion, the cosmological dynamics is basically independent on the numerical values of $\gamma$, but at later stages the cosmological behavior is strongly influenced by the variation of this parameter. The dimensionless scalar field particle number, whose time variation is presented in Fig.~\ref{N1}, shows a rapid monotonic decrease in time, with its dynamic basically independent on the parameter $\gamma$. In the large time limit the scalar field particle number vanishes. The dimensionless scalar field energy density, shown in Fig.~\ref{rf1}, also rapidly decreases in time, with its decay dynamics basically independent on $\gamma$. The radiation fluid energy density, created during the warm inflation period, is depicted in Fig.~\ref{rad1}. The radiation energy density initially increases due to the energy transfer from the scalar field to photons, and it reaches a maximum value $r_{rad}^{(max)}$ after a finite interval $\tau _{max}$. For time intervals $\tau >\tau$ the expansion rate of the Universe becomes larger than the particle creation rate, and the expansionary evolution determines the decrease of the radiation fluid energy density. The generation of the radiation fluid from the scalar field is strongly dependent on the numerical values of the parameter $\gamma$. Finally, the temperature of the radiation fluid, presented in Fig.~\ref{temp1}, shows a similar behavior as the energy density of the radiation fluid. The temperature of the Universe increases rapidly during the early stages of the cosmological evolution, and reaches a maximum value at some finite time $\tau _{max}$. Then the expansion of the Universe takes over the particle production processes, and, due to the cosmological expansion, the temperature of the radiation fluid began to decrease.

The variation of the deceleration parameter $q$, represented in Fig.~\ref{q1}, shows that the Universe field with the Higgs potential scalar field interacting with a radiation fluid begins its expansion from a de Sitter type phase, with $q\approx -1$. The expansion is decelerating, with the numerical values of the deceleration parameter increasing in time. After a finite time interval $\tau _e$, the Universe reaches the marginally accelerating state with $q=0$, and for larger time intervals it enters into a decelerating phase, with $q>0$. The final stages of the decelerating cosmological evolution are strongly dependent on the model parameters, or, in the case of our present numerical analysis, on the values of $\gamma$, describing the coupling between the temperature and the energy density of the scalar field. Presumably a  fundamental (quantum field based) physical theory would be able to estimate the numerical values of the model parameters, thus allowing a precise comparison of the cosmological observations and the theoretical predictions.

\subsubsection{$U(\phi) = \frac{\mu^2}{2}\phi ^2+\frac{\xi }{4}\phi ^4$}
\label{PositiveHiggs}

\begin{figure}
 \centering
 \includegraphics[scale=0.70]{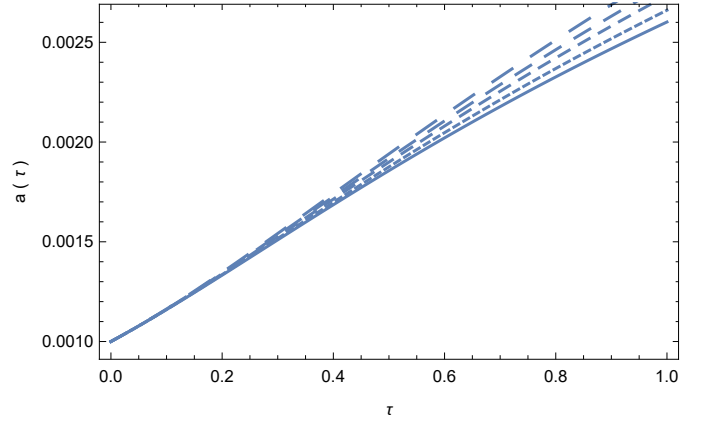}
 \caption{Time variation of the scale factor of the interacting scalar field - radiation fluid filled warm inflationary Universe, for the case of the Higgs type potential of the scalar field with $\mu ^2>0$, and for different values of the parameter
 $\gamma $: $\gamma =0.99$ (solid curve), $\gamma  =1.24$ (dotted curve),
 $\gamma =1.57$ (dashed curve), $\gamma =1.84$ (long dashed curve), and $\gamma =2.17$ (ultra long dashed curve), respectively.
 %The initial conditions used for the numerical integration of the cosmological
 %evolution equations are $a(0)=10^{-3}$, $N_{\phi }(0)=10^5$, $\phi (0)=5$, $u(0)=-1.25$, $r_{rad}(0)=10^{-6}$, and $\theta (0)=10^{-3/2}$, %respectively.
 The numerical value of the parameters $\epsilon  $ and $\sigma $ were fixed to $\epsilon =1.27$ and $\sigma =0.76$, respectively.}
 \label{a2}
\end{figure}

\begin{figure}
 \centering
 \includegraphics[scale=0.70]{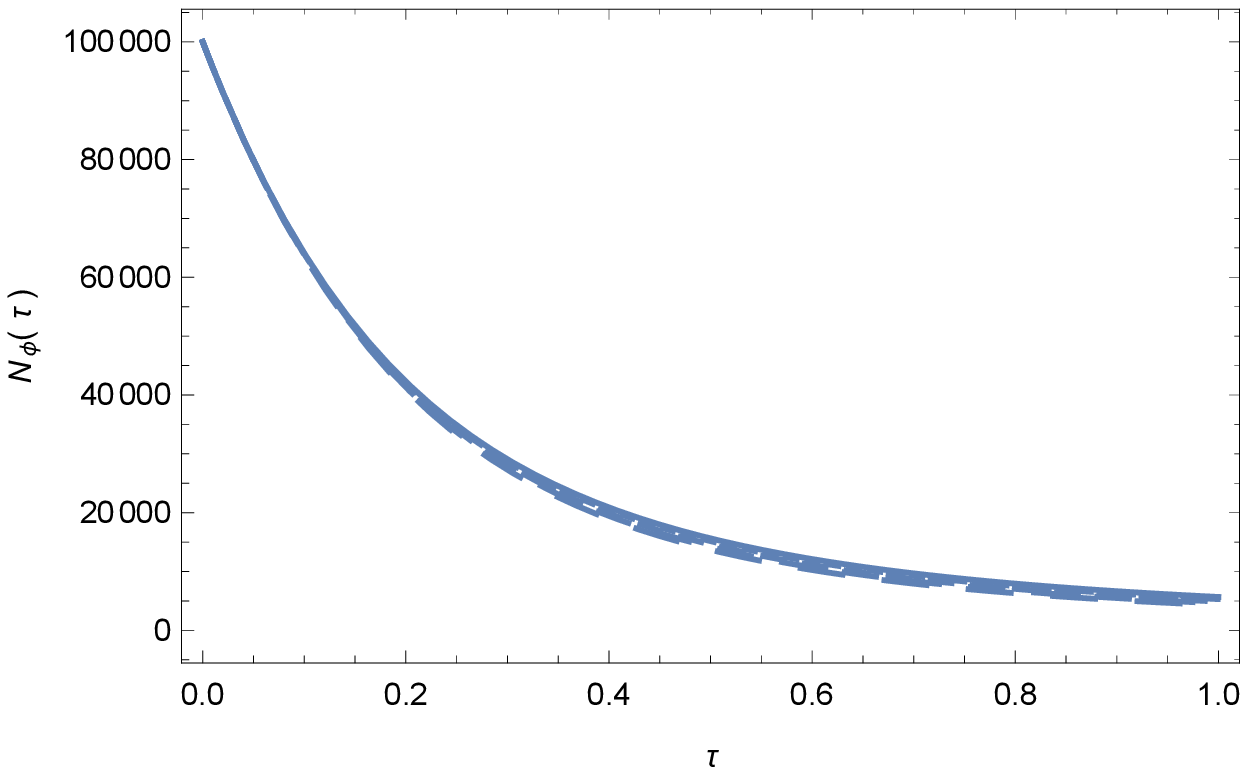}
 \caption{Time variation of the scalar field particle number $N_{\phi}$ in the warm inflationary scenario with
 an interacting scalar field - radiation fluid filled Universe, for the case of the Higgs type potential of the scalar field  with $\mu ^2>0$, and for different values of the parameter  $\gamma $: $\gamma =0.99$ (solid curve), $\gamma  =1.24$ (dotted curve),
 $\gamma =1.57$ (dashed curve), $\gamma =1.84$ (long dashed curve), and $\gamma =2.17$ (ultra long dashed curve), respectively.
 %The initial conditions used for the numerical integration of the cosmological
 %evolution equations are $a(0)=10^{-3}$, $N_{\phi }(0)=10^5$, $\phi (0)=5$, $u(0)=-1.25$, $r_{rad}(0)=10^{-6}$, and $\theta (0)=10^{-3/2}$, %respectively.
 The numerical value of the parameters $\epsilon  $ and $\sigma $ were fixed to $\epsilon =1.27$ and $\sigma =0.76$, respectively.}
 \label{N2}
\end{figure}

\begin{figure}
 \centering
 \includegraphics[scale=0.70]{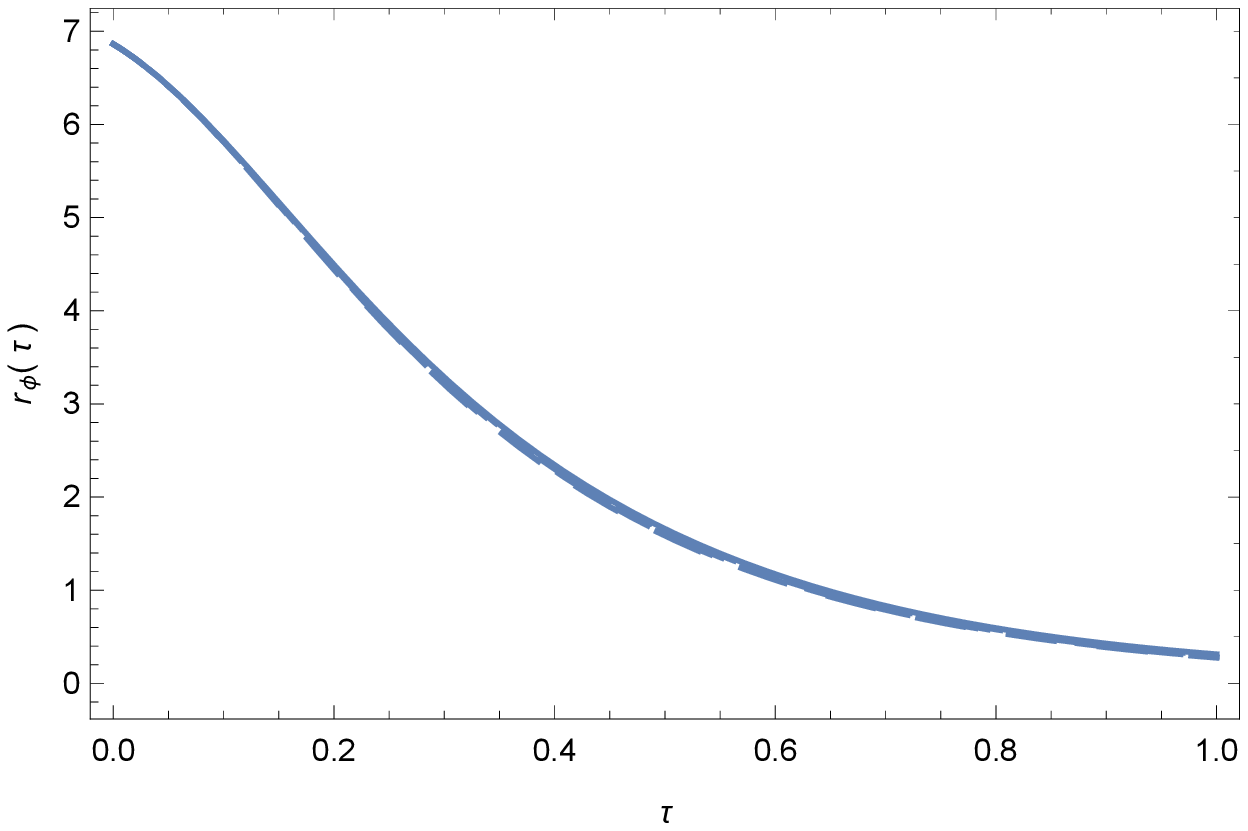}
 \caption{Time variation of the dimensionless scalar field energy $r_{\phi}$  in the warm inflationary scenario with  an interacting scalar field - radiation fluid filled Universe, for the case of the Higgs type potential of the scalar field  with $\mu ^2>0$, and for different values of the parameter  $\gamma $: $\gamma =0.99$ (solid curve), $\gamma  =1.24$ (dotted curve),  $\gamma =1.57$ (dashed curve), $\gamma =1.84$ (long dashed curve), and $\gamma =2.17$ (ultra long dashed curve), respectively.
 %The initial conditions used for the numerical integration of the cosmological  evolution equations are $a(0)=10^{-3}$, $N_{\phi }(0)=10^5$, $\phi %(0)=5$, $u(0)=-1.25$, $r_{rad}(0)=10^{-6}$, and $\theta (0)=10^{-3/2}$, respectively.
 The numerical value of the parameters $\epsilon  $ and $\sigma $ were fixed to $\epsilon =1.27$ and $\sigma =0.76$, respectively.}
 \label{rf2}
\end{figure}

\begin{figure}
 \centering
 \includegraphics[scale=0.70]{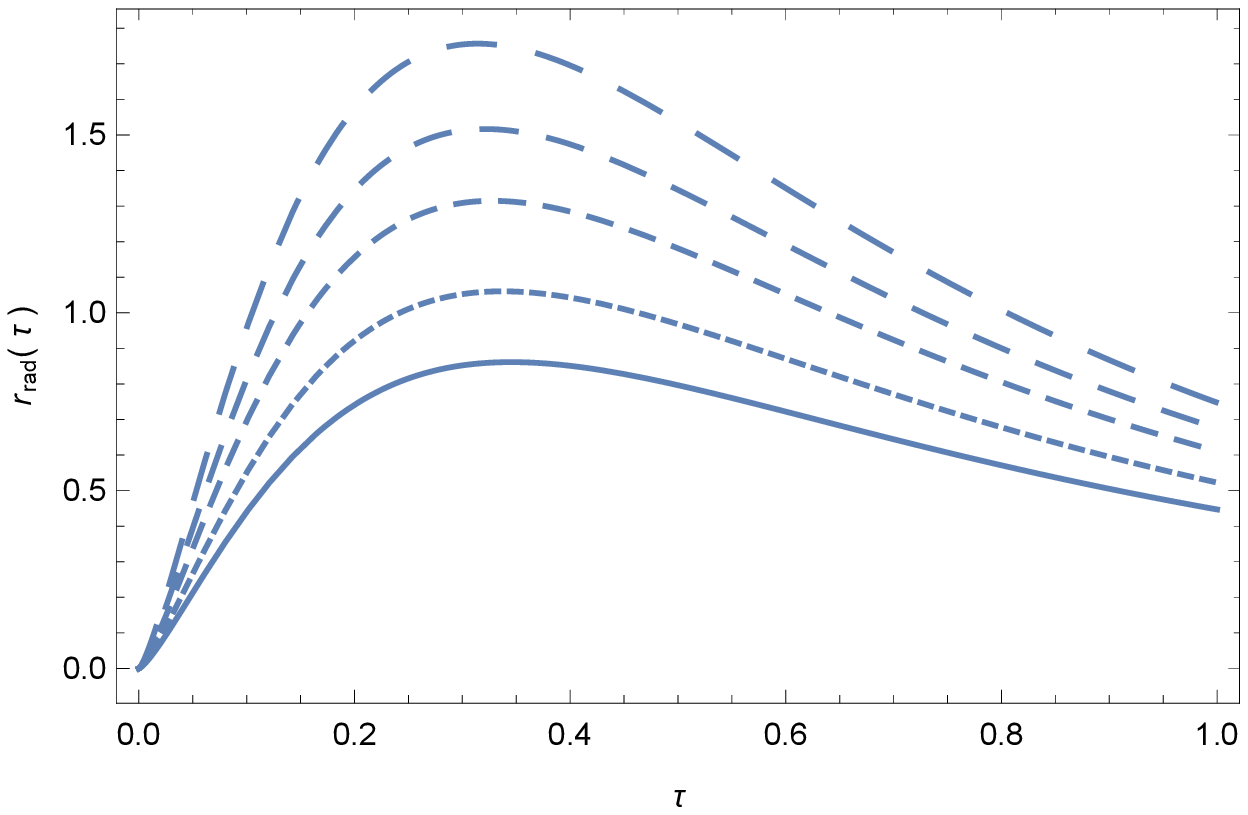}
 \caption{Time variation of the dimensionless radiation fluid  energy density $r_{rad}$ in the warm inflationary scenario with
 an interacting scalar field - radiation fluid filled Universe, for the case of the Higgs type potential of the scalar field  with $\mu ^2>0$, for different values of the parameter  $\gamma $: $\gamma =0.99$ (solid curve), $\gamma  =1.24$ (dotted curve),  $\gamma =1.57$ (dashed curve), $\gamma =1.84$ (long dashed curve), and $\gamma =2.17$ (ultra long dashed curve), respectively.
 %The initial conditions used for the numerical integration of the cosmological
 %evolution equations are $a(0)=10^{-3}$, $N_{\phi }(0)=10^5$, $\phi (0)=5$, $u(0)=-1.25$, $r_{rad}(0)=10^{-6}$, and $\theta (0)=10^{-3/2}$, % respectively.
 The numerical value of the parameters $\epsilon  $ and $\sigma $ were fixed to $\epsilon =1.27$ and $\sigma =0.76$, respectively.}\label{rad2}
\end{figure}

\begin{figure}
 \centering
 \includegraphics[scale=0.70]{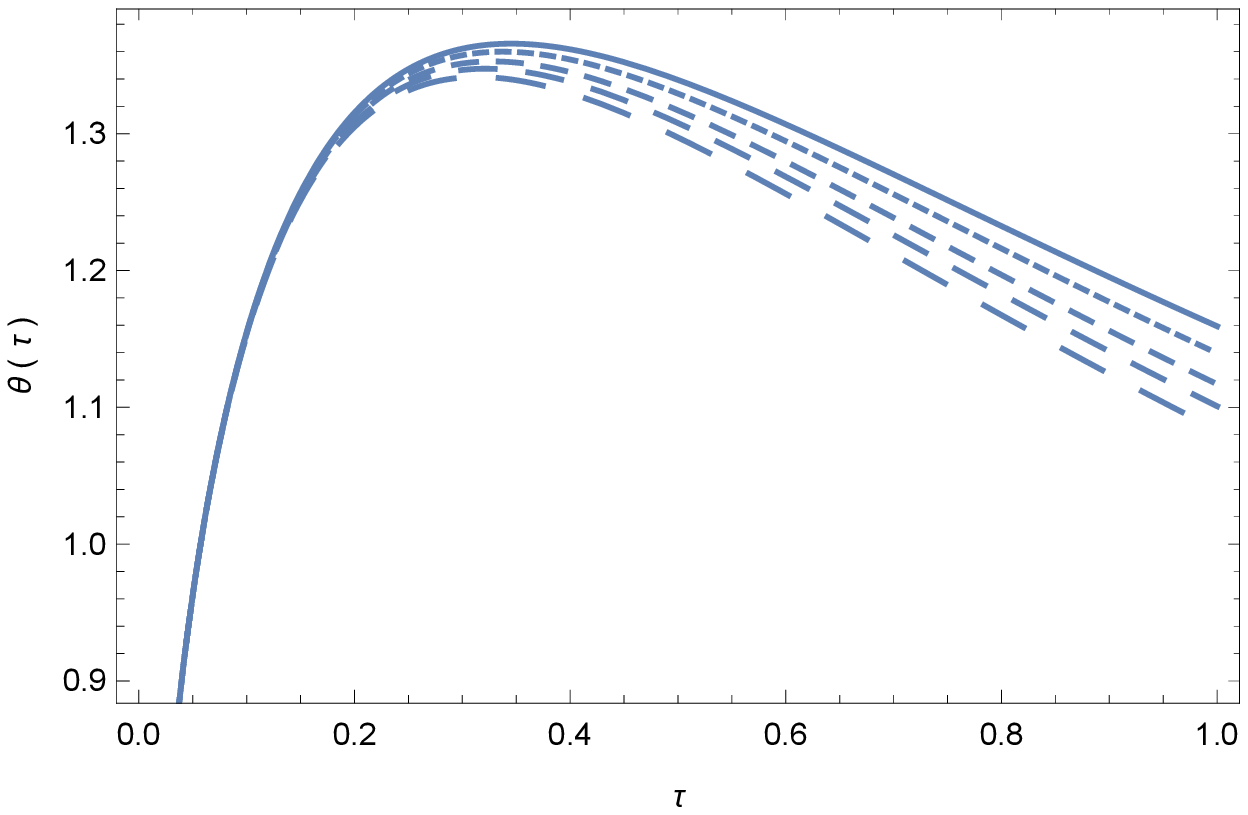}
 \caption{Time variation of the dimensionless temperature $\theta $  in the warm inflationary scenario with
 an interacting scalar field - radiation fluid filled Universe, for the case of the Higgs type potential of the scalar field  with $\mu ^2>0$, for different values of the parameter  $\gamma $: $\gamma =0.99$ (solid curve), $\gamma  =1.24$ (dotted curve),  $\gamma =1.57$ (dashed curve), $\gamma =1.84$ (long dashed curve), and $\gamma =2.17$ (ultra long dashed curve), respectively.
 %The initial conditions used for the numerical integration of the cosmological
 %evolution equations are $a(0)=10^{-3}$, $N_{\phi }(0)=10^5$, $\phi (0)=5$, $u(0)=-1.25$, $r_{rad}(0)=10^{-6}$, and $\theta (0)=10^{-3/2}$, %respectively.
 The numerical value of the parameters $\epsilon  $ and $\sigma $ were fixed to $\epsilon =1.27$ and $\sigma =0.76$, respectively.}\label{temp2}
\end{figure}

\begin{figure}
 \centering
 \includegraphics[scale=0.70]{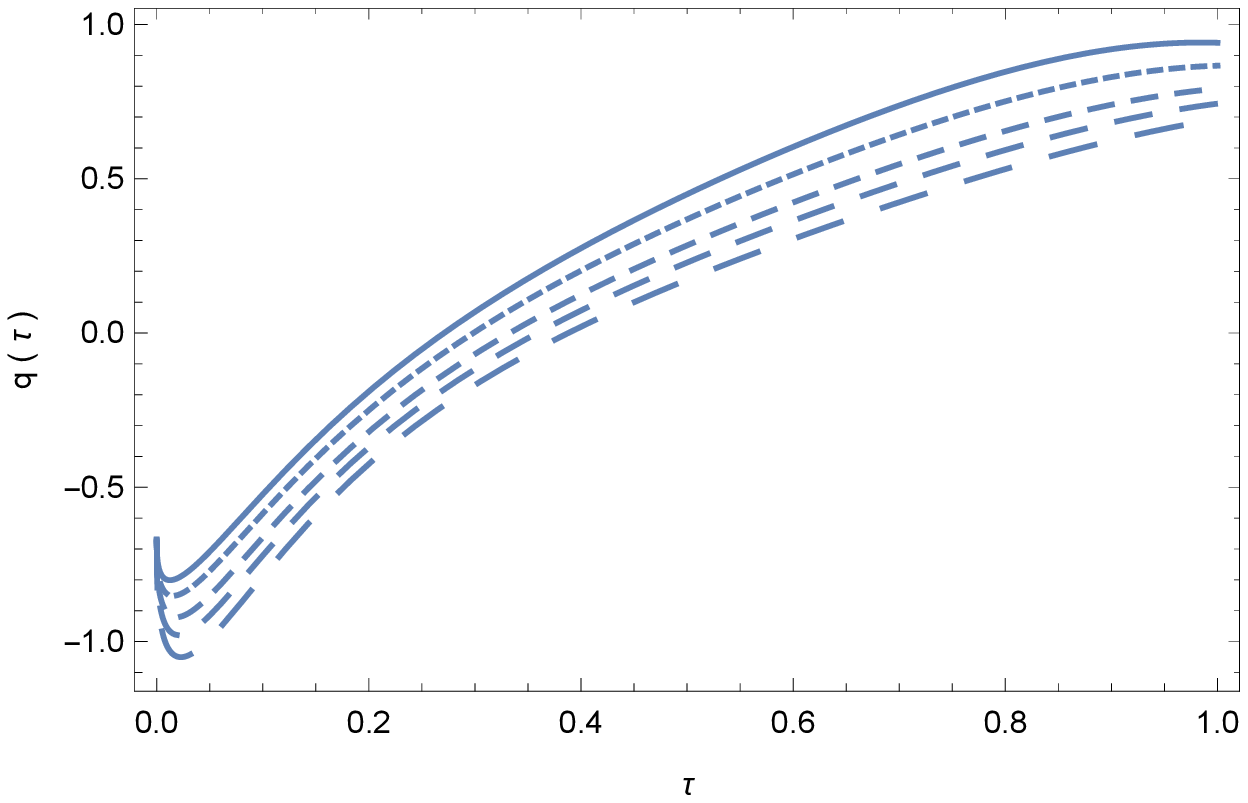}
 \caption{Time variation of the deceleration parameter $q $ in the warm inflationary scenario with
 an interacting scalar field - radiation fluid filled Universe, for the case of the Higgs type potential of the scalar field  with $\mu ^2>0$, for different values of the parameter  $\gamma $: $\gamma =0.99$ (solid curve), $\gamma  =1.24$ (dotted curve),  $\gamma =1.57$ (dashed curve), $\gamma =1.84$ (long dashed curve), and $\gamma =2.17$ (ultra long dashed curve), respectively.
 %The initial conditions used for the numerical integration of the cosmological
 %evolution equations are $a(0)=10^{-3}$, $N_{\phi }(0)=10^5$, $\phi (0)=5$, $u(0)=-1.25$, $r_{rad}(0)=10^{-6}$, and $\theta (0)=10^{-3/2}$, respectively.
 The numerical value of the parameters $\epsilon  $ and $\sigma $ were fixed to $\epsilon =1.27$ and $\sigma =0.76$, respectively.}\label{q2}
\end{figure}

Next, we consider the warm inflationary evolution for a positive $\mu ^2$, that is, we adopt for the Higgs potential the expression $U(\phi) = \mu^2\phi ^2/2+\xi \phi ^4/4$, with $\mu ^2>0$. The time variations of the scale factor, scalar field particle number, scalar field energy density, radiation energy density, and of the temperature of the radiation fluid are represented for this second form of the Higgs potential in Figs.~\ref{a2}-\ref{q2}.

From a qualitative point of view the cosmological dynamics for $\mu ^2>0$ is very similar to the case with $\mu ^2<0$. However, at least for the adopted range of numerical values of the model parameters, important qualitative differences do appear, with the cosmological evolution taking place at a lower pace. The scale factor, presented in Fig.~\ref{a2}, is a monotonically  increasing function of time, whose early evolution os basically independent on the model parameters. However, the late time behavior shows a significant dependence on $\gamma$. The scalar field particle number, depicted in Fig.~\ref{N2}, decreases during the cosmic expansion in an almost $\gamma $ - independent way. The variation of the scalar field energy density, depicted in Fig.~\ref{rf2}, a decreasing function of time,  shows a very small dependence on $\gamma$. However, the parameter $\gamma $ has a significant effect on the evolution of the radiation energy density, shown in Fig.~\ref{rad2}. Due to the decay of the scalar field, the radiation energy density increases in the early periods of the cosmological expansion, and reaches a maximum value at a finite time interval $\tau _{max}$, which is very different as compared to the $\mu ^2<0$ case. For time intervals $\tau >\tau _{max}$, the radiation production significantly decreases, and the evolution of the matter content is governed by the expansion of the Universe. The evolution of the temperature, represented in Fig.~\ref{temp2} has a similar evolution to that of the photon gas. The deceleration parameter, represented in Fig.~\ref{q2}, shows that in its initial stage the Universe was in an accelerating state, with the deceleration parameter having values of the order of $q\approx -0.7$. The expansion did accelerate towards a de Sitter phase, reached in a very short time interval, and followed with a slow decelerating evolution. At the finite moment $\tau =\tau _{max}$, the deceleration parameter becomes zero, indicating the end of inflation. This moment also corresponds to the radiation density reaching its maximum value. For $\tau >\tau _{max}$, the deceleration parameter has positive values, indicating a decelerating dynamics of the Universe.

\section{Constraints from Planck-2018 results}\label{sect5a}

We will now compare the results of the warm inflationary model with irreversible matter creation, developed in the framework of the thermodynamics of open systems in the previous Sections, with the observational data obtained with the Planck satellite. The Planck observational data have been extensively used to test the theoretical predictions of the warm inflationary models.

The reconstruction of a warm inflationary Universe model from the scalar spectral index $n_S(\mathcal{N})$ and the tensor to scalar ratio $r(\mathcal{N})$ as a function of the number of e-folds $\mathcal{N}$ was studied \cite{w36}. The effective potential and the dissipative coefficient were found in terms of the cosmological parameters $n_S$ and $r$ considering the weak and strong dissipative regimes under the slow roll approximation. The warm inflation observational predictions were confronted with the latest CMB data in \cite{w38}, by considering  a linear temperature dependent dissipative coefficient, and the simplest model of inflation, a quartic chaotic potential. Although excluded in the standard cold inflation version, dissipation reduces the tensor-to-scalar ratio and brings the quartic chaotic model within the observable allowed range.
 The constraints that Planck 2015 temperature, polarization and lensing data impose on the parameters of warm inflation were revisited in \cite{w39}, by studying warm inflation driven by a single scalar field with a quartic self interaction potential in the weak dissipative regime. The effects of the parameters of warm inflation, namely, the inflaton self coupling $\lambda$  and the inflaton dissipation parameter $Q_P$ on the CMB angular power spectrum were investigated.  Moreover, $\lambda $ and $Q_P$ were constrained  for 50 and 60 number of e-foldings with the full Planck 2015 data by performing a Markov-Chain Monte Carlo analysis.  Two-field warm inflation models with a double quadratic potential and a linear temperature dependent dissipative coefficient were studied in \cite{w46}. The scalar spectral index $n_s$ and the tensor-to-scalar ratio $r$ was computed for several representative potentials, and the results were compared with the observational data. Warm inflationary scenarios in which the accelerated expansion of the early Universe is driven by chameleon-like scalar fields were investigated in \cite{w47}, and the model was constrained by using Planck 2018 data.

\subsection{Coherent scalar field dominated warm inflation}

As a first example of the comparison of our theoretical predictions  with the cosmological observations, i.e. Planck-2018 results,  we are going to estimate the free parameters of the warm inflationary scenario for the model introduced in Section~\ref{Coherent scalar field-radiation}, corresponding to the case of the coherent scalar field driven cosmological evolution. Even with the inclusion of the extra degrees of freedom, which the irreversible thermodynamic approach  brings into the model,  we still can obtain for this model the exact representations for the most important cosmological parameters, which would allow us to test the cosmological predictions of the theory. For instance  the Hubble parameter $H$, is given by Eq.~(\ref{72}).  From Eq.~(\ref{thetacoh}),  the temperature, $T$ of the photon gas can be obtained as
\begin{eqnarray}\label{Tcoh}
\hspace{-0.9cm}T &=&\left(\frac{\Gamma _1^2M_P^2}{24\pi \zeta(3)m_{\phi}}\right)^{1/3}\theta _{rad}^{1/3}=\nonumber\\
\hspace{-0.9cm}&&T_{0}\left\{ \frac{\left( 1+2\theta
_{rad}^{(0)}/\tilde{h}_{0}^{2}\right) e^{2\left( \tau -\tau _{0}\right) }-1}{%
\left[ \left( 1+1/\tilde{h}_{0}\right) e^{\tau -\tau _{0}}-1\right] ^{2}}%
\right\} ^{1/3},
\end{eqnarray}
where we have assumed $\Gamma _1=\Gamma _2$, and we have denoted  $T_0  = \left(\Gamma _1^2M_P^2/48\pi \zeta (3)m_{\phi}\right)^{1/3}$.

We point out that in the warm inflationary proposal, there are both quantum and thermal fluctuations. Thermal fluctuations depend on $T$, while quantum fluctuations depend on $H$.  One important feature of warm inflation is that the fluid temperature must be bigger than the Hubble parameter, $T>H$, a condition stating that the thermal fluctuations overcome quantum fluctuations, and thus they become the origin of the Universe's large scale structures. Now in order to investigate the observational constraints on the warm inflationary era we introduce the first, and second slow-roll parameters, which are defined as
\be\label{slowroll1}
\epsilon _1 =q+1=  - \frac{\dot {H}}{H^2}=- \frac{3}{2\tilde {h}^2}\frac{d\tilde {h}}{d\tau }=\frac{3}{2}\left(1+\frac{1}{\tilde{h}_0}\right)e^{\tau -\tau _0},
\ee
and
\bea\label{slowroll2}
\epsilon_2&=&\frac{\dot{\epsilon _1}}{H{\epsilon _1}}
\equiv  \frac{\ddot{H}}{H\dot{H}}-\frac{2\dot{H}}{H^2}=\frac{3}{{2{{\tilde h}^2}}}\frac{{{d^2}\tilde h/d{\tau ^2}}}{{d\tilde h/d\tau }} - \frac{3}{{{{\tilde h}^2}}}\frac{{d\tilde h}}{{d\tau }}=\nonumber\\
&&\frac{3}{2}\left[\left(1+\frac{1}{\tilde{h}_0}\right)e^{\tau -\tau _0}-1\right].
\eea

The amount of cosmic expansion during  inflation is measured through the number of e-folds $\mathcal{N}$, defined as
\bea\label{efoldtau}
\hspace{-0.3cm}\mathcal{N} &=& \int_{t_\star}^{t_e} H \; dt = \frac{2}{3}\int_{\tau_\star}^{\tau_e} {\tilde{h}} \; d\tau~=\nonumber\\
\hspace{-0.3cm}&&\frac{2}{3}\left[ \tau _{\ast }-\tau _{e}+\ln \frac{\left( 1+\tilde{h}%
_{0}\right) e^{\tau _{e}-\tau _{0}}-\tilde{h}_{0}}{\left( 1+\tilde{h}%
_{0}\right) e^{\tau _{\ast }-\tau _{0}}-\tilde{h}_{0}}\right].
\eea

 Once the expression of the re-scaled time at the end of inflation, $\tau_{e}$,  is obtained from the relation $\epsilon_1(\tau_e)=1$, Eq. \eqref{efoldtau} is used to derive the re-scaled time at the time of the horizon crossing, i.e. $\tau_\star$, in terms of the number of e-folds. Then, all the perturbation parameters can be expressed in terms of the number of e-folds.

Therefore by means of Eq.~(\ref{slowroll1}) at the end of inflation we obtain
\be\label{tauend}
{\tau _e} = {\tau _{0}} + \ln \left[\frac{2}{{3\left(1 + 1/\tilde {h}_0\right)}}\right].
\ee
Then from Eqs.(\ref{efoldtau}) for the horizon exit time we find
\bea\label{tauend}
{\tau _ \star } = \ln 2 - \ln \Bigg\{ \left[ {\frac{{(1 + {{\tilde h}_0})\left( {2{e^{3\mathcal{N}/2}} + 1} \right)}}{{{{\tilde h}_0}}}} \right] \times {e^{ - \frac{{3N}}{2} - {\tau _0}}}\Bigg\}.  \nonumber\\
\eea

To investigate the accuracy of a theoretical model, one has to perform a comparison between its predictions with the observational data. In this regard, we will obtain some important perturbation parameters, such as the amplitude of scalar perturbations, the scalar spectral index and the tensor-to-scalar ratio,  and they will be compared with the Planck-2018 data \cite{pl2018a,planck2018b}. Following \cite{Bastero-Gil:2016qru,Berera:2018tfc,w16,w47}, the amplitude of the scalar perturbations $\mathcal{P}_s$ can be obtained as
\begin{equation}\label{pswarm}
\mathcal{P}_s = \frac{25}{4} \left( { H \over \dot\phi }\right)^2 \left(\delta\phi \right)^2,\;\;\left(\delta\phi \right)^2=\frac{\sqrt{\Gamma_1 H}T}{2\pi^2}\,.\\
\end{equation}

From Eqs.~(\ref{phicoh}), (\ref{72}), (\ref{Tcoh}) and (\ref{pswarm})  we obtain
\begin{eqnarray}\label{PS}
\mathcal{P}_{s}=\frac{{25{\Gamma _{1}T}}_{0}}{{24 \sqrt{3}{\pi ^{2}M_P^2}}}\frac{\left[ \left( 1+2\theta _{rad}^{(0)}/\allowbreak
\tilde{h}_{0}^{2}\right) e^{2\left( \tau -\tau _{0}\right) }-1\right] ^{1/3}%
}{\left[ \left( 1+1/\tilde{h}_{0}\right) e^{\tau -\tau _{0}}-1\right] ^{7/6}}.\nonumber\\
\end{eqnarray}

The scalar spectral index $n_s$ is obtained from the amplitude of the scalar perturbations, and it is defined as
\begin{eqnarray}\label{nsaswarm}
n_s - 1 &=& {d\ln(\mathcal{P}_s) \over d\ln(k)}={d\ln(\mathcal{P}_s) \over d\ln(a)}\frac{d\ln (a) }{d\ln (k)}=\frac{1}{H}\frac{d}{dt}\ln(\mathcal{P}_s)=\nonumber\\
&&\frac{3}{2}\frac{1}{\tilde{h}}\frac{d}{d\tau}\ln \mathcal{P}_{s} (\tau),
\end{eqnarray}
 where we have assumed that $\frac{{d\ln (a)}}{{d\ln (k)}}=1$. Hence we obtain
 \begin{eqnarray}\label{ns2}
&&\hspace{-0.5cm}n_{s}(\tau )=\nonumber\\
&&\hspace{-0.5cm}\frac{4\tilde{h}_{0}^{3}e^{3\tau _{0}}+3(\tilde{h}_{0}+1)\left(
\tilde{h}_{0}^{2}+2\text{$\theta $}_{rad}^{(0)}\right) e^{3\tau }-7(\tilde{h}%
_{0}+1)\tilde{h}_{0}^{2}e^{\tau +2\tau _{0}}}{4\tilde{h}_{0}^{3}e^{3\tau
_{0}}-4\tilde{h}_{0}\left( \tilde{h}_{0}^{2}+2\text{$\theta $}%
_{rad}^{(0)}\right) e^{2\tau +\tau _{0}}}.\nonumber\\
\end{eqnarray}

Tensor perturbations, known as gravitational waves, are measured indirectly through the  tensor-to-scalar ratio parameter $r=\mathcal{P}_t / \mathcal{P}_s$. The amplitude of the tensor perturbations is given by \cite{Bastero-Gil:2016qru}
\begin{equation}\label{ptwarm}
\mathcal{P}_t = {2 H^2 \over  M_P^2 \pi^2}=\frac{{2\Gamma _1^2}}{{9M_P^2 \pi ^2 {{\left[ {  {{\rm{e}}^{\tau  - \tau 0}}\left( {1 + \frac{1}{{{\tilde {h}_0}}}} \right)}-1 \right]}^2}{}}}\;.
\end{equation}
The tensor spectral index is defined as
\begin{eqnarray}
n_{t} &=&{\frac{d\ln (\mathcal{P}_{t})}{d\ln (k)}}=\frac{1}{H}\frac{d}{dt}%
\ln (\mathcal{P}_{t})=\nonumber\\
&&\frac{2}{H}\frac{d}{dt}\ln H=
\frac{2\dot{H}}{H^{2}}%
=-2\left( q+1\right),
\end{eqnarray}
giving
\be\label{ntwarm}
n_t=-3\left( 1+\frac{1}{\tilde{h}_{0}}\right) e^{\tau -\tau _{0}}\;,
\ee
To compare our results with observations we can considered the consistency relation to obtain the tensor to scalar ratio parameter {\it i.e.} $r=-8n_t$.
The last terms in both Eqs.(\ref{ptwarm}) and (\ref{ntwarm}) are obtained by considering the reparameterized Hubble parameter introduced in Eq.~(\ref{72}).

The constant parameter $\Gamma_1$ can be determined by using the observational data for the amplitude of the scalar perturbations, ${\cal P}_s\left(\tau _\star\right)={\cal P}_s^ \star=2.17 \times 10^{-9}$ at the horizon exit time as
\begin{eqnarray}
&&\hspace{-0.5cm}\Gamma _{1}^{5/3}=\frac{24^{4/3}\sqrt{3}\pi ^{7/3}\left[
\zeta (3)m_{\phi }M_{P}^{4}\right] ^{1/3}\mathcal{P}_{s}^{\star }(1+%
\tilde{h}_{0})^{7/6}}{25\sqrt{(1+\tilde{h}_{0})\left(
2e^{3\mathcal{N}/2}-1\right) }}\times   \nonumber \\
&&\hspace{-0.5cm}\left[ 4(1+\tilde{h}_{0})^{2}e^{3\mathcal{N}/2}+4e^{3\mathcal{N}}(2\theta
_{rad}^{(0)}-2\tilde{h}_{0}-1)-(1+\tilde{h}_{0})^{2}\right] ^{-\frac{1}{3}}.\nonumber \\
\end{eqnarray}

Now with the help of Eqs.~(\ref{tauend}), (\ref{ntwarm}) and (\ref{ns2}), and also by using the expression of  $\Gamma_1$ calculated in the above equation, we can estimate the free parameters of the model. We present the obtained results in Table~\ref{Table1}.

\begin{table}[h]
\begin{center}
  \centering
 \begin{tabular}{|l|l|l|l|l|}
 \hline
$\theta _{rad}^{(0)}$ & $\tilde{h}_{0}$ & $m_{\phi }\;[{\rm GeV}]$ & $r$ & $n_{s}$ \\
\hline
$0.01$ & $ 10$ & $1.1 \times 10^{-46} $ & $0.166$ & $0.9880$ \\
\hline
$0.05$ & $15$ & $2.16 \times 10^{-47}$ & $0.161$ & $0.9884$ \\
\hline
$0.09$ & $20$ & $6.74 \times 10^{-48}$ & $0.158$ & $0.9886$ \\
\hline
$0.10$ & $25$ & $2.75 \times 10^{-48}$ & $0.157$ & $0.9887$ \\
\hline
$0.50$ & $30$ & $1.32 \times 10^{-48}$ & $0.156$ & $0.9887$ \\
\hline
\end{tabular}
\caption{ The comparison of the perturbation parameters $n_s$ and $r$, obtain in the warm inflationary scenario with Barrow-Saich type potential, with the PlanckTT+LowP+Lensing (Planck2015) results  {\cite{{Pl5}}}. To perform this analysis we have fixed the value of the reparameterized time as $\tau=0.02$, and we fixed $\tau_0=5.1$. To avoid imaginary or negative values for cosmological parameters we have adopted for $\Gamma _1$ the value  $\Gamma_1\simeq10^{-2}\;{\rm GeV}$. To estimate the free parameters of the model we have used $\mathcal{N}=65$ for the number of efolds.}\label{Table1}
\end{center}
\end{table}

 The Planck full mission temperature data and a first release of polarization data on large angular scales gives for the spectral index of curvature perturbations the value $n_s = 0.968 \pm 0.006$, while the upper bound on the tensor-to-scalar ratio is $r_{0.002}< 0.11$ (95$\%$ CL) \cite{Pl5}. This upper limit is consistent with the B-mode polarization constraint $r< 0.12$ (95\% CL), obtained from a joint analysis of the BICEP2/Keck Array and Planck data \cite{Pl5}.  As one can see from the Table~\ref{Table1}, the comparison of the theoretical predictions with the observational data show an acceptable concordance between the theoretical model, and the observational results, with $r$ and $n_s$ taking on the edge values for the considered  model parameters. Hence, despite their simplicity, irreversible warm inflationary models with Barrow-Saich potential  can give at least a qualitative description of warm inflation, once the thermodynamic of irreversible processes has been properly included in the general formalism.

\subsection{Higgs potentials and slow-roll conditions}\label{Higgs-slow-rolls}

As a next step in our investigation of the cosmological constraints on the warm inflationary models with irreversible photon fluid production we will investigate the case of the Higgs potential driven inflation. In both the warm and cold inflationary scenarios one can usually explain the accelerated expansion phase of the Universe taking place at very high energy scales, in which the scalar field is the dominant component. Nevertheless,  the very important difference between these two approaches is that besides a scalar field, in the warm inflationary scenario another component of the cosmological fluid is present, usually taken as radiation, with the two components interacting. But a correct thermodynamic description of particle production needs to take into account the irreversible nature of the process, as well as the fact that during inflation the Universe represents an open system. In the case of a scalar field with a Higgs type potential $U(\phi)=\pm \left(\mu ^2\right)/2+\xi \phi ^4/4$, the evolution equation of the scalar field and of the cosmological fluid are given by Eqs.~(\ref{fa})-(\ref{fe}), with the generalized Friedmann equations taking the form
\begin{equation}\label{warmfriedmann}
\hspace{-1.2cm}3H^2 = \frac{1}{M_P^2}\left[{1 \over 2} \dot\phi^2 + U(\phi) + \rho_{rad}\right],
\end{equation}
\begin{equation}
 2\dot{H} = -\frac{1}{M_P^2}\left[\dot\phi^2 + {4 \over 3} \rho_{rad}+\frac{\rho _{\phi}\Gamma _1}{3m_{\phi}H}\left(\frac{\dot{\phi}^2}{n_{\phi}}-\frac{4}{3}\frac{\rho _{rad}}{n_{rad}}\right)\right].
\end{equation}
Due to the irreversible  interaction between scalar field and radiation, there is an energy flux from the scalar field to the photon gas,  and this process is described through the  conservation equations, namely Eqs.~(\ref{fb}) and (\ref{fd}).

 Additionally, the Klein-Gordon equation, Eq. \eqref{scaleq} or \eqref{fc},  known as the equation of motion of the scalar field, can be rewritten as
\begin{equation}\label{warmeom}
\ddot{\phi} + 3H \left[1+Q(\phi)\right] \; \dot{\phi} + U'(\phi) =0\;,
\end{equation}
where the parameter
\bea
Q(\phi)&=&\frac{\Gamma(\phi)}{3H(\phi)}=\frac{\Gamma _1\rho _{\phi}}{3m_{\phi}n_{\phi}H}=\frac{p_c^{(\phi)}}{\dot{\phi}^2}=\nonumber\\
&&\frac{\Gamma _1\left(\dot{\phi}^2/2\pm \mu^2\phi^2/2+\xi \phi ^4/4\right)}{3m_{\phi}n_{\phi}H},
\eea
 is the ratio of the radiation production to the expansion rate.

 At this moment we would like to point out that in warm inflation the slow-roll approximations are still valid, that is,  in order to have a quasi-de Sitter expansion, the rate of the Hubble parameter during a Hubble time must be small.  This condition  is imposed via the first slow-roll parameter, defined  in Eq. \eqref{slowroll1}.
%\begin{equation}\label{epsilon}
%\epsilon_1 = - {\dot{H} \over H^2}.
%\end{equation}\
In the following we will introduce a number of approximations. First of all, we assume that the energy density of the scalar field dominates over the photon fluid energy density. Secondly,  we also suppose that the kinetic term of the scalar field is negligible as compared to its potential term, that is, we have  $\rho_\phi \gg  \rho_{rad}$ and $\rho_\phi \simeq U(\phi)$, respectively. Moreover, similarly to the assumptions that we also have in super cold inflation, in warm inflationary scenario we expect that the photon production rate is quasi-stable during inflationary expansion, so that ${\dot\rho_{rad}} \ll H\rho_{rad}$ and ${\dot\rho_{rad}} \ll \Gamma \dot\phi^2$, respectively. These constraints lead to the result that the friction term in Eq.~\eqref{slowroll1} is much larger than the  term $\ddot{\phi}$.

Then from Eqs. \eqref{warmfriedmann}, \eqref{warmeom}, and \eqref{fd} it follows that the description of the cosmological expansion can be reduced to the following approximate equations
\begin{eqnarray}\label{radiationtemp}
 \hspace{-0.5cm} &&3H^2  \simeq  \frac{1}{M_P^2}U(\phi)\;, \label{vfriedmann} \\
 \hspace{-0.5cm}&&\dot\phi  \simeq  - {U'(\phi) \over 3H\left[1+Q(\phi)\right]}\;, \label{dotphi} \\
 \hspace{-0.5cm}&& \rho _{rad}=\frac{8\pi ^5}{15}T^4 \simeq \frac{\pi^4\Gamma_1}{90\zeta(3)m_{\phi}}\frac{T}{H}\rho_{\phi}=\frac{\pi^2}{30\zeta (3)}Tn_{\phi}Q\;, \label{rhoweak}
\end{eqnarray}
where $T$ is the temperature of the photon gas.

By using the definitions introduced in \eqref{slowroll1} and \eqref{slowroll2}, it follows that in terms of the field potential and of $Q$ the slow-roll parameters can be expressed as
\begin{equation}\label{vsrp}
\epsilon_1 = {M_P^2 \over 2(1+Q)} {U^{\prime 2}(\phi) \over U^2(\phi)}\;, \qquad \epsilon_2 = {\dot{\epsilon}_1 \over H \epsilon_1}\;.
\end{equation}

In calculations related to the power spectrum and spectral indices it is common to introduce two other slow-roll parameters, which are defined in terms of the potential and of $Q$ as,
\begin{equation}\label{etabeta}
\eta = {M_P^2 \over (1+Q)} \; {U''(\phi) \over U(\phi)},\; \beta = {M_P^2 \over (1+Q)} \; {U'(\phi) \Gamma'(\phi) \over U(\phi) \Gamma(\phi)}\;.
\end{equation}

The slow-roll parameter $\epsilon_2$ can be expressed in terms of the above slow-roll parameters through the relation
\begin{equation}
  \epsilon_2 = -2 \eta + 4 \epsilon_1 + {Q \over (1+Q)} \; \left( \beta - \epsilon_1 \right)\;.
\end{equation}

Due to the appearance of the term $(1+Q)$ in the dominator of the slow-roll parameters, it follows that their smallness can be guaranteed for a large range of potentials that satisfy the slow-roll approximations.

The smallness of the slow-roll parameters indicates that first the Universe had a quasi-exponential accelerated expansionary phase, in which it did stand  for an enough high number of e-folds so that the standard problems of the hot Big Bang model can be solved.  The number of e-folds, defined in  Eq.~\eqref{efoldtau}, can be rewritten as functions of the scalar field as,
\begin{equation}\label{efold}
\mathcal{N} = \int_{\phi_\star}^{\phi_e} {H \over \dot\phi} \; d\phi =
- \int_{\phi_\star}^{\phi_e} \frac{(1+Q)}{M_P^2} {U(\phi) \over U'(\phi)} \; d\phi\;,
\end{equation}
where the last equality is obtained by using Eqs.~\eqref{vfriedmann} and \eqref{dotphi}, respectively.

All the results presented above depend on the functional form of the dissipation coefficient $\Gamma _1$, which generally can be assumed to be a function of the thermodynamic parameters. Such a dependence may strongly affect the cosmological dynamics, as well as the particle creation/decay processes. For example, by assuming a power-law dependence of $\Gamma _1$ on the temperature, so that $\Gamma _1=\Gamma _0T^{\alpha}$, where $\Gamma _0$ and $\alpha $ are constants, for the case of the Higgs potential the parameter $Q$ becomes
\be
Q(\phi,T)=\frac{\Gamma _0T^{\alpha}\left(\dot{\phi}^2/2\pm \mu^2\phi^2/2+\xi \phi ^4/4\right)}{3m_{\phi}n_{\phi}H},
\ee
while the radiation energy density varies according to
\be
\rho _{rad}\left(\phi,T\right)\approx\frac{\pi^4\Gamma_0}{90\zeta(3)m_{\phi}}\frac{T^{\alpha +1}}{H}\rho_{\phi}\;.
\ee
Hence for the case of a temperature varying dissipation coefficient the energy density of the radiation may increase much faster as compared to the constant $\Gamma _1$ case. The slow-roll parameters, as well as the number of e-folds will become explicit functions not only of the scalar field, but also of the photon temperature. However, in the present study in order to perform a full comparison of the theoretical model with the observations, in the following we will adopt the simplifying assumption of constant dissipative coefficients.

The primordial curvature spectrum generated during warm inflation is an important observational indicator that can be used to observationally discriminate between different models. The primordial spectrum is dominated by the thermal fluctuations of the radiation
fluid, generated due to the dissipative effects in the scalar field \cite{w9, w9a,w9c,w11a}.  However, since dissipation means departures from
the equilibrium state, other physical effects like the presence of a shear viscous pressure in the photon fluid may also play an important role \cite{w9c}.\emph{}
The scalar power spectrum in warm inflation is given by \cite{w47}
\begin{equation}
{\cal P}_\zeta=\frac{H^2(1+Q)^2{\cal F}}{8\pi^2\epsilon _1M_P^2}\,,
\label{Pz}
\end{equation}
where
\be
{\cal F}\equiv 1+2{\cal N}_*+\frac{T}{H}\left(\frac{2\pi Q}{\sqrt{1+4\pi Q/3}}\right),
\ee
and ${\cal N}_*=\left(e^{H/T}-1\right)^{-1}$ is the statistical distribution of
the scalar field at the horizon crossing. Since in warm inflation the condition $T>> H$ must hold, we can approximate
${\cal N}_*\simeq T/H\gg 1$. Moreover, by assuming $Q<1$, we can approximate ${\cal F}$ as ${\cal F}\simeq 2(1+\pi Q)T/H$. From Eq.~(\ref{rhoweak}) we obtain for the temperature of the photon gas the expression
\be
T=\left(16\pi ^{3}\zeta(3)\right)^{-1/3}n_{\phi}^{1/3}Q^{1/3},
\ee
which allows to obtain the power spectrum in  the irreversible thermodynamic description of warm inflation as
\be
{\cal P}_{\zeta}\approx \frac{1}{8\left(2\zeta (3)\right)^{1/3}\pi^3M_P}\frac{Q^{1/3}\left(1+Q\right)^2\left(1+\pi Q\right)}{\epsilon _1}\frac{H}{M_P}n_{\phi}^{1/3}.
\ee

In the opposite limit of the strong dissipation we can approximate ${\cal F}$ as ${\mathcal F}\approx \sqrt{3\pi Q}\left(T/H\right)$, giving for the primordial power spectrum of the irreversible warm inflation the expression
\be
{\cal P}_{\zeta}\approx\frac{\sqrt{3}}{16\pi ^{5/2}M_P}\frac{Q^{8/3}}{\epsilon _1}\frac{H}{M_P}n_{\phi}^{1/3}.
\ee

The relations are different from those obtained in the standard warm inflation model \cite{w9, w9a,w9c,w11a, w47}, with the differences arising from the different behavior of the photon gas temperature, as obtained from Eq.~(\ref{rhoweak}). The primordial inflationary power spectrum depends explicitly on the scalar field particle number, and such a dependence does appear in both weak and strong dissipation limits.

\subsubsection{Weak regime}

In the weak dissipative regime  we suppose that the dissipative ratio is much smaller than unity, $Q\ll1 (\Gamma \ll 3H)$, and thus we have $(1+Q) \simeq 1$.  By taking into account the relation between the dissipative ratio and the creation pressure for the scalar field, as given by Eq.~(\ref{Qpc}), which gives $Q=p_c^{(\phi)}/\dot{\phi}^2$, the condition for weak dissipation $Q<<1$,  is equivalent to the condition $p_c^{(\phi)}<<\dot{\phi}^2$, showing that in this limit the creation pressure associated to the scalar field is much smaller than the kinetic energy of the field. From a thermodynamic point of view weak dissipation implies the presence of low intensity particle creation processes.

The weak dissipative approximation makes the evolution equations much easier to analyze, since by virtue of Eqs.~\eqref{vfriedmann} and \eqref{dotphi} the time derivative of the scalar field is expressed as
\begin{equation}\label{dotphiweak}
\dot{\phi} \simeq - \frac{{U'(\phi )}}{{3H(\phi )}} = \frac{{ \mp \mu \phi  - \xi {\phi ^3}}}{{\sqrt 3 \sqrt { \pm \frac{{{\mu ^2}}}{2}{\phi ^2} + \frac{\xi }{4}{\phi ^4}} }}\;.
\end{equation}
Also, by taking  this approximation into account, with the use of Eqs.~\eqref{radiationtemp} and (\ref{rhoweak}), the temperature of the fluid is obtained easily as a function of the scalar field as
\begin{equation}\label{tempweak}
T(\phi) = T_{0w}\left(\pm \frac{{{\mu ^2}}}{2}{\phi ^2} + \frac{\xi }{4}{\phi ^4}\right)^{1/6}\;,
\end{equation}
where $T_{0w}  = \left(\sqrt{3} \Gamma _1 M_P /48\pi \zeta (3)m_{\phi}\right)^{1/3}$.

In  the investigations of warm inflation based on the thermodynamics of closed and adiabatic systems, to reduce the degrees of freedom of the model,  usually one introduces an ansatz for the dissipation coefficient function, $\Gamma(\phi)$,  as a mixture of power-law and inverse-power-law functions of the scalar field. But in the present approach of the thermodynamics of open systems we can reduce the degrees of freedom, and we can find the solution of the full set of evolution equations without the need of any strong supplementary assumptions (for the exact results on the Higgs potential see Subsection \ref{PositiveHiggs}).

\paragraph {The case $\mu ^2>0$.} In the following we will restrict our analysis to the case $\mu ^2>0$ only, that is, for the Higgs potential we adopt the expression $U(\phi) = \frac{\mu^2}{2}\phi ^2+\frac{\xi }{4}\phi ^4$.  Then, from Eq.~\eqref{fb}, by assuming $Q>>1$, that is, that the portion of dissipative term should be small in weak regimes, the variation of the particle numbers of the scalar field  $n_{\phi}$ is obtained in the form,

\be\label{nphiweak}
n_{\phi} = n_{0\phi} \exp\left\{ {3 \over 8M_P^2}\Bigg[\phi^2-\frac{\mu^2}{\xi}\ln(\mu^2+\xi \phi^2)\Bigg] \;  \right\}\;,
\ee
where $n_{0\phi}$ is an arbitrary constant of integration.

The slow-roll parameters, introduced in Eqs. \eqref{vsrp} and  \eqref{etabeta}, are obtained as
\begin{eqnarray}
&&\hspace{-0.5cm}\epsilon_1(\phi) = {M_P^2 \over 2} {U^{\prime 2}(\phi) \over U^2(\phi)}=\frac{M_P^2}{2}\frac{{{{\left(\mu^2 \phi  +{\xi}{\phi ^3}\right)}^2}}}{{{{\left(\frac{{{\mu ^2}}}{2}{\phi ^2} + \frac{{{\xi}}}{4}{\phi ^4}\right)}^2}}},\\\label{epsilonweak}
&& \hspace{-0.5cm}\eta(\phi)=M_P^2{U''(\phi) \over  U(\phi)}={M_P^2}\frac{{{{(\mu^2  +3 {\xi}{\phi ^2})}}}}{{{{(\frac{{{\mu ^2}}}{2}{\phi ^2} + \frac{{{\xi}}}{4}{\phi ^4})}}}},\\ \label{etaweak}\nonumber\\
&& \hspace{-0.5cm}\beta(\phi) =2 \epsilon_1(\phi) -\sqrt{2\epsilon_1(\phi)}M_P \frac{n_{\phi}^{\prime}}{ n_{\phi}}\;.\label{betaweak}
\end{eqnarray}

Inflation ends at $\epsilon_1(\phi)=1$ for $\phi_e =\phi_e(\mu, \xi) $, where $\phi _e$ is a solution of the algebraic equation
\be
\sqrt{2}\xi  \phi ^3-4 \xi M_P \phi^2+2 \sqrt{2} \mu^2  \phi  -4M_P\mu^{2}=0\;.
\ee

The scalar field at the horizon crossing is found from the number of e-folds. Hence from Eq. \eqref{efold} one arrives at the expression,
\begin{equation}\label{efoldweak}
\mathcal{N}  =- \frac{1}{M_P^2}\; \int_{\phi_\star}^{\phi_e}\frac{{(\frac{{{\mu ^2}}}{2}{\phi ^2} + \frac{{{\xi}}}{4}{\phi ^4})}}{{({\mu ^2}\phi  + {\xi}{\phi ^3})}}d\phi\;,
\end{equation}
which subsequently gives
\begin{equation}\label{phistarweak}
\phi_\star^2 \simeq 4\mathcal{N} M_P^2+ \phi_e^2\;.
\end{equation}
To obtain the above relation we have supposed that ${\xi}{\phi ^2}$ is much less than $\mu^2$. The comparison with observations will justify this assumption.

As mentioned in the previous Sections, in the weak dissipative regime, the parameter $Q \simeq 1$ at the time of horizon exit. The amplitude of the scalar perturbations in these regime is defined as
\begin{equation}\label{psweak}
\mathcal{P}_s = \frac{25}{4}\left( { H \over \dot\phi } \right)^2 \delta\phi^2=\frac{25T_{0w}}{4\sqrt{3}M_P^5}\frac{U^{8/3}(\phi)}{U'^2(\phi)},
\end{equation}
where  $\delta\phi^2=H\times T$, and it is given by
\be
\mathcal{P}_s=\frac{25 T_{0w} \left(\frac{\mu ^2 \phi ^2}{2}+\frac{\xi  \phi
   ^4}{4}\right)^{8/3}}{4 \sqrt{3}M_P^5 \left(\mu ^2 \phi +\xi  \phi ^3\right)^2}.
\ee

From Eqs.~\eqref{nsaswarm} and {\eqref{psweak} it also follows that the scalar spectral index in the weak regime is obtained as
\be\label{nsweak}
 n_s -1=\left( -{16 \over 3} \epsilon_1 + 2 \eta\right)\;,
\ee
where $\eta$ can be defined through the slow-roll parameter $\epsilon_2$ as $\epsilon_2 =-\eta+\epsilon_1$. Explicitly, for scalar spectral index we obtain
\be
n_s=1-{\frac{8M_P^2}{3} }\frac{ \left(10 \mu ^4+11 \mu ^2 \xi  \phi ^2+7 \xi ^2 \phi ^4\right)}{\left(2 \mu ^2\phi+\xi  \phi
   ^3\right)^2}~.
\ee

Following the tensor spectral index definition
\[n_{t} ={\frac{d\ln (\mathcal{P}_{t})}{d\ln (k)}}=\frac{\dot\phi}{H}\frac{d}{d\phi}%
\ln (\mathcal{P}_{t})~,\]
 the tensor spectral index $n_t$ is found as
\be
n_t=-2 \epsilon_1=- M_P^2\frac{{{{\left( {{\mu ^2}\phi  + \xi {\phi ^3}} \right)}^2}}}{{{{\left( {\frac{{{\mu ^2}{\phi ^2}}}{2} + \frac{{\xi {\phi ^4}}}{4}} \right)}^2}}}~.
\ee

The tensor-to-scalar ratio in this case is also obtained as
\begin{equation}\label{rweak}
r = \frac{{8\sqrt 3 {M_P}}}{{75{T_{0w}}{\pi ^2}}}\frac{{{{U'}^2}(\phi )}}{{{U^{5/3}}(\phi )}}\;.
\end{equation}

To compare the results of the model with observational data, the above perturbation parameters will be computed at the horizon crossing.

From Eq. \eqref{psweak},  the model constant parameter $\Gamma_1$ is found in terms of $\mu$, $\xi$ and $\mathcal{N}$ as
\begin{equation}\label{gammaweak}
\Gamma_1(\mu,\xi,\mathcal{N}) =\frac{9216\pi\zeta(3) m_{\phi}M_P^{14}}{15625}U(\phi_\star)^{-8}U^{\prime}(\phi_\star)^{ 6}\left(\mathcal{P}_s^\star\right) ^3,
\end{equation}
where as before the parameter $\mathcal{P}_s^\star=2.17\times10^{-9}$ is the amplitude of the scalar perturbations at horizon exit.

Using the $r-n_s$ diagram of Planck-2018, one could plot a $\mu-\xi$ diagram as shown in Fig.~\ref{muxiweak}, where the dark blue color indicates an area of $(\mu,\xi)$ where the results for $n_s$ and $r$ stand in $68\%$ CL. The light blue color indicates an area of $(\mu,\xi)$ in which the point $(r,n_s)$ of the model stands in $95\%$ CL.

\begin{figure}
  \centering
  \includegraphics[width=7cm]{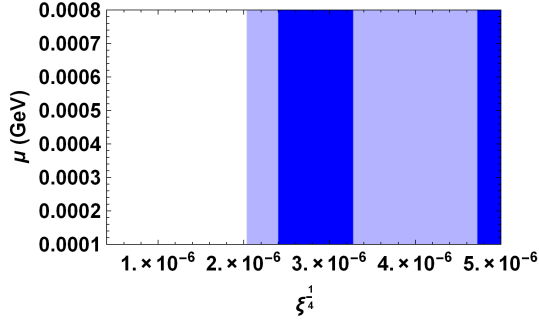}
  \caption{Numerical values of the $(\mu,\xi)$ parameters of the warm inflationary model with irreversible photon fluid creation in the weak dissipative regime for which the point $(r-n_s)$ is located in the observational region.  The dark blue color shows the values of $(\mu,\xi)$ in the $68\%$ CL range of the Planck-2018 data, while the light blue color shows the parameter values  in the range $95\%$ CL. To estimate the free parameters of the model we have used $\mathcal{N}\thickapprox100$ for the number of efolds. Based on Eq.\eqref{gammaweak} the value of $\Gamma_1$ is estimated as $\Gamma _1=0.51\;{\rm GeV}$.  The value of the reduced Planck mass is $M_P=2.4\times 10^{18}\;{\rm GeV}$, and we have fixed the value of the mass of the scalae field as $m_{\phi}=10^{2}\;{\rm GeV}$.}\label{muxiweak}
\end{figure}

An important feature of the warm inflationary scenario is that the thermal fluctuations overcome the quantum fluctuations, since the fluid temperature is bigger than the Hubble parameter. To have a healthy warm inflation model, this condition should be satisfied during the cosmological evolution. Fig.~\ref{fig20} describes the behavior of the ratio of the temperature to the Hubble parameter during this era. From the Figure one can see that the condition is satisfied during this phase of cosmological expansion.
%%%%%%%%%%%%%%%%%%%%%%%%%%%%%%%%%%%%%%%%%%%
\begin{figure}
  \centering
  \includegraphics[width=9cm]{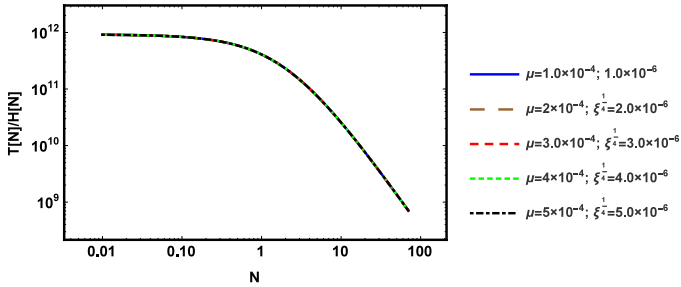}
  \caption{The ratio of the temperature to the Hubble parameter during in the warm inflation model with irreversible matter creation in the weak dissipative regime versus the number of e-folds for different values of $(\mu,\xi)$. As one can see from the plots, based on log-log command, during inflation the temperature is larger than the Hubble parameter, and the condition $T>H$ is satisfied properly.}\label{fig20}
\end{figure}

\subsubsection{Strong regime}

In the strong regime of the warm inflation, the dissipative ratio is usually much larger than unity, so that $(1+Q)\simeq Q$. This condition implies $Q=p_c^{(\phi)}/\dot{\phi}^2>>1$, indicating that the creation pressure is much larger than the kinetic energy of the field. The cosmological evolution is dominated by particle creation processes, and the energy transfer from the field to the radiation fluid is maximal.

In this case, the analyzes of the model becomes more complicated, and the cosmological dynamics is significantly different from the weak dissipative regime. In this approximation, with the use of Eqs.~\eqref{vfriedmann} and \eqref{dotphi}, the time derivative of the scalar field is expressed as
\begin{equation}\label{dotphist}
\dot{\phi} \simeq - {U'(\phi) \over \Gamma} =-\frac{U'(\phi)m_{\phi} n_{\phi}}{\Gamma_1U(\phi)}\;.
\end{equation}

Using Eq. \eqref{radiationtemp}, the form of the temperature of the photon fluid is obtained as same as in the weak regime. One should notice that as compared to the weak regime, in the strong regime  there are some differences originating from the form of the scalar field at the horizon exit.

In the strong regime, to obtain the main perturbation parameters we must first find the function $n_{\phi}$. This can be done by using Eqs.~\eqref{fb} and \eqref{dotphist}, with the  function $n_{\phi}$ obtained as,
\begin{equation}\label{nphistrong}
n_{\phi} \simeq \frac{\Gamma_1\mu}{\sqrt{8}m_{\phi}}\phi^2\;,
\end{equation}
where we have considered the Dirichlet boundary condition, $n_{\phi}(0)=0$, to obtain the integration constant, and we have approximated the potential as $U(\phi)\approx \mu^2 \phi ^2/2$.
The dissipative ratio $Q$ is given by
\begin{equation}\label{Qstrong}
Q(\phi) = {\Gamma \over 3H} =\frac{{2\sqrt {\frac{2}{3}} }M_P}{\mu }\frac{{\sqrt {\frac{{{\mu ^2}{\phi ^2}}}{2} + \frac{{\xi {\phi ^4}}}{4}} }}{{{\phi ^2}}}\;.
\end{equation}

In the strong dissipation regime, the slow-roll parameters introduced in Eqs.~\eqref{vsrp} and  \eqref{etabeta}, are obtained, after substituting Eq.~ \eqref{Qstrong}, as
\begin{eqnarray}
\epsilon_1(\phi) = {M_P^{2} \over 2Q} {U^{\prime 2}(\phi) \over U^2(\phi)}=\frac{{4\sqrt 6 {M_P}\mu {\phi ^4}{{\left( {{\mu ^2} + \xi {\phi ^2}} \right)}^2}}}{{{{\left( {2{\mu ^2}{\phi ^2} + \xi {\phi ^4}} \right)}^{5/2}}}},\\\label{epsilonStrong}
 \eta(\phi)=\frac{M_P^{2}}{ Q}{U''(\phi) \over U(\phi)}=\frac{{2\sqrt 6 {M_P}\mu {\phi ^2}\left( {{\mu ^2} + 3\xi {\phi ^2}} \right)}}{{{{\left( {2{\mu ^2}{\phi ^2} + \xi {\phi ^4}} \right)}^{3/2}}}},\\ \label{etaStrong}\nonumber
\end{eqnarray}
and
\begin{eqnarray}\label{betaStrong}
\beta (\phi )& =& \frac{M_P^{2}}{ Q(\phi )}\frac{{U'(\phi )\Gamma '(\phi )}}{{U(\phi )\Gamma (\phi )}}=\nonumber\\
&& \frac{{4\sqrt 6 {M_P}\left( {{\mu ^3}\xi  + \mu {\xi ^2}{\phi ^3}} \right)}}{{{{\left( {2{\mu ^2} + \xi {\phi ^2}} \right)}^{5/2}}}},
\end{eqnarray}
respectively.

Similarly to the  weak regime case, inflation ends at $\epsilon_1(\phi)=1$ for $\phi_e =\phi_e(\mu, \xi) $, where $\phi _e$ is a solution of the algebraic equation
\be\label{phiend-strong}
 \xi {\phi ^4}+{\mu ^2}{\phi ^2}  - 2{\left(\sqrt {2/3} \frac{1}{{{\mu M_P} }}\right)^{1/2}}\left( \frac{\xi }{4}{\phi ^4}+\frac{{{\mu ^2}}}{2}{\phi ^2}\right) = 0\,.
\ee

From Eq. \eqref{efold}, with $Q\gg1$, the general expression for the e-folds number is obtained as,
\begin{eqnarray}\label{efoldStrong}
\hspace{-0.5cm}\mathcal{N}&=&-\frac{1}{{4\mu M_P\sqrt{6\xi }}}\Bigg\{ {\xi ^{2}}\phi \sqrt{2{\mu ^{2}}+\xi
{\phi ^{2}}}+\nonumber\\
\hspace{-0.5cm}&&2{\mu ^{2}}\arctan \left[ {\frac{{\sqrt{\xi }\phi }}{{\sqrt{2{%
\mu ^{2}}+\xi {\phi ^{2}}}}}}\right] +\nonumber\\
\hspace{-0.5cm}&&4{\mu ^{2}}\ln \left[ {\sqrt{\xi }%
\left( {\sqrt{\xi }\phi +\sqrt{2{\mu ^{2}}+\xi {\phi ^{2}}}}\right) }\right]
\Bigg\} \Bigg|_{\phi _{\star}}^{\phi _{e }}.
\end{eqnarray}

After some algebra, and by expanding $\tan^{-1}$ and the logarithmic functions,  Eq. \eqref{efoldStrong} can be expressed by
\begin{equation}\label{efoldStrong-reduced}
\mathcal{N} = \left[ - \frac{{\mu \ln (\sqrt {2\xi } {\mu })}}{{\sqrt {6\xi } {M_P}}} - \frac{\phi }{{\sqrt 3 {M_P}}} + \frac{{\xi {\phi ^3}}}{{12\sqrt 3 {\mu ^2}{M_P}}} + O({\phi ^4})\right]\Bigg|_{\phi _\star}^{\phi _{e}}\,.
\end{equation}

By substituting the solutions of Eq. \eqref{phiend-strong} for $\phi$ into Eq. \eqref{efoldStrong-reduced} we can obtain the suitable value of the scalar field at the horizon exit.

From Eq.~\eqref{pswarm} the amplitude of the scalar perturbations in the strong regime is obtained as
\begin{equation}\label{psstrong}
\mathcal{P}_s  =\mathcal{P}_s^{(0)}\Gamma _1^{17/6}\frac{{U{{(\phi )}^{41/12}}}}{{n_\phi ^2U'{{(\phi )}^2}}}\;,
\end{equation}
where we have denoted
\be
\mathcal{P}_s^{(0)} = \frac{{25}}{{8 \times {3^{11/12}}M_P^{13/6}{{\left( {16\zeta (3){\pi ^7}m_\phi ^7} \right)}^{1/3}}}}.
\ee

By taking the time derivative of Eq.~(\ref{psstrong}) equation according to Eq.~\eqref{nsaswarm} leads to the scalar spectral index as given by
\begin{eqnarray}\label{nsasstrong}
  n_s -1 &=&   \left(-\frac{17}{6} \epsilon_1  - 2 \beta+ 2 \eta\right).
\end{eqnarray}
To obtain the above equation we have used the relation
\[\frac{{M_P^2}}{{Q(\phi )}}\frac{{U'(\phi ){{n'}_\phi }}}{{U(\phi ){n_\phi }}} = 2{\varepsilon _1} - \beta\;. \]

The amplitude of tensor perturbations in strong dissipative regimes is given by \cite{Bastero-Gil:2016qru,Berera:2018tfc}
\begin{equation}\label{ptstrong}
\mathcal{P}_t = {2 H^2 \over M_P^2 \pi^2} = \frac{{2U(\phi )}}{{3M_P^4{\pi ^2}}}\;.
\end{equation}
Then, the tensor-to-scalar ratio is obtained as follows
\begin{equation}\label{rstrong}
r =\frac{{2U'{{(\phi )}^2}n_\phi ^2}}{{3{\pi ^2}\Gamma _1^{17/6}M_P^4P_s^{(0)}U{{(\phi )}^{29/12}}}}\;.
\end{equation}

%%%%%%%%%%%%%%%%%%%%%%%%%%%%%%%%%%%%%%%%%%%
Now following the same procedure as the one introduced for the weak regime case,  but by using the $r-n_s$ diagram of Planck-2018, we can estimate the $\mu-\xi$ amounts, as shown in Fig.~\ref{muxistrong}. In the Figure the solid line indicates the estimated amounts of the  free parameters $(\mu,\xi)$ are in good agreement with observations arisen from  Planck 2018
results ($TT,TE,EE+{\rm Low}E+{\rm lensing}
+BK14+BAO$)  \cite{pl2018a}.
\begin{figure}
  \centering
  \includegraphics[width=7cm]{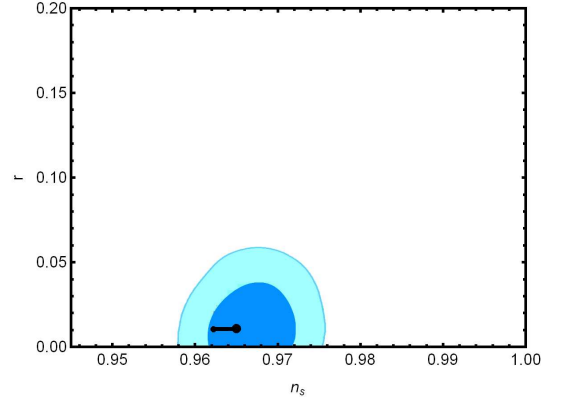}
  \caption{ 1$\sigma$ (dark blue) and 2$\sigma$ (light blue) contours for Planck 2018
results ($TT,TE,EE+{\rm Low}E+{\rm Lensing}
+BK14+BAO$)  \cite{pl2018a}, on $r-n_s$ plane. We depict the predictions of
our scenario, for    the strong regime case with the e-folding value
$\mathcal{N} =56$ (smaller circle), and  $\mathcal{N}=65$ (larger circle). In this case we fixed $\Gamma_1=10^4\;{\rm GeV}$ and the mass of the scalar field at $m_{\phi}=2\times10^{-10}\;{\rm GeV}$. In this diagram the free parameters of the model are estimated for $\mu=65\;{\rm GeV}$ and $\xi^{1/4}=-2.0$. }
 \label{muxistrong}
\end{figure}

To investigate the consistency of the model we have to study the thermal-quantum perturbation ratio, $T/H$.  Fig.~\ref{THStrong}  shows the behavior of the ratio of the temperature to the Hubble parameter in the strong dissipative regime case. From the Figure one can see that the condition is satisfied during this phase of cosmological expansion.
%%%%%%%%%%%%%%%%%%%%%%%%%%%%%%%%%%%%%%%%%%%
\begin{figure}
  \centering
  \includegraphics[width=9cm]{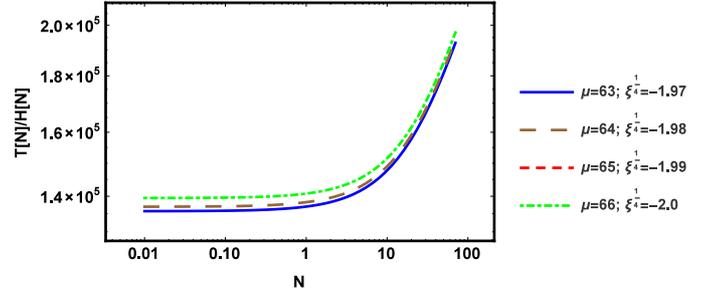}
  \caption{The ratio of the temperature to the Hubble parameter in the warm inflation model with irreversible radiation creation in the strong dissipative regime versus the number of e-folds for different values of $(\mu\; \left({\rm GeV}\right),\xi)$. As one can see from the plots, during inflation the temperature is larger than the Hubble parameter, and the condition $T>H$ is satisfied properly. To plot this diagram we have considered $\Gamma_1=10^4\;{\rm GeV}$. }\label{THStrong}
\end{figure}

\section{Discussions and final remarks}\label{sect5}

In the present paper we have shown that the thermodynamics of the open systems is a valuable tool for describing the interacting scalar field - radiation fluid  phases of the general relativistic cosmological models. As applied to a two component (scalar field and radiation) warm inflationary cosmological model, the thermodynamics of irreversible processes in open systems provides a generalization of the standard theory of scalar field-radiation decay/creation, and of its inflationary implications. During the scalar field energy dominated phase of the expansion of the Universe, the energy contained in the scalar field can generate particles via single particle decay, and such a process is generally not forbidden thermodynamically. Thus, the model presented in this paper gives a rigorous (but still phenomenological) thermodynamical foundation, and a natural generalization, of the theory of warm inflationary models. As required by the thermodynamics of open systems, particle creation (decay) gives rise to a supplementary creation (decay) pressure, which must be included as a distinct component in the energy-momentum tensor of the cosmological scalar field - radiation mixture.

We have considered only photon creation in a single scalar field  dominated Universes, leading to a warm inflationary model in which the particle production rate is extremely high at the beginning of the inflationary period, reaches a maximum at the end of inflation, and afterwards,  when the scalar field energy and pressure become dominated by the scalar field potential,  it tends to zero. At the end of the inflation, and in the post-inflationary phase,  the kinetic energy of the scalar field becomes negligibly small.  As for the matter content of the post-inflationary, we made the simplifying assumption that the Universe did consist entirely of a relativistic, radiation type component only. Of course for the realistic description of the cosmology of the early Universe the presence of more particles is necessary to be taken into account. However, the inclusion of other decay channels for the scalar field can be done in a straightforward manner in the present formalism, with each of such particle creation process generating its own creation/decay pressure, which must be taken into account in the gravitational equations describing the expansionary history of the early Universe.

In our approach we have adopted the fundamental assumption that the early Universe can be described as an open thermodynamic system. However, up to now no clear and consistent definition of the concept of open system in the cosmology of the early Universe has been proposed, and different authors use this thermodynamic concept in different ways, and with different meanings. For example, in the approach pioneered in \cite{prigogine1988thermodynamics}, and further developed in \cite{op0} and \cite{op0a}, an open system is considered as a system in which particle and entropy creation takes place.  The early Universe is then defined as an open system in this sense. The particle creation corresponds to an irreversible energy flow from a source to the newly created matter constituents. An alternative interpretation considers the inflationary Universe as an open system due to the
coupling of the cosmological modes to a bath of comoving momentum modes of the canonically normalized scalar degree of freedom, by considering a single cubic interaction term in the Einstein-Hilbert Lagrangian \cite{bath}.  Hence, in this approach, the short-long mode coupling generates an
effective cosmological fluid - bath interaction. Another definition of the concept of open system in cosmology was proposed in \cite{Hu}, where the classical geometry is treated as the system, and the quantum matter field as the environment. This approach allows the derivation of a fluctuation-dissipation theorem for semiclassical cosmology, which embodies the backreaction effect of matter fields on the dynamics of spacetime, leading to a backreaction equation derivable from the influence action in the form of an Einstein-Langevin equation that contains a dissipative term. For the applications of the theory of open quantum systems to de Sitter geometry see \cite{oq1}, where the dynamics of a freely-falling two-level detector in weak interaction with a reservoir of fluctuating quantized conformal scalar fields in the de Sitter invariant vacuum is considered. The Hawking effect for black holes was considered in the framework of open quantum systems in \cite{oq2}, where a two-atom system was treated as an open quantum system placed in a bath of fluctuating quantized massless scalar fields in vacuum.

By using the fundamental principles of the thermodynamics of open systems and of irreversible processes, and the Einstein field equations, we have developed a consistent and systematic approach that can describe
the time evolution of the two matter components (scalar field and radiation) in the warm inflationary scenario. By adopting an appropriate  set of initial conditions for the cosmological quantities, the general equations describing warm inflation with irreversible  matter creation can be solved numerically. During warm inflation, the particle number density of the scalar field decreases rapidly to (near) zero, while the number of photons increases from (approximately) zero, to a maximum value, thus providing the initial conditions for the
standard Big Bang model of the Universe. The photon fluid reaches a maximum temperature at the end of inflation, with the maximum temperature a function of the physical quantities describing the warm inflationary model (scalar field potential parameters, scalar field decay width, energy density of the scalar field), and of the cosmological parameters, like, for example, the numerical value of the Hubble
function at the time interval when the energy density of the radiation reached its maximum value.

As a direct cosmological applications of our results we have investigated the effects of
particle creation in  warm inflationary scenarios for three different forms of the scalar field potential - the scalar field with Barrow-Saich potential, and scalar fields with constant and Higgs type potentials.  All these potentials  generally lead to similar qualitative results for
the warm inflationary scenario. If for the Barrow-Saich potential a complete analytic solution of the field equations is possible, for the scalar field with constant and Higgs type potentials only numerical solutions are possible. But what we have found from the numerical investigations of the warm inflationary models  is that the choice of the functional form of the scalar field potential has an essential effect on
the warm inflationary cosmological dynamics, and on the associated photon gas creation. This influence can be seen very clearly in the case of the Higgs potential, with $\mu ^2>0$, and $\mu ^2<0$, respectively. Major differences do appear in the time intervals for which the maximum energy density of the radiation fluid is reached, as well as in the decay rate of the particle number and energy density of the scalar field. Significant differences occur in the time evolution of the deceleration parameter, whose behavior also indicate the end of inflation. Moreover, the cosmological dynamics also strongly depends on the numerical values of the Higgs  potential parameters $\mu ^2$ and $\xi$.

Hence, the scalar field potentials have a unique quantitative signature on the properties of radiation after the warm inflationary phase.  By using the latest Planck satellite data we have also performed a general comparison between the theoretical predictions of the model, and the constraint on
the inflationary parameters provided by the study of the Cosmic Microwave Background Radiation. By using the observational data we have
obtained some general estimates on the decay width $\Gamma _1$ of the scalar field, as well as on the parameters $\mu ^2$ and $\xi$
in the Higgs type scalar field potential. Generally, we can obtain an acceptable concordance between theoretical predictions and observations. However, these results on the numerical values of the model parameters are still of a qualitative nature, since a basic cosmological theory is needed
to explain  the physical conditions in the early Universe. The warm inflationary models with Higgs type potentials could accelerate
the radiation fluid production processes, and they may lead  to radiation creation even after the photon fluid energy density has attained its maximum
value. The process is similar to the one induced by the Higgs potentials model.

Photon creation has very important implications on the dynamics and evolution of  the very early Universe. The details of the scalar field-radiation fluid interaction  mechanism depend on the (poorly known)  parameters of the particle physics models necessary to describe the newly created photons. One of the important parameters in the model is the mass of the inflationary scalar field particle, which is an essential physical parameter in the description of the scalar field-radiation interaction process. Unfortunately, presently there is no definite answer giving the value of the mass of the inflaton scalar field, as well as of the physical parameters describing scalar field-radiation fluid interaction in the very early Universe.

An interesting consequence of particle creation is its relation to the so called arrow of time: a process that yields a direction to time, and allows to distinguish the past from the future. One can consider two distinct arrows of time. The first one is the thermodynamical arrow of time, which is given by the direction in which the increase of the entropy takes place. The second arrow of time is the cosmological arrow of time, given by the direction in which the Universe expands, and evolves in time. Particle creation leads to an asymmetry in the Universe's evolution, and permits us to introduce a thermodynamical arrow of time. Interestingly enough, in our model the thermodynamic arrow of time agrees with the cosmological one, both indicating the same direction of thermodynamic evolution. This agreement is naturally determined by the decay of the scalar field, and of its direct consequence, irreversible photon fluid creation.

In our present study  we have introduced some basic theoretical tools that could allow a thermodynamically rigorous formulation of the warm inflationary models. In our approach we have completely neglected the quantum aspects of the radiation production, as well as the intricate aspects of the scalar field decay.  These aspects of the thermodynamic theory of the scalar field - radiation fluid interaction and their implications on warm inflationary models  will be the subject of a future work.

\section*{Acknowledgments}

We would like to thank to the anonymous referee for comments and suggestions that helped us to significantly improve our manuscript. HS thanks A. Starobinsky for very constructive discussions about inflation during Helmholtz International Summer School  2019 in Russia. He is grateful to G. Ellis, A. Weltman, and UCT for arranging his short visit, and for enlightening discussions about cosmological fluctuations and perturbations for both large and local scales. He also thanks  H. Firouzjahi for constructive discussions about inflation and perturbations. His special thanks go to his wife E. Avirdi for her patience during our stay in South Africa.


\begin{thebibliography}{99}
\bibitem{guth1981inflationary} A. Guth, Physical Review {\bf D 23}, 347 (1981).

%\bibitem{B1} A. Linde, Particle Physics and Inflationary Cosmology, Harwood Academic Publishers, London, UK, 1992

%\bibitem{B2} E. Kolb and M. Turner, The Early Universe, Westview Press, New York, USA, 1994

%\bibitem{B3} A. Liddle and D. Lyth, Cosmological Inflation and Large-Scale Structure, Cambridge University Press, Cambridge, UK, 2000

%\bibitem{B4} Y. Fujii and K. Maeda, The Scalar-Tensor Theory of
%Gravitation, Cambridge, Cambridge University Press, 2003

% \bibitem{B5} V. Faraoni, 	Cosmology in scalar-tensor gravity, Dordrecht; Boston, Kluwer Academic Publishers, 2004

\bibitem{B6} V. Mukhanov, Physical Foundations of Cosmology, Cambridge University Press, Cambridge, UK, 2005

\bibitem{linde1982new} A. Linde, Physics Letters {\bf B 108}, 389 (1982).

%\bibitem{linde1982coleman} A. Linde, Physics Letters {\bf B 114}, 431 (1982).

%\bibitem{linde1982scalar} A. Linde, Physics Letters {\bf B 116}, 329 (1982).

\bibitem{albrecht1982cosmology} A. Albrecht and P. J. Steinhardt, Phys. Rev. Lett. {\bf 48}, 1220 (1982).

%\bibitem{linde2000inflationary} A. Linde, Physica Scripta {\bf T85}, 168 (2000).

\bibitem{linde1983chaotic} A. Linde, Physics Letters B 129, 177 (1983).

\bibitem{linde1994hybrid} A. Linde, Physical Review {\bf D 49}, 748 (1994).

%\bibitem{revinfl1} N. Bartolo, E. Komatsu, S. Matarrese, and A. Riotto, Physics Reports {\bf 402}, 103 (2004).

%\bibitem{revinfl2} M. Bartelmann, Rev. Mod. Phys. {\bf 82}, 331 (2010).

%\bibitem{revinfl3} A. Mazumdar and  J. Rocher, Physics Reports {\bf 497}, 85 (2011).

%\bibitem{revinfl4} A. Maleknejad, M. M. Sheikh-Jabbari, and J. Soda, Physics Reports {\bf 528}, 161 (2013).

\bibitem{revinfl5} R. H. Cyburt, B. D. Fields, K. A. Olive, and Tsung-Han Yeh, Rev. Mod. Phys. {\bf 88}, 015004 (2016).

\bibitem{revinfl6} S. Nojiri, S. D. Odintsov, and V. K. Oikonomou, Physics Reports {\bf 692}, 1 (2017).


\bibitem{Pl1}P. A. R. Ade et al. [Planck Collaboration], "Planck 2015 results. I. Overview of products and scientific results", Astron. Astrophys. \textbf{594}, A1 (2016).

\bibitem{Pl2} P. A. R. Ade et al. [Planck Collaboration], "Planck 2015 results. XIII. Cosmological parameters",  Astron. Astrophys. \textbf{594}, A8 (2016). arXiv:1502.01589 [astro-ph.CO].

\bibitem{Pl3} P. A. R. Ade et al. [Planck Collaboration], "Planck 2015 results. XVII. Constraints on primordial non-Gaussianity", Astron. Astrophys. \textbf{594}, A17 (2016).

\bibitem{Pl4} P. A. R. Ade et al. [Planck Collaboration], "Planck 2015 results. XVIII. Background geometry $\&$ topology", Astron. Astrophys. \textbf{594}, A18  (2016).

\bibitem{Pl5} P. A. R. Ade et al. [Planck Collaboration], "Planck 2015 results. XX. Constraints on inflation", Astron. Astrophys. \textbf{594}, A20 (2016).

\bibitem{Haidar2}  H. Sheikhahmadi,  Eur. Phys. J. C  \textbf{79}, 451 (2019).

\bibitem{liddle1992cobe} A. Liddle and D. Lyth, Physics Letters {\bf B 291}, 391 (1992).

\bibitem{dodelson1997cosmic} S. Dodelson, W. H. Kinney and E. W. Kolb, Physical Review {\bf D 56}, 3207 (1997).

\bibitem{kinney1998constraining} W. H. Kinney, Physical Review {\bf D 58}, 123506 (1998).

\bibitem{kinney2000new} W. H. Kinney, A. Melchiorri and A. Riotto, Physical Review {\bf D 63}, 023505 (2001).

\bibitem{kinney2008latest} W. H. Kinney, E. W. Kolb, A. Melchiorri and A. Riotto, Physical Review {\bf D 78}, 087302 (2008).

\bibitem{Haw} S. W. Hawking,  Physics Letters {\bf B, 115},  295 (1982).

\bibitem{Muk} V. F. Mukhanov, Journal of Experimental and Theoretical Physics Letters {\bf  41}, 493 (1985).

 \bibitem{An}  C. J. Copi, D. Huterer, D. J. Schwarz, and G. D. Starkman, Adv. Astron. {\bf 2010}, 847541 (2010).

\bibitem{Alb} A. Albrecht, P. J. Steinhardt, M. S. Turner, and F. Wilczek,  Phys. Rev. Lett. {\bf 48}, 1437 (1982).

\bibitem{Kof} L. Kofman, A. Linde, and A. A. Starobinsky, Phys. Rev. Lett. {\bf 73}, 3195 (1994).

\bibitem{kofman1997towards} L. Kofman, A. Linde and A. A.Starobinsky, Phys. Rev. {\bf D 56}, 3258 (1997).

\bibitem{Pre1} J. Repond and J. Rubio, Journal of Cosmology and Astroparticle Physics {\bf 07}, 043 (2016).

\bibitem{Pre2} K. D. Lozanov and M. A. Amin, Journal of Cosmology and Astroparticle Physics {\bf 06}, 032 (2016).

\bibitem{Pre3} S. Antusch, D. Nolde, and S. Orani, Journal of Cosmology and Astroparticle Physics {\bf 06},  009 (2015).

\bibitem{Pre4} I. Rudenok, Y. Shtanov, and S. Vilchinskii, Phys. Lett. {\bf B 733}, (2014).

%\bibitem{scherrer1985decaying} R. Scherrer and M. Turner, Phys. Rev. {\bf D 31}, 681 (1985).

%\bibitem{giudice2001largest} G. Giudice, E. Kolb and A. Riotto, Phys. Rev. {\bf D 64}, 023508 (2001).

\bibitem{Re11} T. Harko, W. F. Choi, K. C. Wong, and K. S. Cheng, Journal of Cosmology and Astroparticle Physics {\bf 06}, 002 (2008).

\bibitem{Re10} R. K. Jain, P. Chingangbam, and L. Sriramkumar, Nucl. Phys. {\bf B 852}, 366 (2011).

\bibitem{Re9} V. Demozzi and C. Ringeval, Journal of Cosmology and Astroparticle Physics {\bf 05},  009 (2012).

\bibitem{Re8} H. Motohashi and A. Nishizawa, Phys. Rev. {\bf D 86}, 083514 (2012).

\bibitem{Re7} K. Mukaida and K. Nakayama, Journal of Cosmology and Astroparticle Physics {\bf 03}, 002 (2013).

\bibitem{Re6} A. Nishizawa and H. Motohashi, Phys. Rev. {\bf D 89}, 063541 (2014).

\bibitem{Re5} L. Dai, M. Kamionkowski, and J. Wang, Phys. Rev. Lett. {\bf 113}, 041302 (2014).

\bibitem{Re4} J. Martin, C. Ringeval, and V. Vennin, Phys. Rev. Lett. {\bf 114}, 081303 (2015).

\bibitem{Re3} C. Gross, O. Lebedev, and M. Zatta, Phys. Lett. {\bf B 753}, 178 (2016).

\bibitem{Re2} Y. S. Myung and T. Moon, Journal of Cosmology and Astroparticle Physics {\bf 07},  014 (2016).

\bibitem{Re1} M. Rinaldi and L. Vanzo, Phys. Rev. {\bf D 94}, 024009 (2016).

\bibitem{Reh1} T. Rehagen and G. B. Gelmini, Journal of Cosmology and Astroparticle Physics {\bf 06},  039, (2015).

\bibitem{allahverdi2010reheating} R. Allahverdi, R. Brandenberger, F. Cyr-Racine and A. Mazumdar, Annu. Rev. Nucl. Part. Sci. {\bf 60}, 27 (2010).

\bibitem{RehRev} M. A. Amin, M. P. Hertzberg, D. I. Kaiser, and J. Karouby, Int. J. Mod. Phys. {\bf D 24},  1530003 (2015).

\bibitem{w1} A. Berera and L. -Z. Fang, Phys. Rev. Lett. {\bf 74}, 1912 (1995).

\bibitem{w2} A. Berera, Phys. Rev. Lett. {\bf 75}, 3218 (1995).

\bibitem{w3}  A. Berera, Phys. Rev. {\bf D 54}, 2519 (1996).

\bibitem{w4} A. Berera, M. Gleiser and R. O. Ramos, Phys. Rev. {\bf D 58} 123508 (1998).

\bibitem{w5} A. Berera and T. W. Kephart, Phys. Rev. Lett. {\bf 83}, 1084 (1999).

\bibitem{w6} A. Berera, Nucl. Phys. {\bf B 585}, 666 (2000).

\bibitem{w7}  A. Berera, M. Gleiser and R. O. Ramos, Phys. Rev. Lett. {\bf 83}, 264 (1999).

\bibitem{w8} H. P. De Oliveira and S. E. Jor\'{a}s, Phys.Rev. {\bf D 64},  063513 (2001).

\bibitem{w9} L. M. H. Hall, I. G. Moss, and A. Berera, Phys. Rev. {\bf D 69},  083525 (2004).

\bibitem{w9a} I. G. Moss and C. M. Graham, Phys. Rev. {\bf D 78},  123526 (2008).

\bibitem{w9a} Y. Zhang, JCAP {\bf 03}, 023 (2009).

\bibitem{w9b} N. Barnaby and Z. Huang, Phys. Rev. {\bf D 80}, 126018 (2009).

\bibitem{w9c} M. Bastero-Gil, A. Berera, and R. O. Ramos, J. Cosmol. Astropart. Phys. {\bf 1107},  030 (2011).

\bibitem{w10} R. Herrera, M. Olivares, and N. Videla, Phys. Rev. {\bf D 88}, 063535 (2013).

\bibitem{w11} M. R. Setare and  V. Kamali, Phys. Lett. {\bf B 726},  56 (2013).

\bibitem{w11a} R. O. Ramos, L. A. da Silva, J. Cosmol. Astropart. Phys. {\bf 1303},  032 (2013).

\bibitem{w12} X.-M. Zhang and J.-Y. Zhu, JCAP {\bf 02}, 005 (2014).

\bibitem{w12a} M. Bastero-Gil, A. Berera, R. O. Ramos, and J. G. Rosa, JCAP {\bf 1410}, 053 (2014).

\bibitem{w13} G. Piccinelli, A. Sanchez, A. Ayala, and A. J. Mizher, Phys. Rev. {\bf D 90}, 083504 (2014).

\bibitem{w14} X.-M. Zhang and J.-Y. Zhu, Phys. Rev. {\bf D 91}, 063510 (2015).

\bibitem{w15} G. Panotopoulos and N. Videla, European Physical Journal {\bf C 75}, 525 (2015).

\bibitem{w15a} M. Bastero-Gil, A. Berera, R.- O. Ramos, and J. G. Rosa, Phys. Rev. Lett. {\bf 117}, 151301 (2016).

\bibitem{w15b} M. Motaharfar and H.- R. Sepangi, European Physical Journal  {\bf C 76}, 646 (2016).

\bibitem{w16} K. Sayar, A. Mohammadi, L. Akhtari, and K. Saaidi, Phys. Rev. {\bf D 95}, 023501 (2017).

\bibitem{w17} M. Benetti and R.-O. Ramos, Phys. Rev. {\bf D 95}, 023517 (2017).

\bibitem{w18} A. Jawad, N. Videla, and F. Gulshan,  The European Physical Journal {\bf C 77}, 271 (2017).

\bibitem{w19} M. Motaharfar, E. Massaeli, and H.-R. Sepangi,  Phys. Rev. {\bf D 96}, 103541 (2017).

\bibitem{w20} A. Jawad, S. Chaudhary, and N. Videla,  European Physical Journal {\bf C 77}, 808 (2017).


\bibitem{w21} A. Jawad, S. Hussain, S. Rani, and N. Videla,  European Physical Journal {\bf C 77}, 700 (2017).


\bibitem{w22} P. Goodarzi and H.-M. Sadjadi, European Physical Journal {\bf C 77}, 463 (2017).


\bibitem{w23} R. Herrera, Journal of Cosmology and Astro-Particle Physics {\bf 5}, 029 (2017).


\bibitem{w24} K. Li, X.-M. Zhang, H.-Y. Ma, and J.-Y. Zhu, Phys. Rev. {\bf D 98}, 123528 (2018).


\bibitem{w25} V. Kamali,  European Physical Journal {\bf C 78}, 975 (2018).


\bibitem{w26} M. Bastero-Gil, A. Berera, R. Hern{\'a}ndez-Jim{\'e}nez, and J.-G. Rosa,  Phys. Rev. {\bf D 98}, 083502 (2018).


\bibitem{w27} M. Motaharfar, E. Massaeli, and H.-R. Sepangi, Journal of Cosmology and Astro-Particle Physics {\bf 10}, 002 (2018).


\bibitem{w28} X.-B.  Li, Y.-Y. Wang, H.  Wang, and J.-Y. Zhu,  Phys. Rev. {\bf D 98}, 043510 (2018).


\bibitem{w29} L.-L. Graef and R.-O. Ramos, Phys. Rev. {\bf D 98}, 023531 (2018).


\bibitem{w30} A. Berera, J. Mabillard, M. Pieroni, and R.-O. Ramos,  Journal of Cosmology and Astro-Particle Physics {\bf 7}, 021 (2018).


\bibitem{w31} N. Videla and G. Panotopoulos, Phys. Rev. {\bf D 97}, 123503 (2018).


\bibitem{w32} X. Tong, Y. Wang, and S. Zhou,  Journal of Cosmology and Astro-Particle Physics {\bf 6}, 013 (2018).


\bibitem{w33} Z.-P. Peng, J.-N. Yu, X.-M. Zhang, and J.-Y.  Zhu, Phys. Rev. {\bf D 97}, 063523 (2018).


\bibitem{w34} X.-B. Li, H. Wang, and J.-Y. Zhu,  Phys. Rev. {\bf D 97}, 063516 (2018).


\bibitem{w35} Y.-Y.  Wang, J.-Y. Zhu, and X.-M. Zhang,  Phys. Rev. {\bf D 97}, 063510 (2018).


\bibitem{w36} R. Herrera, European Physical Journal {\bf C 78}, 245 (2018).


\bibitem{w37} M. Bastero-Gil, A. Berera, R.-O. Ramos, and J.-G. Rosa, Journal of High Energy Physics {\bf 1802}, 063 (2018).


\bibitem{w38} M. Bastero-Gil, S. Bhattacharya, K. Dutta, and M.-R. Gangopadhyay,  Journal of Cosmology and Astro-Particle Physics {\bf 2}, 054 (2018).


\bibitem{w39} R. Arya, A. Dasgupta, G. Goswami, J. Prasad, and R. Rangarajan,  Journal of Cosmology and Astro-Particle Physics {\bf 2}, 043 (2018).

\bibitem{w40} Y.-Y. Wang, J.-Y. Zhu, and X.-M. Zhang,   Phys. Rev. {\bf D 99}, 103529 (2019).


\bibitem{w41} M. Bastero-Gil, A. Berera, R. Hern{\'a}ndez-Jim{\'e}nez, and J.-G. Rosa, Phys. Rev. {\bf D 99}, 103520 (2019).


\bibitem{w42} J.-G. Rosa and L.-B. Ventura,  Phys. Rev. Lett. {\bf 122}, 161301 (2019).


\bibitem{w43} S. Das, Phys. Rev. {\bf D 99}, 063514 (2019).


\bibitem{w44} M. Motaharfar, V. Kamali, and R.-O.  Ramos, Phys. Rev. {\bf D 99}, 063513 (2019).


\bibitem{w45} X.-B. Li, X.-G. Zheng, and J.-Y. Zhu, Phys. Rev. {\bf D 99}, 043528 (2019).


\bibitem{w46} S. Rasouli, K. Rezazadeh, A. Abdolmaleki, and K.  Karami, European Physical Journal {\bf C 79}, 79 (2019).

\bibitem{w47} K. Dimopoulos and  L. Donaldson-Wood, Phys. Lett. {\bf B 796}, 26 (2019).

%\bibitem{rev} M. A. Amin, M. P. Hertzberg, D. I. Kaiser, and J. Karouby, Int. J. Mod. Phys. {\bf D 24}, 1530003 (2015).

\bibitem{w48} H. Sheikhahmadi, A. Mohammadi, A. Aghamohammadi, T. Harko, R. Herrera, C. Corda, A. Abebe, and K. Saaidi, European Physical Journal {\bf C 79}, 1038 (2019).

\bibitem{Zim} W. Zimdahl, J. Triginer, and D. Pavon, Phys. Rev. {\bf D 54}, 6101 (1996).

\bibitem{prigogine1988thermodynamics} I. Prigogine, J. Geheniau, E. Gunzig and P. Nardone, Proceedings of the National Academy of Sciences {\bf 85}, 7428 (1988).

\bibitem{Calv} M. Calv$\tilde{{\rm a}}$o, J. Lima and I. Waga, Physics Letters {\bf A 162}, 223 (1992).

\bibitem{Calv1}  J. A. S. Lima and A. S. M. Germano, Physics Letters {\bf A 170}, 373 (1992).

\bibitem{op0} I. Prigogine, Int. J. Theor. Phys. {\bf 28}, 927 (1989).

\bibitem{op0a} I. Prigogine, J. Geheniau, E. Gunzig and P. Nardone, Gen. Rel. Grav. {\bf 21}, 767 (1989).

\bibitem{op1} T. Harko and M. K. Mak, Astrophys. Space Science {\bf 253}, 161 (1997).

\bibitem{op2} T. Harko and M. K. Mak, Class. Quantum Grav. {\bf 16}, 2741 (1999).

\bibitem{op3} T. Harko and M. K. Mak, Gen. Relativ. Grav. {\bf 31}, 849 (1999).

\bibitem{op4} M. K. Mak and T. Harko, Class. Quantum Grav. {\bf 16}, 4085 (1999).

\bibitem{op5} M. K. Mak and T. Harko, Aust. J. Phys. {\bf 52}, 659 (1999).

\bibitem{op6} M. P. Freaza, R. S. de Souza, and I. Waga,  Phys. Rev. {\bf D 66}, 103502 (2002).

\bibitem{op7} V. V. Papoyan, V. N. Pervushin, and D. V.  Proskurin, Astrophysics {\bf  46}, 92 (2003).

\bibitem{op8} R. Brandenberger and A. Mazumdar, JCAP {\bf 0408},  015 (2004).

\bibitem{op9} Y. Qiang, T.-J. Zhang, and Z.-L. Yi, Astrophys. Space Sci. {\bf 311}, 407 (2007).

\bibitem{op10} J. A. S. Lima, J. F. Jesus, and F. A. Oliveira, Journal of Cosmology and Astroparticle Physics {\bf 11},  027 (2010).

 \bibitem{op11} S. K. Modak and D. Singleton, Phys. Rev. {\bf D 86}, 123515 (2012).

\bibitem{op12} S. K. Modak and D. Singleton, Int. J. Mod. Phys. {\bf D 21}, 1242020 (2012).

\bibitem{op13} 	J. Chen, F. Wu, H. Yu, and Z. Li, The European Physical Journal {\bf C 72}, 1861 (2012).

 \bibitem{op14} T. Harko and F. S. N. Lobo, Phys. Rev. {\bf D 87}, 044018 (2013).

 \bibitem{op15} R. O. Ramos, M. Vargas dos Santos, and I. Waga, Phys. Rev. {\bf D 89}, 083524 (2014).

 \bibitem{op16} S. Chakraborty, Phys. Lett. {\bf B 732}, 81 (2014).

 \bibitem{op17} J. Quintin, Y.-F. Cai, and R. H. Brandenberger, Phys. Rev. {\bf D 90}, 063507 (2014).

 \bibitem{op18} S. Chakraborty and S. Saha, Phys. Rev. {\bf D 90}, 123505 (2014).

\bibitem{op19} J. C. Fabris, J. A. de Freitas Pacheco, and O. F. Piattella, JCAP {\bf 06}, 038 (2014).

\bibitem{op20} J. A. S. Lima and I. Baranov, Phys. Rev. {\bf D 90}, 043515 (2014).

\bibitem{op21} T. Harko, Phys. Rev. {\bf D 90}, 044067 (2014).

\bibitem{op22} R. O. Ramos, M. V. dos Santos, and I. Waga, Phys. Rev. {\bf D 89}, 083524 (2014).

\bibitem{op23} S. Chakraborty and S. Saha, Phys. Rev. {\bf D 90}, 123505 (2014).

\bibitem{op24} S. Pan and S. Chakraborty, Adv. High Energy Phys. {\bf 2015} 654025 (2015).

\bibitem{op25} N. Bilic and D. Tolic, Phys. Rev. {\bf D 91}, 104025 (2015).

\bibitem{op26} T. Harko, F. S. N. Lobo, J. P. Mimoso, and D. Pav\'{o}n, Eur. Phys. J. {\bf C 75}, 386 (2015).

\bibitem{op27} 	R. C. Nunes and S. Pan, Monthly Notices of the Royal Astronomical Society {\bf 459}, 673 (2016).

\bibitem{op28} V. Singh and C. P.  Singh, International Journal of Theoretical Physics {\bf 55}, 1257 (2016).

\bibitem{op29} J. A. S. Lima, S. Basilakos, and J Sol$\grave{{\rm a}}$, Eur. Phys. J. {\bf C 76}, 228 (2016).

\bibitem{op29a} M. de Campos, Int. J. Geom. Meth. Mod. Phys. {\bf 13}, 1650059 (2016).

\bibitem{op30} C. Pigozzo, S. Carneiro, J. S. Alcaniz, H. A. Borges, and J. C.  Fabris, Journal of Cosmology and Astroparticle Physics {\bf 05}, 022 (2016).

\bibitem{op31} D. C. F. Celani, N. Pinto-Neto, and S. D. P. Vitenti, Phys. Rev. {\bf D 95}, 023523 (2017).

\bibitem{op32} J. Su, T. Harko, and S.-D. Liang, Advances in High Energy Physics {\bf 2017}, 7650238 (2017).

\bibitem{op33} S. Shandera, N. Agarwal, and A. Kamal, Phys. Rev. D 98, 083535 (2018).

\bibitem{op34} C. Fey, T. Schaetz, and R. Sch\"{u}tzhold, Phys. Rev. {\bf A 98}, 033407 (2018).

\bibitem{op35} S. Pan, J. D. Barrow, and A. Paliathanasis, European Physical Journal {\bf C 79},  115 (2019).

\bibitem{op36} R. I. Ivanov and E. M. Prodanov, European Physical Journal {\bf C 79}, 973 (2019).

\bibitem{statphys} L. D. Landau and E. M. Lifshitz, Statistical Physics, Part 1, Pergamon Press, Oxford, 1980

\bibitem{Bar} J. D. Barrow and P. Saich,  Class. Quantum Grav. {\bf 10}, 279 (1993).

\bibitem{Kolb} E. W. Kolb and M. S. Turner,  {\it The Early Universe}, Addison Wesley, Redwood City (1990).

\bibitem{Higgs} G. Aad et al. [ATLAS and CMS Collaborations], Phys. Rev. Lett. {\bf 114}, 191803 (2015).

\bibitem{Bastero-Gil:2016qru}
  M.~Bastero-Gil, A.~Berera, R.~O.~Ramos and J.~G.~Rosa,
  Phys.\ Rev.\ Lett.\  {\bf 117}, 151301 (2016).

\bibitem{Berera:2018tfc}
  A.~Berera, J.~Mabillard, M.~Pieroni and R.~O.~Ramos, JCAP {\bf 1807}, 021 (2018).

\bibitem{pl2018a} Y. Akrami et al. [Planck Collaboration], "Planck 2018 results. X. Constraints on inflation",
arXiv:1807.06211 [astro-ph.CO]


\bibitem{planck2018b}  N. Aghanim et al. [Planck Collaboration], "Planck 2018 results. VI. Cosmological parameters",
arXiv:1807.06209 [astro-ph.CO]

\bibitem{bath}  S. Shandera, N. Agarwal, and A. Kamal, Phys. Rev. {\bf D 98}, 083535 (2018).

\bibitem{Hu} B. L. Hu and S. Sinha, Phys. Rev. {\bf D 51}, 1587 (1995).

\bibitem{oq1} H. Yu, Phys. Rev. Lett. {\bf 106}, 061101 (2011).

\bibitem{oq2} J. Hu and H. Yu, JHEP {\bf 08}, 137 (2011).

\end{thebibliography}
\end{document}